\documentclass[aps,prd,preprintnumbers,showpacs,showkeys,nofootinbib,
superscriptaddress,fleqn,floatfix,tightenlines,10pt]{revtex4-1}
\usepackage{amsmath,amsfonts,amssymb,amscd,amsxtra,amsthm}
\usepackage{graphicx}  
\usepackage{epstopdf}
\usepackage{dcolumn}  
\usepackage{bm}          
\usepackage{slashed}
\usepackage{cancel} 
\usepackage{float}  
\usepackage{mathtools}
\usepackage{amsbsy}
\usepackage{amstext}
\usepackage{threeparttable}

\usepackage[utf8]{inputenc}  
\usepackage{booktabs}
\usepackage[normalem]{ulem} 
\usepackage[dvipsnames,table,xcdraw]{xcolor} 
\usepackage{tabularx}
\usepackage{enumitem}  
\usepackage{array} 
\usepackage{slashed}
\usepackage{tikz}
\usepackage{float}
\usepackage{multirow}
\usepackage[symbol]{footmisc}
\usepackage{rotating}
\renewcommand\sout{\bgroup \color{red} \ULdepth=-.5ex \ULset}
\newcommand{\com}[1]{{\sf\color[rgb]{0,0,1}{#1}}}

\makeatletter

\begin{document}  
\preprint{INHA-NTG-03/2020}
\title{Axial-vector form factors of the baryon decuplet with flavor 
SU(3) symmetry breaking}
\author{Yu-Son Jun}
\email[E-mail: ]{ysjun@inha.edu}
\affiliation{Department of Physics, Inha University, Incheon 22212,
Republic of Korea}

\author{Jung-Min Suh}
\email[E-mail: ]{suhjungmin@inha.edu}
\affiliation{Department of Physics, Inha University, Incheon 22212,
Republic of Korea}

\author{Hyun-Chul Kim}
\email[E-mail: ]{hchkim@inha.ac.kr}
\affiliation{Department of Physics, Inha University, Incheon 22212,
Republic of Korea}
\affiliation{School of Physics, Korea Institute for Advanced Study
(KIAS), Seoul 02455, Republic of Korea}
\date{\today}
\begin{abstract}
The axial-vector form factors and axial-vector constants of the baryon
decuplet are investigated within a pion mean-field approach, which is
also known as the chiral quark-soliton model. Given an axial-vector
current with a specified flavor, there are four different form factors
of a decuplet baryon. When we consider the singlet, triplet, and octet
axial-vector currents, we have twelve different form factors for each
member of the baryon decuplet. We compute all these axial-vector form
factors of the baryon decuplet, taking into account the rotational
$1/N_c$ corrections and effects of flavor SU(3) symmetry breaking. We
find that, for a given flavor, two of the form factors for a decuplet
baryon are only independent within the present approach. We first
examine properties of the axial-vector form factors of the $\Delta^+$
isobar and $\Omega^-$ hyperon. We also compare the results of the
triplet axial-vector form factors of $\Delta^+$ with those from lattice
QCD and those of the present work for the axial-vector constants of
the baryon decuplet with the lattice data. All the results for other
members of the baryon decuplet are then presented. The results of the
axial charges are compared with those of other works. The axial masses
and axial radii are also discussed.      
\end{abstract}
\maketitle
\section{Introduction}
The axial-vector current probes multi-faceted structures of a
baryon. For example, the flavor singlet axial-vector constant of the
proton provides its spin content, which is identified as the first
moment of the longitudinally polarized spin structure 
function of the proton~\cite{Aidala:2012mv}. On the other hand, it is
well known that the triplet axial-vector constant or the axial charge
contains essential information on the neutron beta decay. Moreover,
this axial charge can be related to the $\pi NN$ coupling 
constant by the Goldberger-Treiman
relation~\cite{Goldberger:1958tr}. It indicates  
that the axial-vector constants play a very important role in
understanding the structure of a baryon both in strong and weak
interactions. While the axial-vector structures of the baryon octet
are relatively well known by their semileptonic decays, those of the
baryon decuplet are still much less understood, since almost all
members of the decuplet decay strongly except for the $\Omega^{-}$
baryon. Thus, it is very difficult to get access to the internal structure
of them. However, since the lattice data on the axial-vector form
factors and the axial-vector constants of the baryon decuplet are now 
available~\cite{Alexandrou:2013opa, Alexandrou:2016xok}, we anticipate
that lattice QCD will provide more information on the axial-vector
structure of the baryon decuplet in the near future. While it is
rather difficult to measure the axial-vector properties of the baryon
decuplet experimentally, there have been various theoretical
works. For instance, the axial charge of the $\Delta$ isobar was
studied in chiral perturbation theory~\cite{Jenkins:1991es,
  Jiang:2008we}. In Refs.~\cite{Choi:2010ty, Choi:2013ysa}, the axial
charges of the $\Delta$, $\Sigma^*$, and $\Xi^*$ were computed within
the Goldstone-boson-exchange relativistic constituent quark model
(RCQM). Recently, the axial-vector form factors and the
axial-vector constants of the baryon decuplet were derived from lattice
QCD~\cite{Alexandrou:2013opa, Alexandrou:2016xok}. The axial charge of
$\Delta^+$ was also studied in the light-cone sum rules
(LCSR)~\cite{Kucukarslan:2014bla}. Very recently, the axial charges of
the baryon decuplet except for the $\Omega^{-}$ baryon were also
calculated in a pertubative chiral quark model 
(PCQM)~\cite{Liu:2018jiu}. 

In the present work, we want to investigate the axial-vector form
factors of the baryon decuplet within the framework of the chiral
quark-soliton model ($\chi$QSM)~\cite{Diakonov:1987ty, 
Diakonov:1997sj, Blotz:1992pw}. The model is based on the pion
mean-field approach that was proposed ingeniously by
Witten~\cite{Witten:1979kh, Witten:1983tx}. In the limit of the
large number of colors ($N_c\to\infty$), a baryon can be viewed as a
bound state of the $N_c$ valence quarks by a pion mean field. The
presence of the $N_c$ valence quarks brings about the vacuum
polarization that creates a pion mean field. Then the pion mean field 
influences the valence quarks self-consistently. As a result, they 
are bound by the pion mean field, so that a baryon emerges as a 
bound state of the $N_c$ valence quarks in the form of a chiral
soliton. This $\chi$QSM has been successfully applied to 
describe various properties of the lowest-lying baryons including both
the light and singly heavy  baryons. For example, the model explains
very well the electromagnetic form factors of the baryon octet and
decuplet~\cite{Kim:1995mr, Ledwig:2008es, Kim:2019gka}, the
axial-vector form factors of the nucleon~\cite{Silva:2005fa},  
the scalar form factor~\cite{Kim:1995hu}, tensor charges
and tensor form factors~\cite{Kim:1995bq, Kim:1996vk, Ledwig:2010tu,
  Ledwig:2010zq}, the gravitational form factors~\cite{Goeke:2007fp},
and so on. Very recently, the model was extended to the description of
singly heavy baryons. For example, the electromagnetic form
factors of the singly heavy baryons both with spin 1/2 and 3/2 were
investigated~\cite{Kim:2018nqf, Kim:2019wbg}. Thus, we want to examine
the axial-vector form factors of the baryon decuplet within the
same framework, focusing on the comparison of the present results with
those from the lattice QCD~\cite{Alexandrou:2013opa,
  Alexandrou:2016xok}.  

The present paper is organized as follows: In Section II, we 
recapitulate the axial-vector form factors of the baryon
decuplet. In Section III, we show succinctly how to compute them
within the framework of the $\chi$QSM. In Section IV, we first present
the results of the axial-vector form factors of the $\Delta^+$ isobar 
and $\Omega^{-}$ hyperon, scrutinizing the effects of flavor SU(3)
symmetry breaking. In order to compare the present results with those
from the lattice data, we first derive the form factors with the pion
mass varied from the physical value to unphysical ones. The results
are then compared with those from the lattice QCD with the
corresponding values of the pion mass employed. 
We show the results of the axial-vector form factors of all other
members of the baryon decuplet, emphasizing the effects of flavor
SU(3) symmetry breaking. Finally, we show the results for the axial
charges in comparison with those from other approaches. The results of
the axial masses and axial radii are also presented.  
In the last Section we summarize the present work and give outlook for
future works.  

\section{Axial-vector form factors of the baryon decuplet}
\label{sec:2}
The axial-vector current is defined as 
\begin{align}
A^{a}_\mu (x) = \bar{\psi} (x) \gamma_\mu \gamma_{5} 
\frac{\lambda^{a}}{2} \psi(x),
\label{eq:AxialCUR}
\end{align}
where $\psi(x)$ denotes the quark field $\psi=(u,\,d,\,s)$ in flavor
space. The $\lambda^{a}$ stand for the well-known the flavor SU(3)
Gell-Mann matrices. The superscript $a$ represents one of
$a=0,\,3,\,8$ that correspond to the singlet, triplet, and octet
currents, respectively. 
By using the Lorentz structure and the consideration of spin, parity,
and charge conjugation, we can parametrize the matrix element of the
axial-vector current between the baryon 
decuplet with spin 3/2 in terms of four different real form 
factors~\cite{Alexandrou:2013opa,Scadron:1968zz}:
\begin{align}
\langle B(p',J_3') | A_\mu^{a}(0) | B(p,J_3) \rangle 
&= - \overline{u}^{\alpha}(p',J_3') \left[ \gamma_{\mu} \left 
  \{g^{(a)B}_{1}(q^2) \eta_{\alpha \beta} + h^{(a)B}_{1}(q^2) 
  \frac{ q_{\alpha} q_{\beta}}{4M_{B}^{2}} \right \}\right. \cr
&\hspace{2cm} \left. + \,\frac{q_{\mu}}{2M_{B}}
  \left \{ g^{(a)B}_{3}(q^2) \eta_{\alpha \beta} + h^{(a)B}_{3} (q^2)
  \frac{q_{\alpha}q_{\beta}}{4 M_{B}^2} \right \} \right ] 
  \gamma^{5} {u}^{\beta}(p,J_3), 
\label{eq:MatrixEl1}
\end{align}
where $M_B$ is the mass of the baryon involved. $\eta_{\alpha \beta}$
represents the metric tensor of Minkowski space, expressed as
$\eta_{\alpha \beta} =\mathrm{diag}(1,\,-1,\,-1,\,-1)$. $q_\alpha$
designates the momentum transfer $q_\alpha=p'_\alpha-p_\alpha$ and its
square is given as $q^2=-Q^2$ with $Q^2 >0$. $J_3$ ($J_3'$) is the
eigenvalue of the third component of the spin operator $\bm{J}$
($\bm{J}'$), which is projected along the direction of the momentum
$\bm{p}$ ($\bm{p}'$). $u^\alpha (p,\,J_3)$ is the Rarita-Schwinger
spinor that describes a decuplet baryon with spin
3/2~\cite{Rarita:1941mf}, carrying the momentum $p$ and $J_3$, which
can be described by the combination of the polarization vector and the
Dirac spinor, $u^\alpha (p,\,J_3)= 
\sum_{i,s} C^{\frac{3}{2} J_3}_{1i\,\frac{1}{2}s} \epsilon^{\alpha}_{i}
(p) u_{s}(p)$. It satisfies the Dirac equation and the auxiliary 
equations $p_\alpha u^\alpha(p,J_3)=0$ and $\gamma_\alpha 
u^\alpha(p,J_3)=0$.

In the Breit frame, the form factors defined in Eq
~\eqref{eq:MatrixEl1} are expressed in terms of the 
the spatial parts of the axial-vector current projected by the 
spherical basis vectors $\bm{e}_{n}$ ~\cite{Jones:1972ky, Weber:1978dh}
\begin{align}
g^{(a)B}_{1}(Q^2) &= -\sqrt{\frac{3}{2}} \frac{M_{B}}{E_{B}}
\langle B\left (p',3/2) | \bm{e}_{1} \cdot
    \bm{A}^{a}(0) | B(p, 1/2\right)\rangle , \cr  
h^{(a)B}_{1}(Q^2) &= -\sqrt{\frac{3}{2}} \frac{4M^{5}_{B}}{E_{B}^{3}Q^{2}}
 \left[ \frac{2M^{2}_{B} + Q^{2}}{2M^{2}_{B}} 
 \langle B(p',3/2) | \bm{e}_{1} \cdot \bm{A}^{a}(0) | B(p,1/2) 
  \rangle \right. \cr
&\hspace{2.6cm}-\left. \frac{\sqrt{3}}{2} \langle B(p',1/2) |
  \bm{e}_{1} \cdot \bm{A}^{a}(0) | B(p,-1/2) \rangle \right], \cr
g^{(a)B}_{3}(Q^2)&= - \frac{4M^{2}_{B}}{Q^{2}} 
  \left[ \langle B(p',3/2) | \bm{e}_{0} \cdot \bm{A}^{a}(0) |
  B(p,3/2) \rangle - \sqrt{\frac{3}{2}} \frac{M_{B}}{E_{B}}  
  \langle B(p',3/2) | \bm{e}_{1} \cdot \bm{A}^{a}(0) |
  B(p,1/2) \rangle \right] , \cr
h^{(a)B}_{3}(Q^2) &= \frac{8 M^{6}_{B}}{E_{B}^{2} Q^{4}} \left[ 
  3 \langle B(p',1/2) | \bm{e}_{0} \cdot \bm{A}^{a}(0) | B(p,1/2) 
  \rangle -\frac{\sqrt{3}(2M^{2}_{B} + Q^{2})}{\sqrt{2} E_{B} M_{B}} 
  \langle B(p',3/2) | \bm{e}_{1} \cdot \bm{A}^{a}(0) | B(p,1/2) 
  \rangle \right. \cr
&\hspace{1.cm} \left. + \frac{3 M_{B}}{\sqrt{2} E_{B}} 
  \langle B(p',1/2) | \bm{e}_{1} \cdot \bm{A}^{a}(0) | B(p,-1/2) 
  \rangle - \frac{M^{2}_{B} + Q^{2}}{M^{2}_{B}} 
  \langle B(p',3/2) | \bm{e}_{0} \cdot \bm{A}^{a}(0) | B(p,3/2) 
  \rangle \right] ,
\label{eq:MatrixEl2}
\end{align}
where $E_{B}$ denotes the energy of the corresponding baryon,
i.e. $E_{B}=\sqrt{M^{2}_{B} +Q^{2}/4}$, and $\bm{e}_{n}$ are expressed
explicitly in terms of the Cartesian basis vectors
$\bm{e}_{0}=\hat{\bm{z}}$, $\bm{e}_{1}=   
-(\hat{\bm{x}}+i\hat{\bm{y}})/ \sqrt{2}$, $\bm{e}_{-1}=
(\hat{\bm{x}}-i\hat{\bm{y}})/ \sqrt{2}$.
We want to mention that the form factors $h_{1,3}^{(a)B} (Q^{2})$ 
are in fact the same as $g_{1,3}^{(a)B} (Q^{2})$ apart 
from the kinematical factors. 

\section{Axial-vector form factors in the chiral quark-soliton model} 
\label{sec:3}
The $\chi$QSM is constructed by the effective chiral action as a
functional of the pseudo-Nambu-Goldstone (pNG) field $\pi^a$ given as 
\begin{align}
S_{\mathrm{eff}}[\pi^{a}] \;=\; -N_{c}\mathrm{Tr}\ln D, 
\label{eq:echa}
\end{align}
where $\mathrm{Tr}$ represents the functional trace running over
four dimensional Euclidean space-time, spin, flavor and color
spaces. The $N_c$ is the number of colors. $D$ designates the
one-body Dirac operator defined by 
\begin{align}
D := i\slashed{\partial} + i M U^{\gamma_5} + i \hat{m},
\label{eq:Dirac}
\end{align}
where $M$ stands for the dynamical quark 
mass and $U^{\gamma_5}(x)$ denotes the flavor SU(3) chiral field
defined by 
\begin{align}
U^{\gamma_5}(x) := \frac{1+\gamma_5}{2} U(x) + \frac{1-\gamma_5}{2}
  U^\dagger (x)
\end{align}
where $U(x) = \exp(i\lambda^a \pi^a (x) /f_\pi)$ with 
the pion decay constant $f_\pi$. $\hat{m}$ in Eq.~\eqref{eq:Dirac}
represents the current quark mass matrix given as 
$\hat{m}=\mathrm{diag}(m_{\mathrm{u}},\, m_{\mathrm{d}},\,
m_{\mathrm{s}})$ in flavor space. 
We assume the isospin symmetry in this work, so that the current quark
mass of the up and down quarks are set equal to eath other,
i.e. $m_{\mathrm{u}}=m_{\mathrm{d}}$ with their average mass 
$\overline{m}=(m_{\mathrm{u}} + m_{\mathrm{d}})/2$. Then, the current 
quark mass matrix is written as $\hat{m} =
\mathrm{diag}(\overline{m},\, \overline{m},\, m_{\mathrm{s}}) =
\overline{m} +\delta m$.
$\delta m$ includes the mass of the strange current quark, which can be 
decomposed as 
\begin{align}
\delta m \;=\; m_{1} \bm{1} + m_{8} \lambda^{8}, 
\label{eq:deltam}
\end{align}
where $m_1$ and $m_8$ represent the singlet and octet components of 
the current quark masses respectively:
$m_1=(-\overline{m}+m_{\mathrm{s}})/3$ and
$m_8=(\overline{m}-m_{\mathrm{s}})/\sqrt{3}$. 
The Dirac operator~\eqref{eq:Dirac} with $\gamma_4$ can be written as
\begin{align}
\gamma_4 D = -i\partial_{4} + h(U(\pi^{a})) - \delta m,
\end{align}
where $\partial_4$ stands for the time derivative in Euclidean space. 
$h(U)$ is called the one-body Dirac Hamiltonian written as 
\begin{align}
h(U) \;=\;
i\gamma_{4}\gamma_{i}\partial_{i}-\gamma_{4}MU^{\gamma_{5}} -
\gamma_{4} \overline{m}\, .
\label{eq:DiracHam}  
\end{align}
The presence of the $\overline{m}$ in the Hamiltonian is required to
reproduce correctly the Yukawa tail of the pion field, which plays an
essential role in describing the isovector charge radii of the
proton~\cite{Beg:1973sc}. 

In flavor SU(3), we need to incorporate the hedgehog structure of the
pion field~\cite{Pauli:1942kwa, Witten:1983tx} by embedding the SU(2) 
$U_{\mathrm{SU(2)}}(x)$ field into SU(3) such that the hedgehog symmetry
is preserved. The pion field with hedgehog symmetry is expressed as 
\begin{align}
\pi^i = n^i P(r), \; i=1,\,2,\,3,
\end{align}
where $n^i=x^i/r$ with $r=|\bm{x}|$ and $P(r)$ represents the profile
function of the chiral soliton. All other components of $\pi^a$ are
set equal to zero. Thus, to preserve this hedgehog symmetry, the
SU(3) $U(x)$ field can be constructed by the trivial
embedding~\cite{Witten:1983tx} 
\begin{align}
U(x) = \exp(i\pi^a \lambda^a/f_\pi) = 
  \begin{pmatrix}
    \exp(i\bm{n}\cdot \bm{\tau}P(r)/f_\pi) & 0 \\ 0 & 1
  \end{pmatrix}.
\end{align}
In the pion mean-field approximation, the pion mean field arises as the
solution of the classical equation of motion, which is derived from 
$\delta S_{\mathrm{eff}}/\delta P(r) =0$. The equation of motion can
be solved self-consistently, which resembles the Hartree approximation
in many-body problems. 

We can derive the matrix elements of the axial-vector
current~\eqref{eq:MatrixEl1} by using the functional integral 
\begin{align}
&\langle B(p',\, J_3') | A^{a}_\mu(0) |B(p,\,J_3)\rangle =
  \frac1{\mathcal{Z}} \lim_{T\to\infty} \exp\left(i p_4\frac{T}{2} 
  - i p_4' \frac{T}{2}\right) \int d^3x d^3y \exp(-i \bm{p}'\cdot 
  \bm{y} + i \bm{p}\cdot \bm{x}) \cr
&\hspace{1cm}\times \int \mathcal{D} \pi^a \int \mathcal{D} 
  \psi \int \mathcal{D} \psi^\dagger 
J_{B}(\bm{y},\,T/2) \psi^\dagger(0) 
  \gamma_4\gamma_\mu \gamma_{5} \frac{\lambda^{a}}{2} \psi(0) 
  J_B^\dagger (\bm{x},\,-T/2) \exp\left[-\int d^4 r \psi^\dagger 
  iD(\pi^a) \psi\right],  
\label{eq:correlftn}
\end{align}
where the baryon states $|B(p,\,J_3)\rangle$ and $\langle
B(p',\,J_3')|$ are respectively written in terms of Ioffe-type
baryonic currents
\begin{align}
|B (p,\,J_3)\rangle &= \lim_{x_4\to-\infty} \exp(i p_4 x_4)
  \frac1{\sqrt{\mathcal{Z}}} \int d^3 x \exp(i\bm{p}\cdot \bm{x}) 
  J_B^\dagger (\bm{x},\,x_4)|0\rangle,\cr
\langle B(p',\,J_3')| &= \lim_{y_4\to\infty} \exp(-i p_4' y_4)
  \frac1{\sqrt{\mathcal{Z}}} \int d^3 y \exp(-i\bm{p}'\cdot \bm{y}) 
  \langle 0| J_B (\bm{y},\,y_4),
\label{eq:correl}
\end{align}
where $J_B(x)$ denotes the Ioffe-type current consisting of $N_c$ valence
quarks~\cite{Ioffe:1981kw}  
\begin{align}
J_B(x) = \frac1{N_c!} \epsilon_{i_1\cdots i_{N_c}} \Gamma_{JJ_3
  TT_3 Y}^{\alpha_1\cdots \alpha_{N_c}} \psi_{\alpha_1 i_1} (x)
  \cdots \psi_{\alpha_{N_c} i_{N_c}}(x),  
\end{align}
with spin-flavor and color indices $\alpha_1\cdots \alpha_{N_c}$
and $i_1\cdots i_{N_c}$, respectively. The matrices
$\Gamma_{JJ_3 TT_3 Y}^{\alpha_1\cdots \alpha_{N_c}}$ secure the
baryon state with pertinent quantum numbers $JJ_3TT_3Y$ by projecting
out. Similarly, we can express the creation current operator
$J_B^\dagger(x)$~\cite{Diakonov:1987ty, Christov:1995vm}.

In order to quantize the chiral soliton, we have to perform the
functional integral over the pNG fields. Since we use the pion
mean-field approximation or the saddle-point approximation, we neglect
the $1/N_c$ pion-loop corrections. However, we have to take into
account the zero modes that do not change the energy of the
soliton. Thus, the functional integral over the $U$ field is replaced
by those over both the rotational and translational zero modes. We
refer to Ref.~\cite{Kim:1995mr} for details. The integral over
the translational zero modes yields naturally the Fourier transform,
which indicates that the baryon state has the proper translational
symmetry. On the other hand, by performing the rotational zero-mode
quantization, we can restore the  rotational symmetries. Thus, the
zero-mode quantization leads to the collective Hamiltonian 
\begin{align}
H_{\mathrm{coll}} = H_{\mathrm{sym}} + H_{\mathrm{sb}},
\end{align}
where
\begin{align}
  \label{eq:Hamiltonian}
H_{\mathrm{sym}} = M_{\mathrm{cl}} + \frac1{2I_1} \sum_{i=1}^3
  \hat{J}_i^2 + \frac1{2I_2} \sum_{p=4}^7 \hat{J}_p^2,\quad
H_{\mathrm{sb}} = \alpha D_{88}^{(8)} + \beta \hat{Y} +
  \frac{\gamma}{\sqrt{3}} \sum_{i=1}^3 D_{8i}^{(8)} \hat{J}_i.
\end{align}
Here, $I_1$ and $I_2$ represent the moments of inertia for the soliton
and $D^{(8)}_{ab}$ denote SU(3) Wigner $D$ functions. 
The inertial parameters $\alpha$, $\beta$ and $\gamma$, which break
flavor SU(3) symmetry explicitly, are expressed in terms of the
moments of inertia $I_1$ and $I_2$, and the anomalous moments of
inertia $K_1$ and $K_2$
\begin{align}
\alpha=\left (-\frac{{\Sigma}_{\pi N}}{3\overline{m}}
  +\frac{K_{2}}{I_{2}} \right )m_{\mathrm{s}},
  \quad \beta=-\frac{ K_{2}}{I_{2}}m_{\mathrm{s}}, 
  \quad \gamma=2\left ( \frac{K_{1}}{I_{1}}-\frac{K_{2}}{I_{2}}
  \right ) m_{\mathrm{s}},
\label{eq:alphaetc}  
\end{align}
where $\Sigma_{\pi N}$ stands for the pion-nucleon $\Sigma$ term. 
In the presence of the flavor SU(3) symmetry breaking term
$H_{\mathrm{sb}}$, the collective wavefunctions of the baryon decuplet
are no more in pure states but are mixed with states in higher
representations. The states of the baryon decuplet are then obtained
by the standard second-order perturbation theory:
\begin{align}
|B_{{\bm10}_{3/2}}\rangle = |{\bm{10}}_{3/2},B\rangle + 
  a^{B}_{{27}}|{{\bm{27}}}_{3/2},B\rangle + 
  a^{B}_{{35}}|{{\bm{35}}}_{3/2},B\rangle
\label{eq:mixedWF1}
\end{align}
with the mixing coefficients
\begin{eqnarray}
a_{{27}}^{B}
\;=\;
a_{{27}}\left[\begin{array}{c}
\sqrt{15/2}\\
2 \\
\sqrt{3/2} \\
0
\end{array}\right], 
& 
a_{35}^{B}
\;=\; 
a_{35}\left[\begin{array}{c}
5/\sqrt{14}\\
2\sqrt{{5}/{7}} \\
3\sqrt{{5}/{14}} \\
2\sqrt{{5}/{7}}
\end{array}\right], 
\label{eq:pqmix}
\end{eqnarray}
respectively, in the basis
$\left[\Delta,\;\Sigma^{*},\;\Xi^{*},\;\Omega\right]$. The 
parameters $a_{{27}}$ and $a_{35}$ are written as
\begin{eqnarray}
a_{27} \;=\;
{\displaystyle -\frac{{I}_{2}}{8} \left ( \alpha + \frac{5}{6}
  \gamma \right)}, & 
a_{35} \;=\; {\displaystyle -\frac{{I}_{2}}{24} \left( \alpha 
  -\frac{1}{2}\gamma \right)}, 
\label{eq:pqmix2}
\end{eqnarray}
which have been already determined numerically in
Ref.~\cite{Kim:2018xlc}: $a_{27}=0.126$ and $a_{35}=0.035$. Each state
in Eq.~\eqref{eq:mixedWF1} is given in terms of the SU(3) Wigner $D$
functions in such a way that they satisfy the quantization
condition~\cite{Blotz:1992pw}.

Having calculated Eq.~\eqref{eq:correlftn} with the zero-mode
quantizations, we can derive the final expressions of the axial-vector
form factors
\begin{align}
g^{(0)B}_{1} (Q^{2}) =& \frac{\langle \hat{J}_{3} \rangle}{3 I_{1}} 
  \{ \mathcal{B}^{B}_{0}(Q^{2})-\mathcal{B}^{B}_{2}(Q^{2}) \} 
  +\frac{2m_{\mathrm{s}}}{3 \sqrt{3}} \langle D^{(8)}_{83} \rangle 
  \left[ \frac{K_{1}}{ I_{1}} \{ \mathcal{B}^{B}_{0}(Q^{2})
  -\mathcal{B}^{B}_{2}(Q^{2}) \} - \{ \mathcal{I}^{B}_{0}(Q^{2})
  -\mathcal{I}^{B}_{2}(Q^{2}) \} \right], 
\label{eq:gasinglet1}\\
g^{(0)B}_{3} (Q^{2}) =& \frac{\langle \hat{J}_{3} \rangle}{3 I_{1}} 
  \{ \mathcal{B}^{\prime B}_{0}(Q^{2}) +\mathcal{B}^{\prime B}_{2}(Q^{2}) 
  \} +\frac{2m_{\mathrm{s}}}{3 \sqrt{3}} \langle D^{(8)}_{83} \rangle 
  \left[ \frac{K_{1}}{ I_{1}} \{ \mathcal{B}^{\prime B}_{0}(Q^{2})
  +\mathcal{B}^{\prime B}_{2}(Q^{2}) \} - \{\mathcal{I}^{\prime B}_{0}
  (Q^{2}) +\mathcal{I}^{\prime B}_{2}(Q^{2}) \} \right],
\label{eq:gasinglet3}
\end{align}
for the flavor singlet,
\begin{align}
g^{(a)B}_{1} (Q^{2}) =& \frac{ \langle D^{(8)}_{a3} \rangle}{3} 
  \{\mathcal{A}^{B}_{0}(Q^{2}) -\mathcal{A}^{B}_{2}(Q^{2})\} 
  +\frac{1}{3\sqrt{3} I_{1}} \left[ \langle D^{(8)}_{a8} \hat{J}_{3} 
  \rangle +\frac{2m_{\mathrm{s}}}{\sqrt{3}} K_{1} \langle D^{(8)}_{83} 
  D^{(8)}_{a8} \rangle \right] 
  \{\mathcal{B}^{B}_{0}(Q^{2})-\mathcal{B}^{B}_{2}(Q^{2})\} \cr 
& +\frac{d_{pq3}}{3 I_{2}} \left[ \langle D^{(8)}_{ap} \hat{J}_{q} 
  \rangle +\frac{2m_{\mathrm{s}}}{\sqrt{3}} K_{2} \langle D^{(8)}_{ap} 
  D^{(8)}_{8q} \rangle \right]
  \{\mathcal{C}^{B}_{0}(Q^{2}) -\mathcal{C}^{B}_{2}(Q^{2})\}
  -\frac{i \langle D^{(8)}_{a3} \rangle}{6I_{1}} 
  \{\mathcal{D}^{B}_{0}(Q^{2})-\mathcal{D}^{B}_{2}(Q^{2})\} \cr
& +\frac{2 m_{\mathrm{s}}}{9} ( \langle D^{(8)}_{a3} \rangle - \langle 
  D^{(8)}_{88}D^{(8)}_{a3} \rangle) \{\mathcal{H}^{B}_{0}(Q^{2})
  -\mathcal{H}^{B}_{2}(Q^{2})\} -\frac{2 m_{\mathrm{s}}}{9} 
  \langle D^{(8)}_{83} D^{(8)}_{a8} \rangle 
  \{\mathcal{I}^{B}_{0}(Q^{2})-\mathcal{I}^{B}_{2}(Q^{2})\} \cr 
& -\frac{2 m_{\mathrm{s}}}{3\sqrt{3}} d_{pq3} \langle D^{(8)}_{ap} 
  D^{(8)}_{8q} \rangle \{\mathcal{J}^{B}_{0}(Q^{2})
  -\mathcal{J}^{B}_{2}(Q^{2})\},
\label{eq:gatrioct1} \\
g^{(a)B}_{3} (Q^{2}) =& \frac{ \langle D^{(8)}_{a3} \rangle}{3} 
  \{\mathcal{A}^{\prime B}_{0}(Q^{2}) +\mathcal{A}^{\prime B}_{2}
  (Q^{2})\} +\frac{1}{3\sqrt{3} I_{1}} \left[ \langle D^{(8)}_{a8} 
  \hat{J}_{3} \rangle +\frac{2m_{\mathrm{s}}}{\sqrt{3}} K_{1} \langle 
  D^{(8)}_{83} D^{(8)}_{a8} \rangle \right] \{\mathcal{B}^{\prime B}_{0}
  (Q^{2})+\mathcal{B}^{\prime B}_{2}(Q^{2})\} \cr 
& +\frac{d_{pq3}}{3 I_{2}} \left[ \langle D^{(8)}_{ap} \hat{J}_{q} \rangle
  +\frac{2m_{\mathrm{s}}}{\sqrt{3}} K_{2} \langle D^{(8)}_{ap} 
  D^{(8)}_{8q} \rangle \right]
  \{\mathcal{C}^{\prime B}_{0}(Q^{2}) +\mathcal{C}^{\prime B}_{2}(Q^{2})\}
  -\frac{i \langle D^{(8)}_{a3} \rangle}{6I_{1}} 
  \{\mathcal{D}^{\prime B}_{0}(Q^{2})+\mathcal{D}^{\prime B}_{2}
  (Q^{2})\} \cr
& +\frac{2 m_{\mathrm{s}}}{9} ( \langle D^{(8)}_{a3} \rangle - \langle 
  D^{(8)}_{88}D^{(8)}_{a3} \rangle)  
  \{\mathcal{H}^{\prime B}_{0}(Q^{2})+\mathcal{H}^{\prime B}_{2}
  (Q^{2})\} -\frac{2 m_{\mathrm{s}}}{9} 
  \langle D^{(8)}_{83} D^{(8)}_{a8} \rangle 
  \{\mathcal{I}^{\prime B}_{0}(Q^{2})+\mathcal{I}^{\prime B}_{2}
  (Q^{2})\} \cr 
& -\frac{2 m_{\mathrm{s}}}{3\sqrt{3}} d_{pq3} \langle D^{(8)}_{ap} 
  D^{(8)}_{8q} \rangle
  \{\mathcal{J}^{\prime B}_{0}(Q^{2})+\mathcal{J}^{\prime B}_{2}(Q^{2})\},
\label{eq:gatrioct3}
\end{align}
for the non-singlet. The indices $p$ and $q$ run over
4 to 7. The $d_{apq}$ stand for the SU(3) symmetric tensors.
$\mathcal{A}^{B}_{0(2)}$ to $\mathcal{J}^{B}_{0,(2)}$ and 
$\mathcal{A}^{\prime B}_{0(2)}$ 
to $\mathcal{J}^{\prime B}_{0,(2)}$ represent components of the 
axial-vector form factors, 
of which the explicit expressions can be found in Appendix~\ref{app:A} 
and Ref.~\cite{Silva:2005fa,Ledwig:2008rw}. The $\langle \cdots \rangle$ 
are just the short-handed notations for the matrix elements of the SU(3) 
Wigner $D$ function between the decuplet baryons. The explicit results 
for the matrix elements of the SU(3) Wigner $D$ function can be found in 
Appendix~\ref{app:B}.
We have also considered the symmetry-conserving 
quantization~\cite{Praszalowicz:1998jm}, which makes it possible to
remove redundant terms by using the limit of the nonrelativistic quark
model. 

The contributions coming from the flavor SU(3) symmetry breaking consist
of two different terms, i.e. that from the effective chiral action and
that from the collective wavefunctions, which are decomposed as
\begin{align}
g^{(a)B}_{1(3)} (Q^{2}) &= (g^{(a)B}_{1(3)} (Q^{2}))^{(\mathrm{sym})} +
(g^{(a)B}_{1(3)} (Q^{2}))^{(\mathrm{op})} + (g^{(a)B}_{1(3)} 
(Q^{2}))^{(\mathrm{wf})}, 
\label{eq:ga_decompose}
\end{align}
where $(g^{(a)B}_{1(3)} (Q^{2}))^{(\mathrm{sym})}$, $(g^{(a)B}_{1(3)}
(Q^{2}))^{(\mathrm{op})}$, 
and $(g^{(a)B}_{1(3)} (Q^{2}))^{(\mathrm{wf})}$ correspond
respectively to the flavor SU(3) symmetric term, the flavor SU(3)
symmetry-breaking term from the effective chiral action and the
collective wavefunctions. The singlet axial-vector form factors
($a=0$) can be written as
\begin{align}
(g^{(0)B}_{1} (Q^{2}))^{(\mathrm{sym})} =& \frac{J_{3}}{3 I_{1}} 
  \{\mathcal{B}^{B}_{0}(Q^{2})-\mathcal{B}^{B}_{2}(Q^{2})\},
\label{eq:ga10leading} \\
(g^{(0)B}_{3} (Q^{2}))^{(\mathrm{sym})} =& \frac{J_{3}}{3 I_{1}} 
  \{\mathcal{B}^{\prime B}_{0}(Q^{2})+\mathcal{B}^{\prime B}_{2}
  (Q^{2})\}, 
\label{eq:ga30leading} \\
(g^{(0)B}_{1} (Q^{2}))^{(\mathrm{op})} =& -\frac{m_{\mathrm{s}} Y}{12}
  \left[ \frac{K_{1}}{I_{1}} \{\mathcal{B}^{B}_{0}(Q^{2})
  -\mathcal{B}^{B}_{2}(Q^{2})\}
  -\{\mathcal{I}^{B}_{0}(Q^{2})-\mathcal{I}^{B}_{2}(Q^{2})\} \right], 
\label{eq:ga10opcorr}\\
(g^{(0)B}_{3} (Q^{2}))^{(\mathrm{op})} =& -\frac{m_{\mathrm{s}} Y}{12}
  \left[ \frac{K_{1}}{I_{1}} \{\mathcal{B}^{\prime B}_{0}(Q^{2})
  +\mathcal{B}^{\prime B}_{2}(Q^{2})\}
  -\{\mathcal{I}^{\prime B}_{0}(Q^{2})+\mathcal{I}^{\prime B}_{2}
  (Q^{2})\} \right], 
\label{eq:ga30opcorr} \\
(g^{(0)B}_{1} (Q^{2}))^{(\mathrm{wf})} =& 0.
\label{eq:ga10wfcorr}\\
(g^{(0)B}_{3} (Q^{2}))^{(\mathrm{wf})} =& 0.
\label{eq:ga30wfcorr}
\end{align}
Note that there are no $N_c$ leading-order contributions to the singlet 
axial-vector form factors, which is the well-known fact from any
chiral solitonic models. For example, a simple chirally-symmetric
version of the Skyrme model yields the null result of 
$g_A^{(0)}$~\cite{Brodsky:1988ip}.
Since the singlet axial-vector constant of
the proton is just its quark spin content, chiral solitonic models
explain rather well the reason why the quark spin content of the
proton turns out very small experimentally. 
Moreover, the linear $m_{\mathrm{s}}$ corrections from
the collective wavefunctions also vanish. However, as we will show
later explicitly, the linear $m_{\mathrm{s}}$ corrections
contribute most dominantly to the $g_3^{(0)} (Q^2)$ form 
factors of the baryon decuplet except for those with hypercharge $Y=0$.
As written in Eqs.~\eqref{eq:ga10leading} and~\eqref{eq:ga30leading},
the rotational $1/N_c$ corrections are flavor independent, i.e. they
contribute equally to $g_3^{(0)} (Q^2)$ form factors 
for any hyperons in the decuplet.

The triplet axial-vector form factors ($a=3$) are expressed as 
\begin{align}
(g^{(3)B}_{1} (Q^{2}))^{(\mathrm{sym})} =&- \frac{T_{3}}{24}
  \left[ 2 \{\mathcal{A}^{B}_{0}(Q^{2})-\mathcal{A}^{B}_{2}(Q^{2})\}
  -\frac{\mathcal{B}^{B}_{0}(Q^{2})-\mathcal{B}^{B}_{2}(Q^{2})}{I_{1}} 
  \right. \cr
&\hspace{1.2cm}
   \left. -\frac{\mathcal{C}^{B}_{0}(Q^{2})
   -\mathcal{C}^{B}_{2}(Q^{2})}{I_{2}}  
  -\frac{i\{\mathcal{D}^{B}_{0}(Q^{2})
  -\mathcal{D}^{B}_{2}(Q^{2})\}}{I_{1}} \right], 
\label{eq:ga13leading} \\
(g^{(3)B}_{3} (Q^{2}))^{(\mathrm{sym})} =& -\frac{T_{3}}{24}
  \left[ 2\{\mathcal{A}^{\prime B}_{0}(Q^{2})
  +\mathcal{A}^{\prime B}_{2}(Q^{2})\} 
  -\frac{\mathcal{B}^{\prime B}_{0}(Q^{2})
  +\mathcal{B}^{\prime B}_{2}(Q^{2})}{I_{1}} \right. \cr
&\hspace{1.2cm} \left. -\frac{\mathcal{C}^{\prime B}_{0}(Q^{2}) 
  +\mathcal{C}^{\prime B}_{2}(Q^{2})}{I_{2}}  
  -\frac{i \{\mathcal{D}^{\prime B}_{0}(Q^{2})
  +\mathcal{D}^{\prime B}_{2} (Q^{2})\} }{I_{1}} \right],
\label{eq:ga33leading}
\end{align}
\begin{align}
(g^{(3)B}_{1} (Q^{2}))^{(\mathrm{op})} =& -\frac{m_{\mathrm{s}} T_{3}}{378} 
  \Bigg[ \left( \begin{array}{c} 5 \\ 3 \\ 1 \\ 0 \end{array} \right)
  \left\{ \frac{K_{1}}{I_{1}}\{\mathcal{B}^{B}_{0}(Q^{2})
  -\mathcal{B}^{B}_{2}(Q^{2})\}
  -\{\mathcal{I}^{B}_{0}(Q^{2})-\mathcal{I}^{B}_{2}(Q^{2})\} \right\} \cr 
& \hspace{1.5cm} +\left( \begin{array}{c} 11 \\ 15 \\ 19 \\ 0 \end{array}
 \right) \left\{ \frac{K_{2}}{I_{2}} \{\mathcal{C}^{B}_{0}(Q^{2})
 -\mathcal{C}^{B}_{2}(Q^{2})\} -\{\mathcal{J}^{B}_{0}(Q^{2})
 -\mathcal{J}^{B}_{2}(Q^{2})\} \right\} \cr
& \hspace{1.5cm} + \left( \begin{array}{c} 16 \\ 18 \\ 20 \\ 21 
\end{array} \right) \{\mathcal{H}^{B}_{0}(Q^{2})-\mathcal{H}^{B}_{2}
(Q^{2})\} \Bigg],  
\label{eq:ga13opcorr} \\
(g^{(3)B}_{3} (Q^{2}))^{(\mathrm{op})} =& -\frac{m_{\mathrm{s}} T_{3}}{378} 
  \Bigg[ \left( \begin{array}{c} 5 \\ 3 \\ 1 \\ 0 \end{array} \right)
  \left\{ \frac{K_{1}}{I_{1}} \{\mathcal{B}^{\prime B}_{0}(Q^{2})
  +\mathcal{B}^{\prime B}_{2}(Q^{2})\} -\{\mathcal{I}^{\prime B}_{0}
  (Q^{2}) +\mathcal{I}^{\prime B}_{2}(Q^{2})\} \right\} \cr 
& \hspace{1.5cm} +\left( \begin{array}{c} 11 \\ 15 \\ 19 \\ 0 \end{array}
  \right) \left\{ \frac{K_{2}}{I_{2}} \{\mathcal{C}^{\prime B}_{0}(Q^{2})
  +\mathcal{C}^{\prime B}_{2}(Q^{2})\} 
  -\{\mathcal{J}^{\prime B}_{0}(Q^{2}) +\mathcal{J}^{\prime B}_{2}
  (Q^{2})\} \right\} \cr 
& \hspace{1.5cm} + \left( \begin{array}{c} 16 \\ 18 \\ 20 \\ 21 
  \end{array} \right) \{\mathcal{H}^{\prime B}_{0}(Q^{2}) 
  +\mathcal{H}^{\prime B}_{2}(Q^{2})\} \Bigg], 
\label{eq:ga33opcorr}
\end{align}
\begin{align}
(g^{(3)B}_{1} (Q^{2}))^{(\mathrm{wf})} =& -\frac{T_{3}}{24}
  \Bigg[ \frac{a_{27}}{3} \left( \begin{array}{c} 5 \\ 6 \\ 7 \\ 0 
  \end{array} \right) \left\{ 2\{\mathcal{A}^{B}_{0}(Q^{2})
  -\mathcal{A}^{B}_{2}(Q^{2})\} +\frac{3\{\mathcal{B}^{B}_{0}(Q^{2})
  -\mathcal{B}^{B}_{2}(Q^{2})\}}{I_{1}} \right. \cr
& \hspace{3.1cm} \left. +\frac{\mathcal{C}^{B}_{0}(Q^{2})
  -\mathcal{C}^{B}_{2}(Q^{2})}{I_{2}} -\frac{i\{\mathcal{D}^{B}_{0}(Q^{2})
  -\mathcal{D}^{B}_{2}(Q^{2})\}}{I_{1}} \right\} \cr 
& \hspace{0.8cm} +\frac{a_{35}}{7} \left( \begin{array}{c} 1 \\ 2 \\ 3 \\ 0 
  \end{array} \right) \left\{ 2\{\mathcal{A}^{B}_{0}(Q^{2})
  -\mathcal{A}^{B}_{2}(Q^{2})\} -\frac{5\{\mathcal{B}^{B}_{0}(Q^{2})
  -\mathcal{B}^{B}_{2}(Q^{2})\}}{I_{1}} \right. \nonumber
\end{align}
\begin{align}
& \hspace{3.1cm} \left. +\frac{5\{\mathcal{C}^{B}_{0}(Q^{2})
  -\mathcal{C}^{B}_{2}(Q^{2})\}}{I_{2}} -\frac{i\{\mathcal{D}^{B}_{0}(Q^{2})
  -\mathcal{D}^{B}_{2}(Q^{2})\}}{I_{1}} \right\} \Bigg], 
\label{eq:ga13wfcorr} \\
(g^{(3)B}_{3} (Q^{2}))^{(\mathrm{wf})} =& -\frac{T_{3}}{24}
  \Bigg[ \frac{a_{27}}{3} \left( \begin{array}{c} 5 \\ 6 \\ 7 \\ 0 
  \end{array} \right) \left\{ 2\{\mathcal{A}^{\prime B}_{0}(Q^{2})
  +\mathcal{A}^{\prime B}_{2}(Q^{2})\} +\frac{3\{\mathcal{B}^{\prime B}_{0}(Q^{2})
  +\mathcal{B}^{\prime B}_{2}(Q^{2})\}}{I_{1}} \right. \cr
& \hspace{3.1cm} \left. +\frac{\mathcal{C}^{\prime B}_{0}(Q^{2})
  +\mathcal{C}^{\prime B}_{2}(Q^{2})}{I_{2}}
  -\frac{i\{\mathcal{D}^{\prime B}_{0}(Q^{2})
  +\mathcal{D}^{\prime B}_{2}(Q^{2})\}}{I_{1}} \right\} \cr
& \hspace{0.8cm} +\frac{a_{35}}{7} \left( \begin{array}{c} 1 \\ 2 \\ 3 \\ 0 
  \end{array} \right) \left\{ 2\{\mathcal{A}^{\prime B}_{0}(Q^{2})
  +\mathcal{A}^{\prime B}_{2}(Q^{2})\} -\frac{5\{\mathcal{B}^{\prime B}_{0}(Q^{2})
  +\mathcal{B}^{\prime B}_{2}(Q^{2})\}}{I_{1}} \right. \cr
& \hspace{3.1cm} \left. +\frac{5\{\mathcal{C}^{\prime B}_{0}(Q^{2})
  +\mathcal{C}^{\prime B}_{2}(Q^{2})\}}{I_{2}} -\frac{i\{\mathcal{D}^{\prime B}_{0}(Q^{2})
  +\mathcal{D}^{\prime B}_{2}(Q^{2})\}}{I_{1}} \right\} \Bigg]. 
\label{eq:ga33wfcorr}
\end{align}
Note that they are proportional to the eigenvalues of the third
component of the isospin operator, $T_3$. The octet axial-vector 
form factors ($a=8$) are obtained as
\begin{align}
(g^{(8)B}_{1} (Q^{2}))^{(\mathrm{sym})} =& -\frac{Y}{16\sqrt{3}} 
  \left[ 2\{\mathcal{A}^{B}_{0}(Q^{2})-\mathcal{A}^{B}_{2}(Q^{2})\}
  -\frac{\mathcal{B}^{B}_{0}(Q^{2})-\mathcal{B}^{B}_{2}(Q^{2})}{I_{1}} 
  \right. \cr
& \left. \hspace{1.5cm} -\frac{\mathcal{C}^{B}_{0}(Q^{2})
  -\mathcal{C}^{B}_{2}(Q^{2})}{I_{2}} -\frac{i\{\mathcal{D}^{B}_{0}(Q^{2})
  -\mathcal{D}^{B}_{2}(Q^{2})\}}{I_{1}} \right], 
\label{eq:ga18leading}\\
(g^{(8)B}_{3} (Q^{2}))^{(\mathrm{sym})} =& -\frac{Y}{16\sqrt{3}} 
  \left[ 2\{\mathcal{A}^{\prime B}_{0}(Q^{2})
  +\mathcal{A}^{\prime B}_{2}(Q^{2})\}
  -\frac{\mathcal{B}^{\prime B}_{0}(Q^{2})
  +\mathcal{B}^{\prime B}_{2}(Q^{2})}{I_{1}} \right.\cr
&\left. \hspace{1.5cm} -\frac{\mathcal{C}^{\prime B}_{0}(Q^{2})
  +\mathcal{C}^{\prime B}_{2}(Q^{2})}{I_{2}} 
  -\frac{i\{\mathcal{D}^{\prime B}_{0}(Q^{2})
  +\mathcal{D}^{\prime B}_{2}(Q^{2})\}}{I_{1}} \right], 
\label{eq:ga38leading}
\end{align}
\begin{align}
(g^{(8)B}_{1} (Q^{2}))^{(\mathrm{op})} =& 
  \frac{m_{\mathrm{s}}}{252\sqrt{3}} 
  \Bigg[ \left( \begin{array}{c} 3 \\ 2 \\ -3 \\ -12 \end{array} \right)
  \left\{ \frac{K_{1}}{I_{1}}\{\mathcal{B}^{B}_{0}(Q^{2})
  -\mathcal{B}^{B}_{2}(Q^{2})\} -\{\mathcal{I}^{B}_{0}(Q^{2})
  -\mathcal{I}^{B}_{2}(Q^{2})\} \right\} \cr
& \hspace{1.3cm} +\left( \begin{array}{c} 15 \\ -4 \\ -15 \\ -18 
  \end{array} \right) 
  \left\{ \frac{K_{2}}{I_{2}}\{\mathcal{C}^{B}_{0}(Q^{2})
  -\mathcal{C}^{B}_{2}(Q^{2})\} -\{\mathcal{J}^{B}_{0}(Q^{2})
  -\mathcal{J}^{B}_{2}(Q^{2})\} \right\} \cr
& \hspace{1.3cm} -2\left( \begin{array}{c} 12 \\ 1 \\ -12 \\ -27 
  \end{array} \right) \{\mathcal{H}^{B}_{0}(Q^{2})
  -\mathcal{H}^{B}_{2}(Q^{2})\} \Bigg], 
\label{eq:ga18opcorr}
\end{align}
\begin{align}
(g^{(8)B}_{3} (Q^{2}))^{(\mathrm{op})} =& 
  \frac{m_{\mathrm{s}}}{252\sqrt{3}} 
  \Bigg[ \left( \begin{array}{c} 3 \\ 2 \\ -3 \\ -12 \end{array} \right)
  \left(\frac{K_{1}}{I_{1}}\{\mathcal{B}^{\prime B}_{0}(Q^{2})
  +\mathcal{B}^{\prime B}_{2}(Q^{2})\} 
  -\{\mathcal{I}^{\prime B}_{0}(Q^{2})
  +\mathcal{I}^{\prime B}_{2}(Q^{2})\} \right) \cr
& \hspace{1.3cm} +\left( \begin{array}{c} 15 \\ -4 \\ -15 \\ -18 
  \end{array} \right)  
  \left( \frac{K_{2}}{I_{2}}\{\mathcal{C}^{\prime B}_{0}(Q^{2})
  +\mathcal{C}^{\prime B}_{2}(Q^{2})\} 
  -\{\mathcal{J}^{\prime B}_{0}(Q^{2})
  +\mathcal{J}^{\prime B}_{2}(Q^{2})\} \right) \cr
& \hspace{1.3cm} -2\left( \begin{array}{c} 12 \\ 1 \\ -12 \\ -27 
  \end{array} \right) \{\mathcal{H}^{\prime B}_{0}(Q^{2})
  +\mathcal{H}^{\prime B}_{2}(Q^{2})\} \Bigg], 
\label{eq:ga38opcorr}
\end{align}
\begin{align}
(g^{(8)B}_{1} (Q^{2}))^{(\mathrm{wf})} =& \frac{1}{16\sqrt{3}}
  \Bigg[ \frac{a_{27}}{3} \left( \begin{array}{c} 15 \\ 8 \\ 3 \\ 0 
  \end{array} \right) \left\{ 2\{\mathcal{A}^{B}_{0}(Q^{2})
  -\mathcal{A}^{B}_{2}(Q^{2})\} +\frac{3\{\mathcal{B}^{B}_{0}(Q^{2})
  -\mathcal{B}^{B}_{2}(Q^{2})\}}{I_{1}} \right. \cr
& \hspace{3.1cm} \left. +\frac{\mathcal{C}^{B}_{0}(Q^{2})
  -\mathcal{C}^{B}_{2}(Q^{2})}{I_{2}} -\frac{i\{\mathcal{D}^{B}_{0}(Q^{2})
  -\mathcal{D}^{B}_{2}(Q^{2})\}}{I_{1}} \right\} \cr
& \hspace{0.8cm} -\frac{a_{35}}{7} \left( \begin{array}{c} 5 \\ 8 \\ 9 \\ 8 
  \end{array} \right) \left\{ 2\{\mathcal{A}^{B}_{0}(Q^{2})
  -\mathcal{A}^{B}_{2}(Q^{2})\} -\frac{5\{\mathcal{B}^{B}_{0}(Q^{2})
  -\mathcal{B}^{B}_{2}(Q^{2})\}}{I_{1}} \right. \cr
& \hspace{3.1cm} \left. +\frac{5\{\mathcal{C}^{B}_{0}(Q^{2})
  -\mathcal{C}^{B}_{2}(Q^{2})\}}{I_{2}} -\frac{i\{\mathcal{D}^{B}_{0}(Q^{2})
  -\mathcal{D}^{B}_{2}(Q^{2})\}}{I_{1}} \right\} \Bigg], 
\label{eq:ga18wfcorr} \\
(g^{(8)B}_{3} (Q^{2}))^{(\mathrm{wf})} =& \frac{1}{16\sqrt{3}}
  \Bigg[ \frac{a_{27}}{3} \left( \begin{array}{c} 15 \\ 8 \\ 3 \\ 0 
  \end{array} \right) \left\{ 2\{\mathcal{A}^{\prime B}_{0}(Q^{2})
  +\mathcal{A}^{\prime B}_{2}(Q^{2})\} +\frac{3\{\mathcal{B}^{\prime B}_{0}(Q^{2})
  +\mathcal{B}^{\prime B}_{2}(Q^{2})\}}{I_{1}} \right. \cr
& \hspace{3.1cm} \left. +\frac{\mathcal{C}^{\prime B}_{0}(Q^{2})
  +\mathcal{C}^{\prime B}_{2}(Q^{2})}{I_{2}}
  -\frac{i\{\mathcal{D}^{\prime B}_{0}(Q^{2})
  +\mathcal{D}^{\prime B}_{2}(Q^{2})\}}{I_{1}} \right\} \cr
& \hspace{0.8cm} -\frac{a_{35}}{7} \left( \begin{array}{c} 5 \\ 8 \\ 9 \\ 8 
  \end{array} \right) \left\{ 2\{\mathcal{A}^{\prime B}_{0}(Q^{2})
  +\mathcal{A}^{\prime B}_{2}(Q^{2})\} -\frac{5\{\mathcal{B}^{\prime B}_{0}(Q^{2})
  +\mathcal{B}^{\prime B}_{2}(Q^{2})\}}{I_{1}} \right. \cr
& \hspace{3.1cm} \left. +\frac{5\{\mathcal{C}^{\prime B}_{0}(Q^{2})
  +\mathcal{C}^{\prime B}_{2}(Q^{2})\}}{I_{2}} -\frac{i\{\mathcal{D}^{\prime B}_{0}(Q^{2})
  +\mathcal{D}^{\prime B}_{2}(Q^{2})\}}{I_{1}} \right\} \Bigg]. 
\label{eq:ga38wfcorr}
\end{align}
\section{Results and discussion}
\label{sec:4}
In the $\chi$QSM, the only free parameter is the dynamical quark mass,
$M$. Though it is determined by the saddle-point approximation from
the instanton vacuum, it is fixed by reproducing the electric form
factor of the proton. The pion decay constant is determined by using
the experimental data $f_\pi=93$ MeV. The current-quark masses are
fixed by the pion and kaon masses. However, we will take the value of
the strange current quark mass to be $m_{\mathrm{s}}=180$ MeV that is
larger than those taken in chiral perturbation theory. The reason is
that with this value of $180$ MeV we are able to reproduce the mass
splittings of the hyperons and singly heavy baryons. Thus, we have no
free parameter to fit in the present calculation. 

\begin{figure}[ht]
  \includegraphics[scale=0.285]{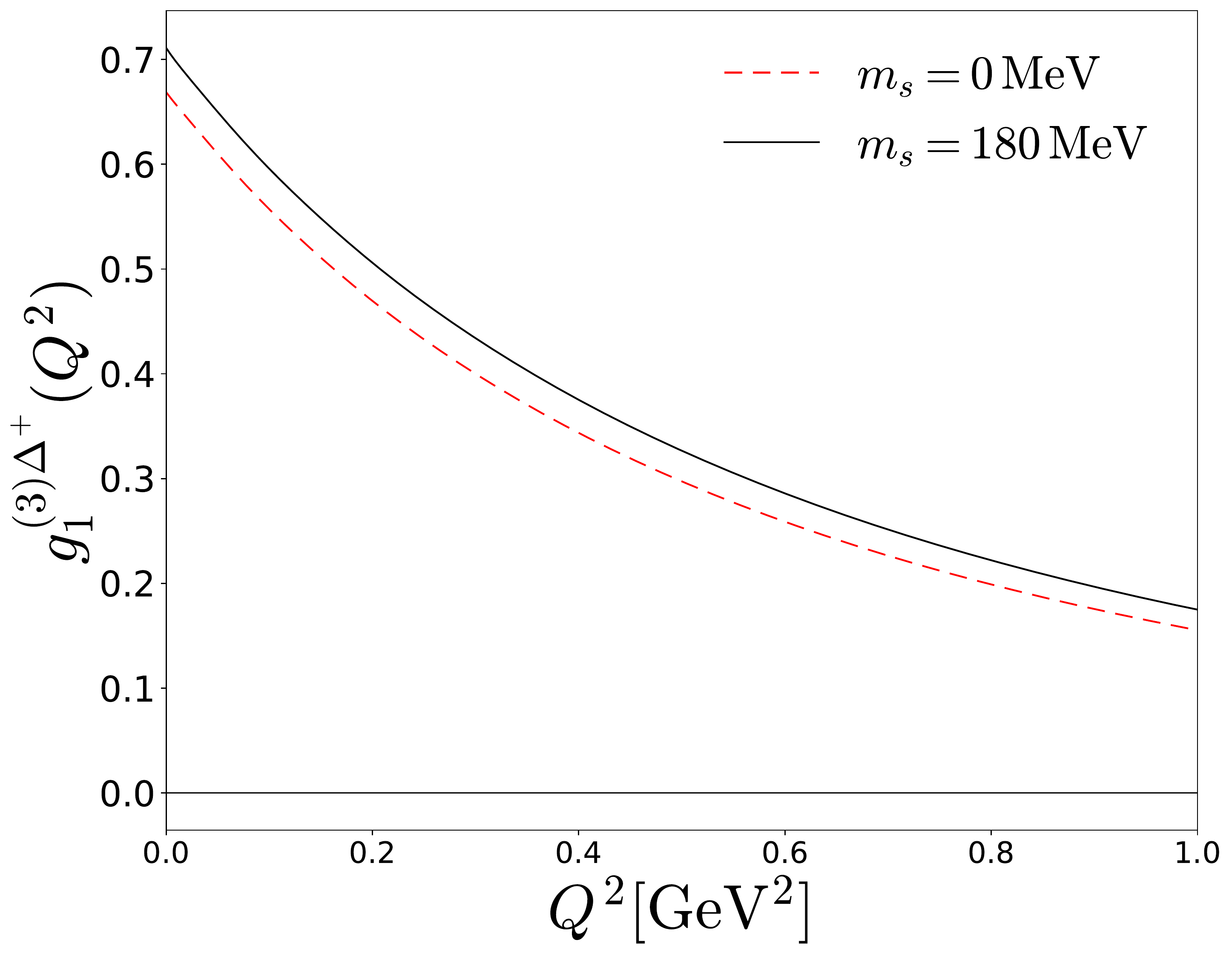}\hspace{0.5cm}
  \includegraphics[scale=0.285]{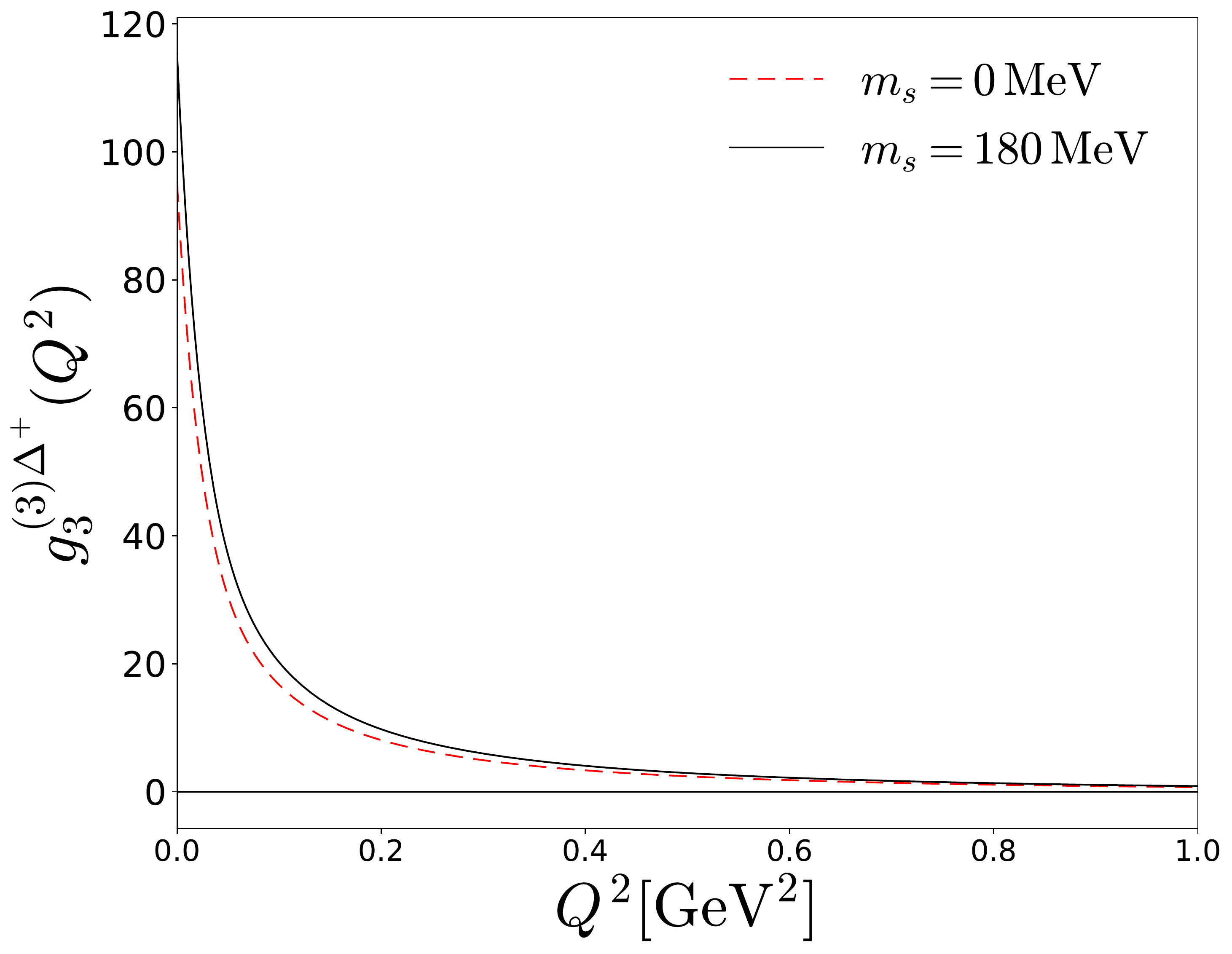}
\caption{Effects of the explicit flavor SU(3) symmetry breaking on 
the triplet axial-vector form factors $g_1^{(3)}(Q^{2})$ and 
$g_3^{(3)}(Q^{2})$ of the $\Delta^+$ isobar. In the right panel, 
the results of $g_1^{(3)}(Q^{2})$ are drawn whereas in the left panel, 
those of $g_3^{(3)}(Q^{2})$ are depicted. The solid and dashed curves 
represent the total results and those in the SU(3) symmetric case, 
respectively.}
\label{fig:1}
\end{figure}
\begin{figure}[ht]
  \includegraphics[scale=0.26]{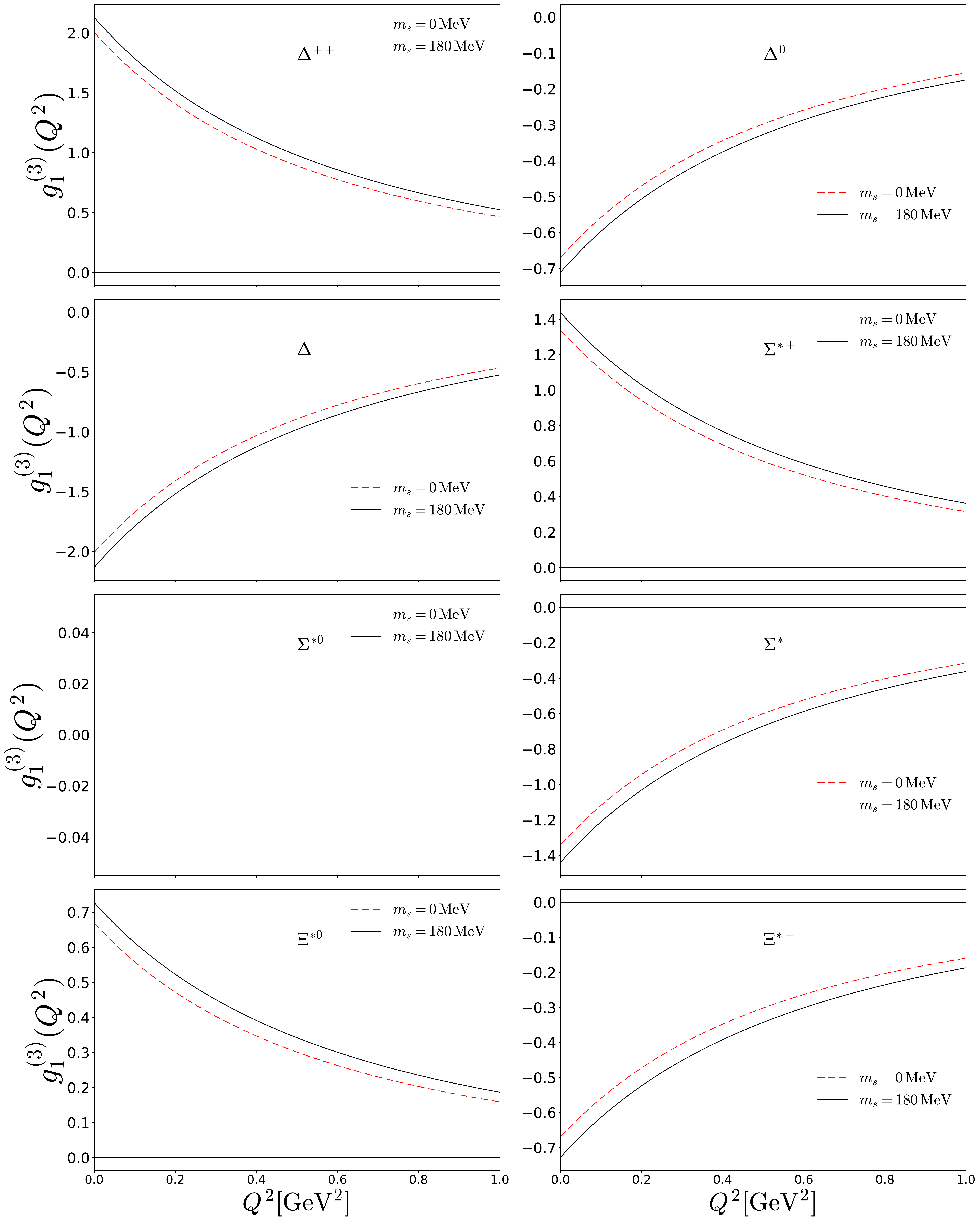}
\caption{Effects of the explicit flavor SU(3) symmetry breaking on
  $g^{(3)B}_{1}(Q^{2})$ of the baryon decuplet except for the 
  $\Delta^+$ and $\Omega^{-}$ baryons. Notations are the same as 
  in Fig.~\ref{fig:1}.}
\label{fig:5}
\end{figure}
\begin{figure}[ht]
  \includegraphics[scale=0.26]{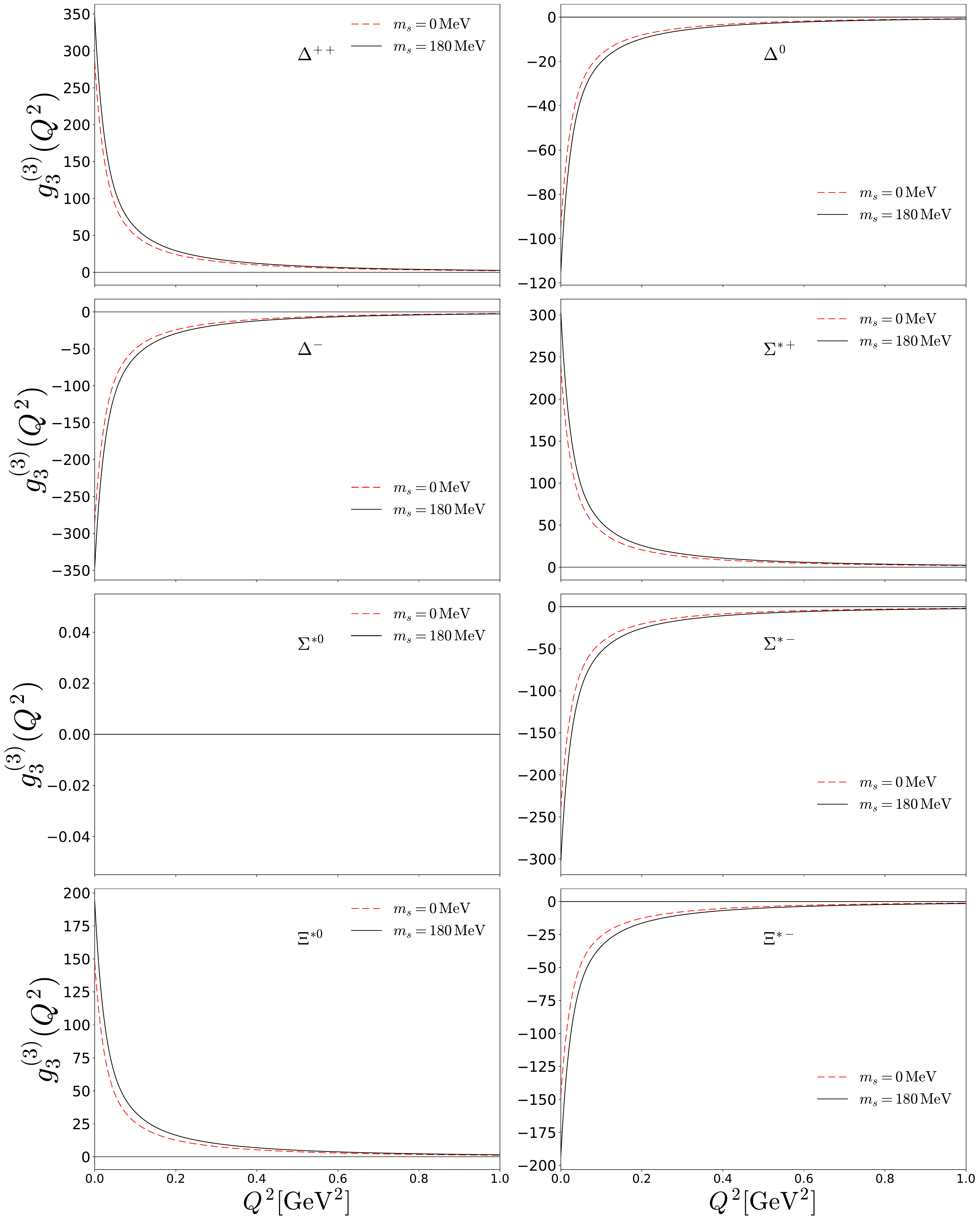}
\caption{Effects of the explicit flavor SU(3) symmetry breaking on
  $g^{(3)B}_{3}(Q^{2})$ of the baryon decuplet except for the 
  $\Delta^+$ and $\Omega^{-}$ baryons. Notations are the same as 
  in Fig.~\ref{fig:1}.}
\label{fig:6}
\end{figure}

Since there are numerous form factors of the baryon decuplet, we will
first concentrate on those of the $\Delta^+$ isobar and
$\Omega^-$ hyperon. In the left panel of
Fig.~\ref{fig:1}, we draw the results of the  
triplet axial-vector form factor $g_1^{(3)}(Q^{2})$ of the $\Delta^+$ 
whereas in the right panel we show those of $g_3^{(3)}(Q^{2})$. As 
for those for other members of the baryon decuplet, we will display 
them on Fig.~\ref{fig:5} and Fig.~\ref{fig:6} and will discuss them
later on. Note that as defined in Eq.~\eqref{eq:MatrixEl1} 
$g_1^{(3)}(Q^{2})$ form factor is the most well-known axial-vector 
form factor. A great deal of theoretical works have considered this one, 
because its value at $Q^2=0$ gives the axial charge of the $\Delta^+$.
The Goldberger-Treiman relation connects it to the strong coupling 
constant $g_{\pi \Delta \Delta}$. As depicted in the left panel of 
Fig.~\ref{fig:1}, the results of $g_1^{(3)}(Q^{2})$ decrease 
monotonically and slowly as $Q^2$ increases. The effects of the linear 
$m_{\mathrm{s}}$ corrections are marginal. They provide approximately 
overall $10~\%$ correction to the form factor $g_1^{(3)}(Q^2)$. The 
right panel of Fig.~\ref{fig:1} exhibits the numerical results for the 
second axial-vector form factor $g_3^{(3)}(Q^{2})$ of the $\Delta^+$. 
In constrast with $g_1^{(3)}(Q^{2})$, the $Q^2$ dependence of 
$g_3^{(3)}(Q^{2})$ is prominent. The result of $g_3^{(3)}(Q^{2})$ 
falls off drastically as $Q^2$ increases. One can understand this 
behavior as follows: as shown in Appendix~\ref{app:A}, all 
components of $g_3^{(3)}(Q^{2})$ are proportional to $Q^{-2}$, which 
cause such strong $Q^2$ dependence. Moreover, the kinematical 
prefactor makes the magnitude of $g_3^{(3)}(Q^{2})$ much larger 
than that of $g_1^{(3)}(Q^{2})$. On the other hand, the effects of 
flavor SU(3) symmetry breaking are rather small on $g_3^{(3)}(Q^{2})$.

\begin{figure}[ht]
  \includegraphics[scale=0.285]{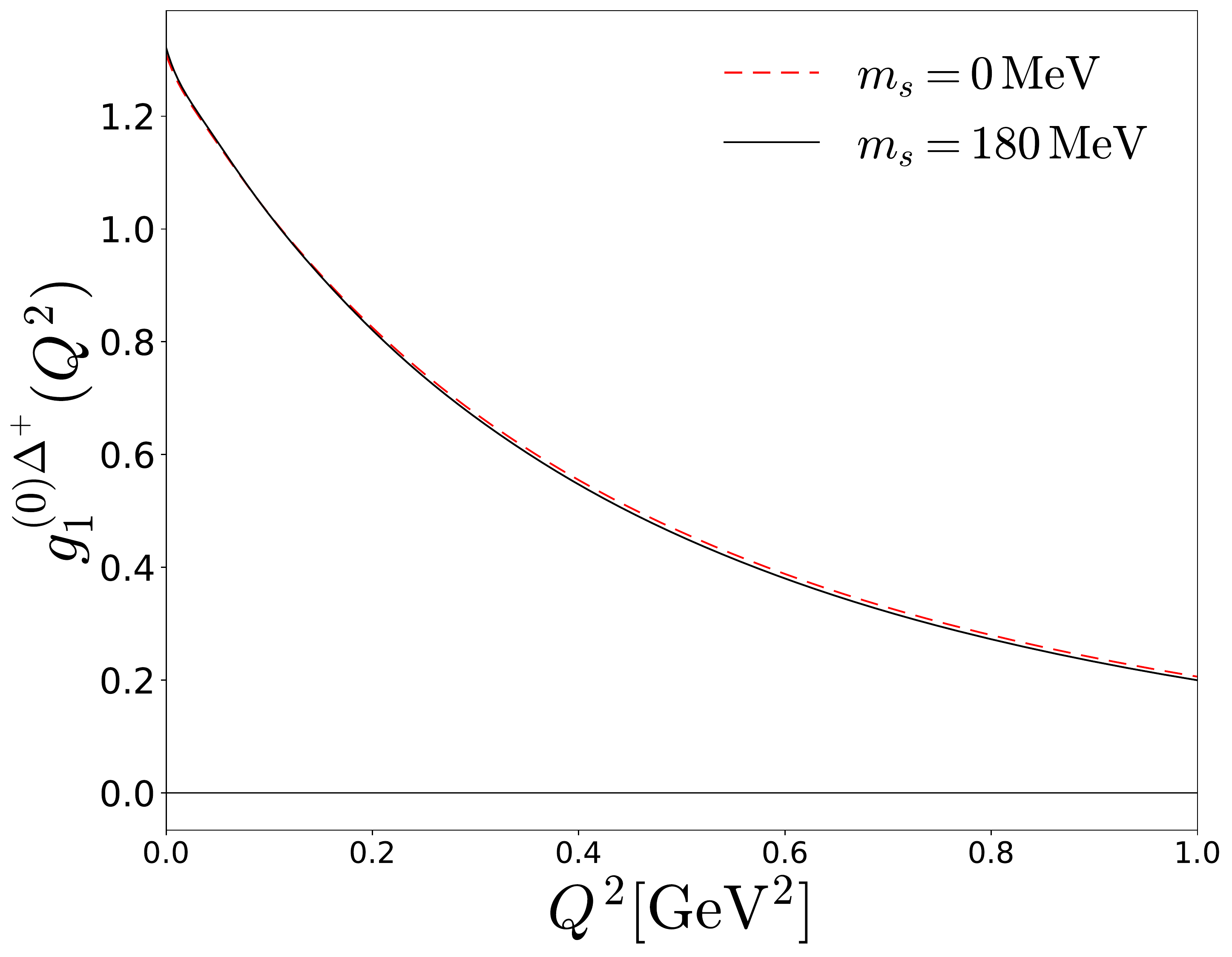}\hspace{0.5cm}
  \includegraphics[scale=0.285]{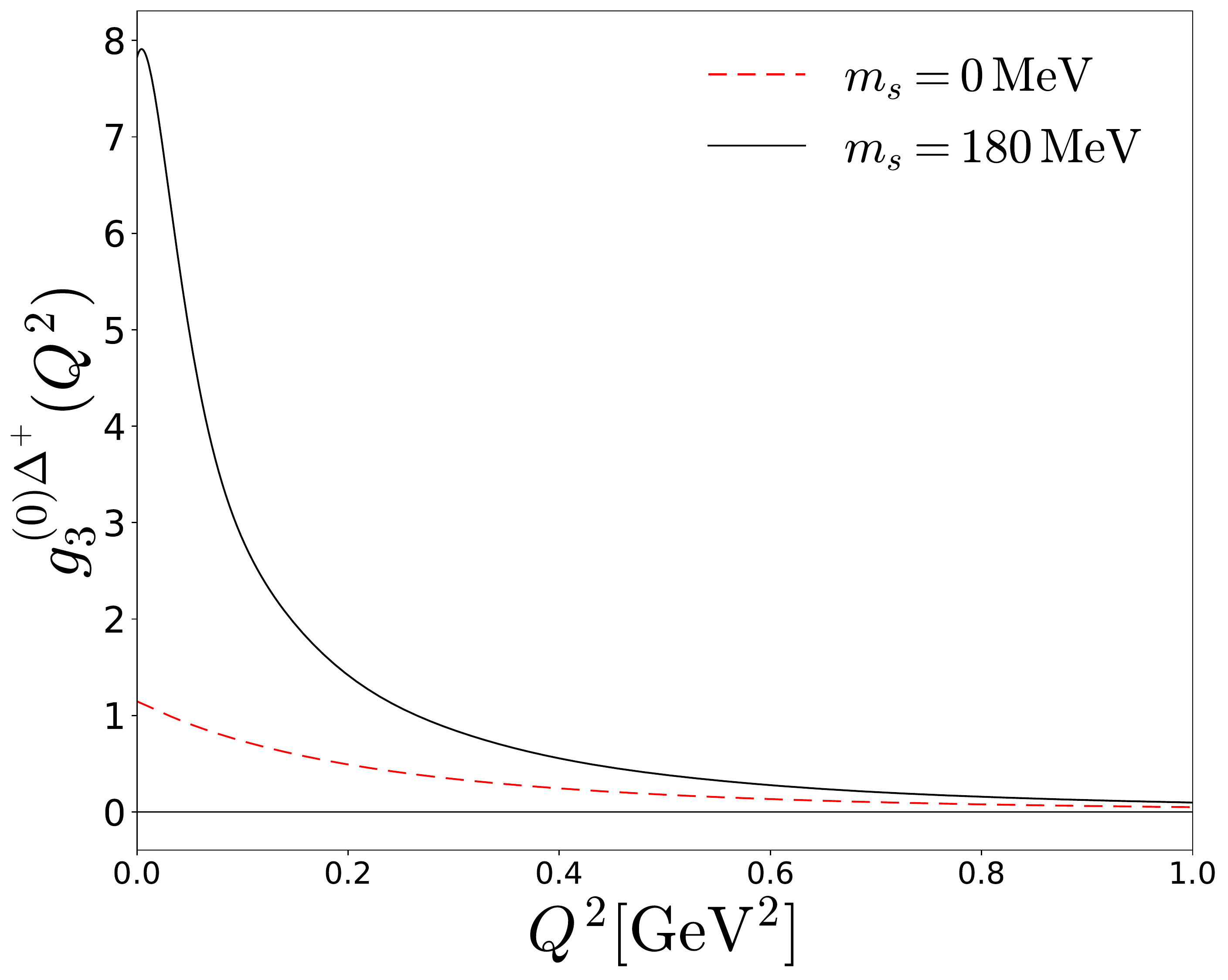}
  \includegraphics[scale=0.285]{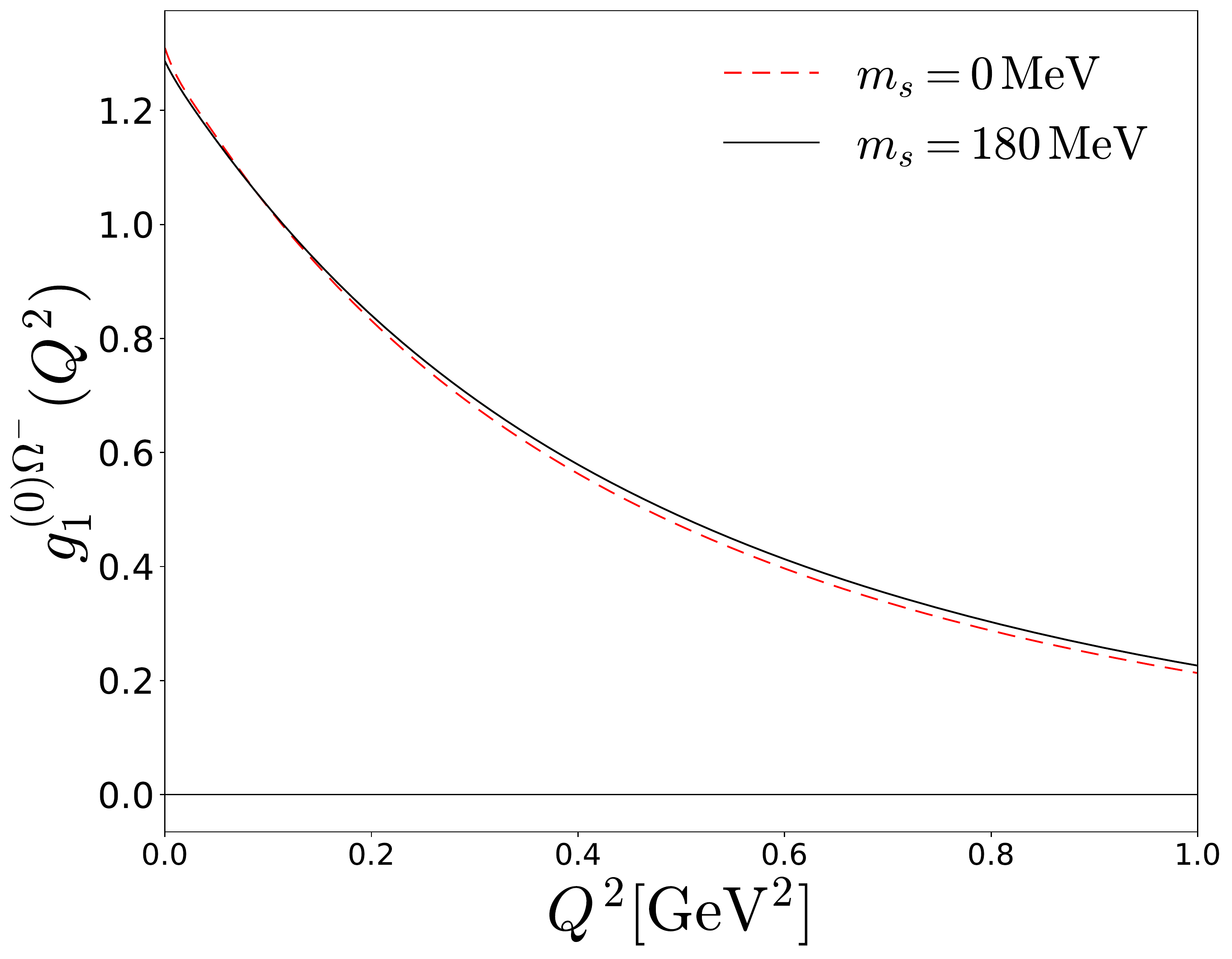}\hspace{0.5cm}
  \includegraphics[scale=0.285]{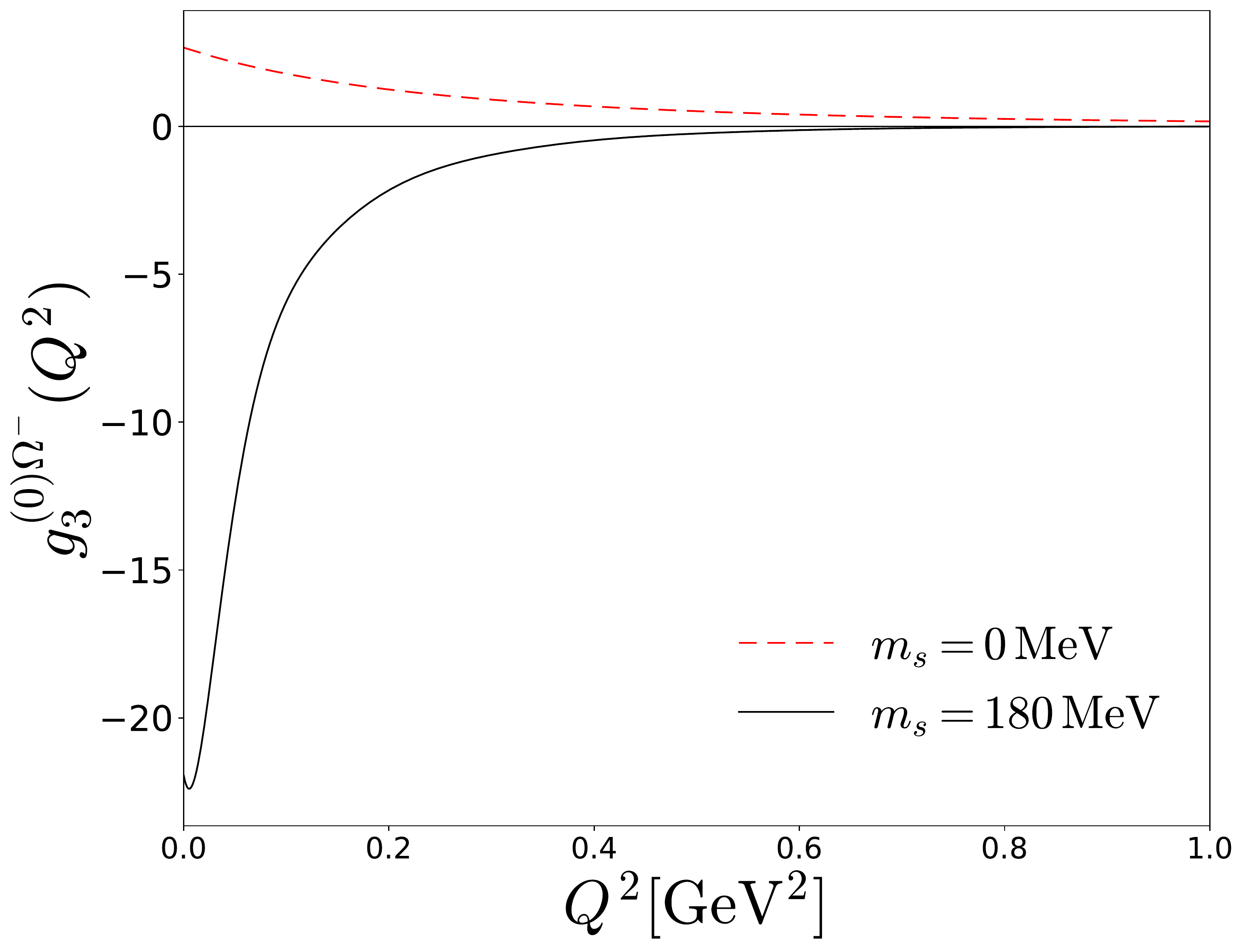}
\caption{Effects of the explicit flavor SU(3) symmetry breaking on 
  the singlet axial-vector form factors $g_1^{(3)}(Q^{2})$ and 
  $g_3^{(3)}(Q^{2})$ of the $\Delta^+$ isobar and $\Omega^-$. In the 
  upper right panel, the results of $g_1^{(0)}(Q^{2})$ of $\Delta^+$ 
  are drawn whereas in the upper left panel, those of 
  $g_3^{(0)\Delta^+}(Q^{2})$ are depicted. The lower left panel shows 
  the results of the first $\Omega^-$ singlet axial-vector form 
  factor $g_1^{(0)}(Q^{2})$, while the lower right one illustrates 
  those of the $g_3^{(0)\Omega^-}(Q^{2})$. The solid and dashed 
  curves represent the total results and those in the SU(3) symmetric 
  case, respectively. Notations are the same as in Fig.~\ref{fig:1}.} 
\label{fig:2}
\end{figure}
In Fig.~\ref{fig:2}, we show the results of the singlet axial-vector
form factors. As shown in the left panel of Fig.~\ref{fig:2}, the
effects of the linear $m_{\mathrm{s}}$ are almost negligible on the
$g_1^{(0)}(Q^{2})$ form factors of both $\Delta^+$ and $\Omega^-$.
Note that the singlet axial-vector constant, 
$g_1^{(0) B}(0)$, explains the quark spin content of the 
corresponding baryon. If one neglects the effects of flavor SU(3) 
symmetry breaking, $g_1^{(0)B}(Q^{2})$ is independent of the flavor 
content of a decuplet baryon, as explicitly expressed in 
Eq.~\eqref{eq:ga10leading}. $g_3^{(0)B}(Q^{2})$ is also flavor-independent
(see Eq.~\eqref{eq:ga30leading}) without $m_{\mathrm{s}}$ corrections. 
However, when it comes to the $g_3^{(0)}(Q^{2})$ form factors, the 
linear $m_{\mathrm{s}}$ corrections contribute remarkably large to 
them. This can be understood by scrutinizing 
Eq.~\eqref{eq:ga_decompose} for the singlet axial-vector form factors, 
i.e. when $a=0$. The flavor SU(3) symmetric part 
$(g_3^{(0) B}(Q^{2}))^{(\mathrm{sym})}$ contains only the rotational 
$1/N_c$ corrections. Since there are no wavefunction corrections, 
we have only the linear $m_{\mathrm{s}}$ corrections from 
the current quark mass term of the effective chiral action.  
The tensor contribution $\mathcal{I}_2^{'B}(Q^2)$ is the most dominant 
one that governs the behavior of $g_{3}^{(0)\Delta^+}(Q^2)$. Even 
though we ignore all other terms, the result is not much changed. 
If one looks into the expression for $\mathcal{I}_2^{'B}(Q^2)$, which 
is given in Appendix~\ref{app:A}, one can easily see that in general 
the tensor contributions are dominant over the scalar ones kinematically. 
We can find a similar tendency in the $\Omega^-$ singlet axial-vector 
form factors as shown in the lower panel of Fig.~\ref{fig:2}. In the 
case of $\Omega^-$, the results of $g_1^{(0)}(Q^2)$ are very similar 
to those of $\Delta^+$. The rotational $1/N_c$
corrections give the positive values of $g_3^{(0)}(Q^2)$ for
$\Omega^-$, which are the same as in the case of $\Delta^+$.
This is due to the fact that as explained previously, the rotational
$1/N_c$ corrections do not depend on the flavor of the baryon
decuplet. On the other hand, the linear $m_{\mathrm{s}}$ contributions
become negative to the $g_3^{(0)\Omega^-}(Q^{2})$ form factor. This is
due to the fact that the hypercharge is present in
Eq.~\eqref{eq:ga30opcorr}. One could suspect that the second-order 
$m_{\mathrm{s}}$ corrections might come into play in describing the
$g_3^{(0)}(Q^{2})$ form factors of the baryon decuplet. However, the
second-order $m_{\mathrm{s}}$ corrections should be suppressed at
least by two reasons: Firstly, the parameter
$(m_{\mathrm{s}}/\Lambda)^2$ is much smaller than the leading term,
where $\Lambda$ is the cutoff mass or 
the normalization scale of the $\chi$QSM, which is of order 1
GeV. Secondly, expressions for the second-order $m_{\mathrm{s}}$
corrections contain doubly-summed energy denominators, which lead to
further suppression. So, we expect that the second-order 
$m_{\mathrm{s}}$ corrections should be much smaller than
the linear $m_{\mathrm{s}}$ corrections. 

\begin{figure}[ht]
  \includegraphics[scale=0.285]{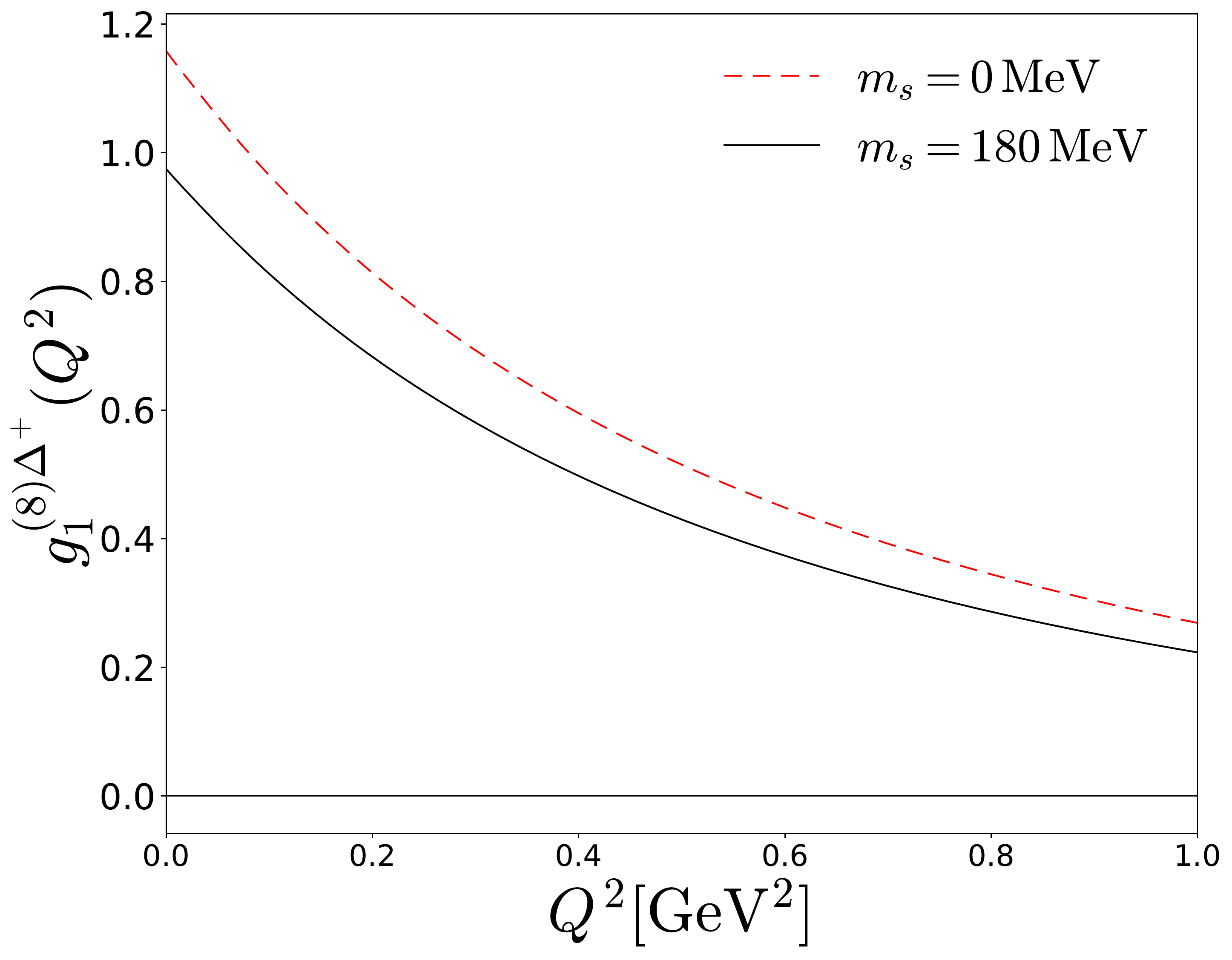}\hspace{0.5cm}
  \includegraphics[scale=0.285]{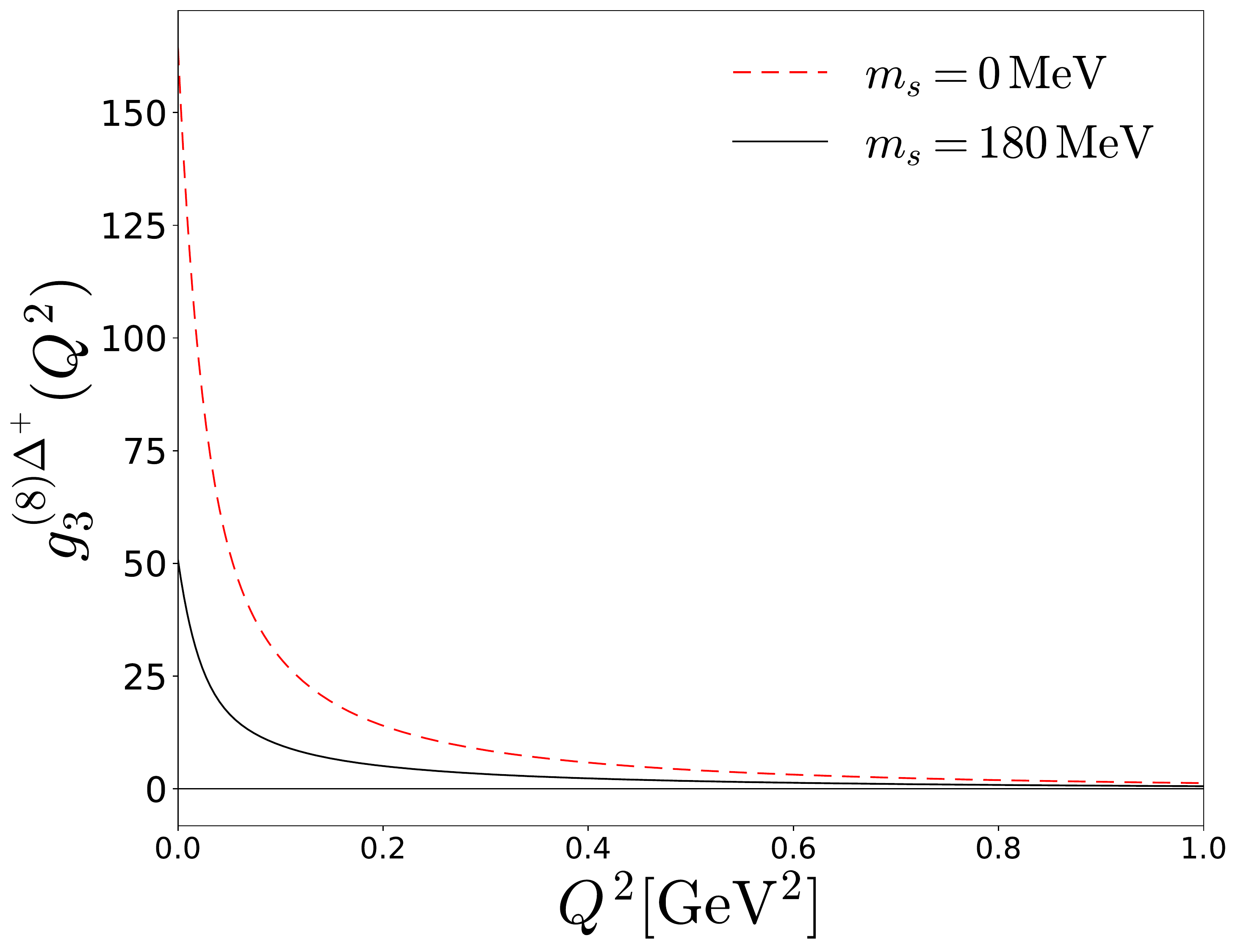}
  \includegraphics[scale=0.285]{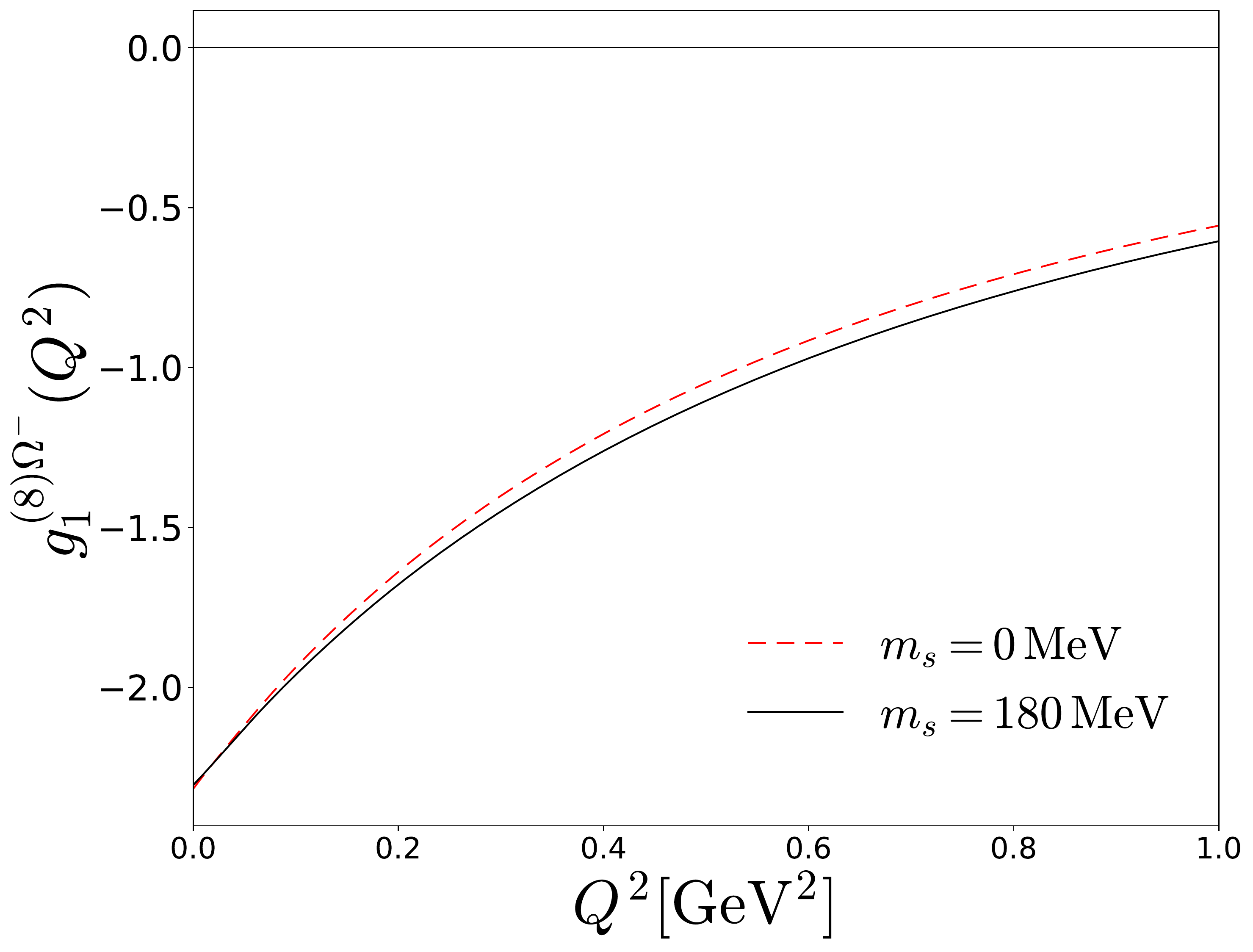}\hspace{0.5cm}
  \includegraphics[scale=0.285]{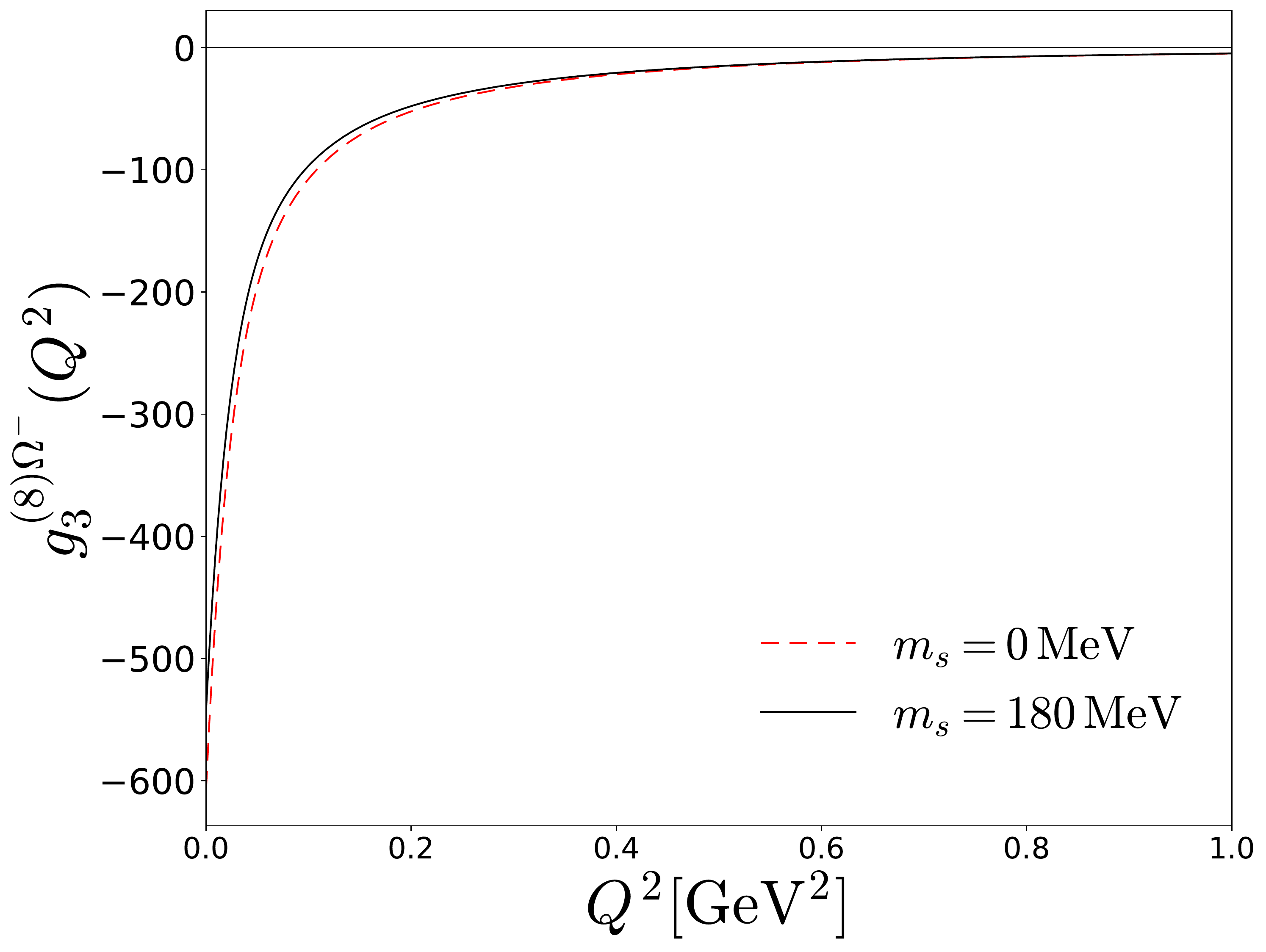}
\caption{Effects of the explicit flavor SU(3) symmetry breaking on
  the octet axial-vector form factors of the $\Delta^+$ and
  $\Omega^{-}$ baryon. Notations are the same as in Fig.~\ref{fig:1}.} 
\label{fig:3}
\end{figure}
Figure~\ref{fig:3} depicts the octet axial-vector form factors of
$\Delta^+$ and $\Omega^-$. As illustrated in the upper left panel of
Fig.~\ref{fig:3}, the linear $m_{\mathrm{s}}$ corrections suppress
$g_1^{(8)\Delta^+}(Q^2)$ by about $17~\%$, while they are negligible to
$g_1^{(8)\Omega^-}(Q^2)$ as shown in the lower left panel of
fig.~\ref{fig:3}. Interestingly, the linear $m_{\mathrm{s}}$
corrections to $g_1^{(8)\Omega^-}(Q^2)$ become visible as $Q^2$
increases, though they are still very small. On the other hand, the
results for $g_3^{(8) \Delta^+}(Q^2)$ show peculiar 
behavior. While it shows a similar tendency of the linear
$m_{\mathrm{s}}$ corrections to $g_1^{(8)\Delta^+}(Q^2)$, their
magnitude is rather large. Interestingly, the wavefunction corrections
come into play in this case. As shown in Eq.~\eqref{eq:ga38wfcorr},
the contribution from the 27-plet to $g_3^{(8)\Delta^+}(Q^2)$ dominates
over all other contributions, since it contains
$\mathcal{A}_2'^{\Delta^+}(Q^2)$ and $i\mathcal{D}_2'^{\Delta^+}(Q^2)$, 
which are the most contributive ones. The linear $m_{\mathrm{s}}$ corrections
arising from the effective chiral action are much smaller than those
of the wavefunction corrections. As a result, the $g_3^{(8)\Delta^+}(Q^{2})$
form factor is much reduced by the effects of flavor SU(3) symmetry
breaking. The lower right panel of Fig.~\ref{fig:3} draws the
numerical results for $g_3^{(8)\Omega^-}(Q^2)$. In contrast to the
$g_3^{(8)\Delta^+}(Q^{2})$, the linear $m_{\mathrm{s}}$ corrections are
almost negligible. The reason can be also found in
Eq.~\eqref{eq:ga38wfcorr}, where the contribution of the 27-plet
vanishes.

\begin{figure}[ht]
  \includegraphics[scale=0.29]{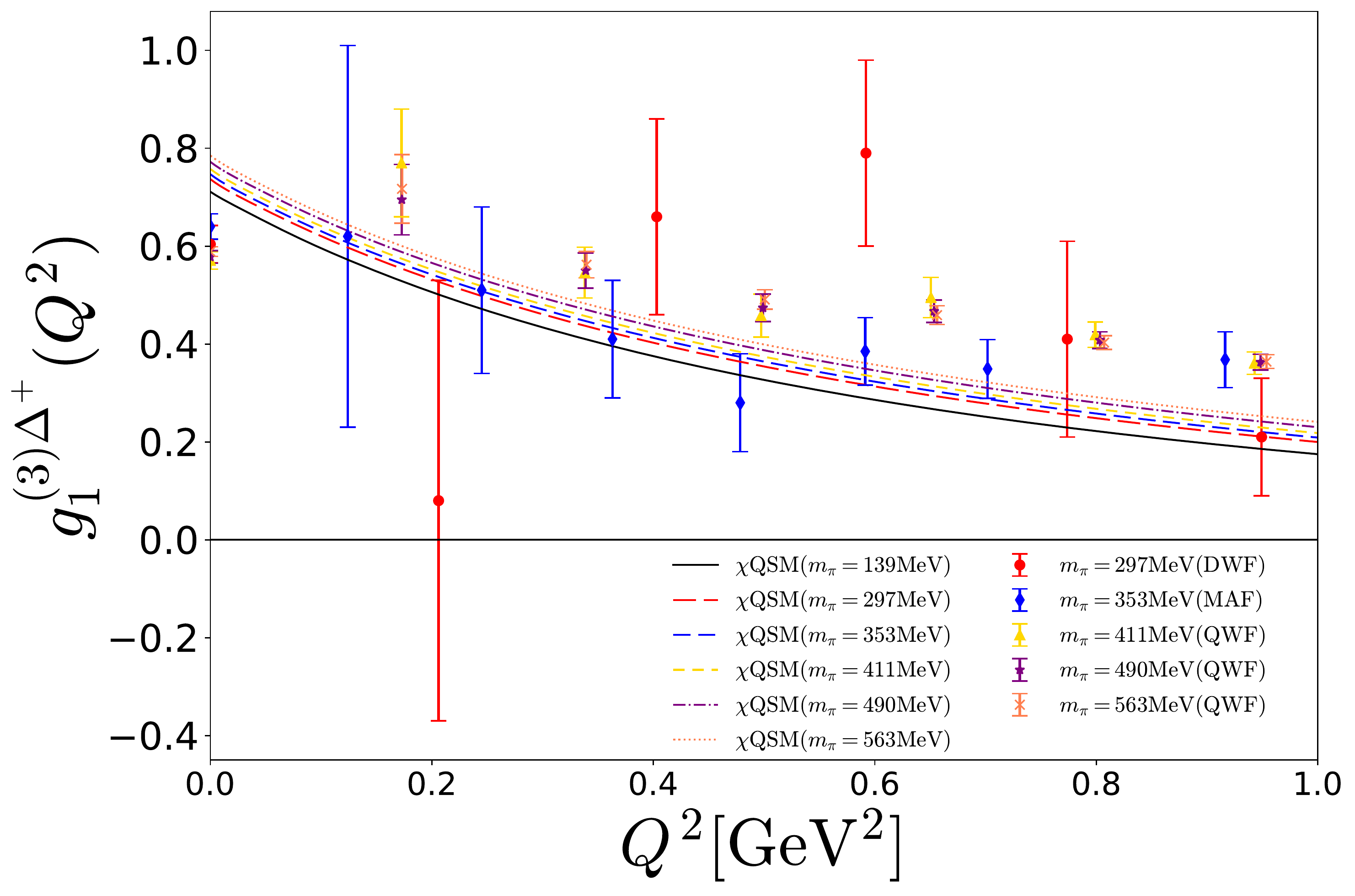}\hspace{0.3cm}
  \includegraphics[scale=0.29]{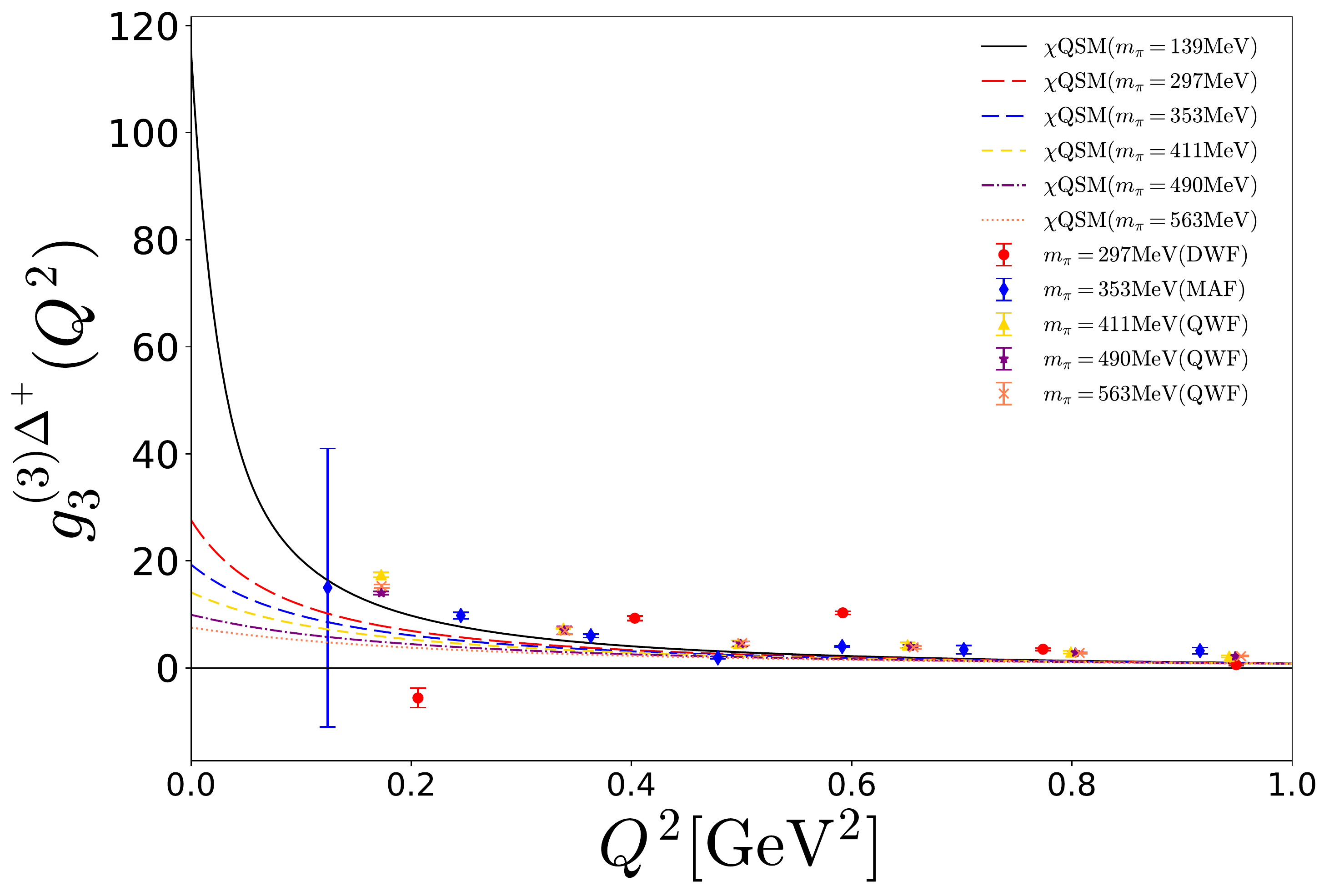}
  \includegraphics[scale=0.29]{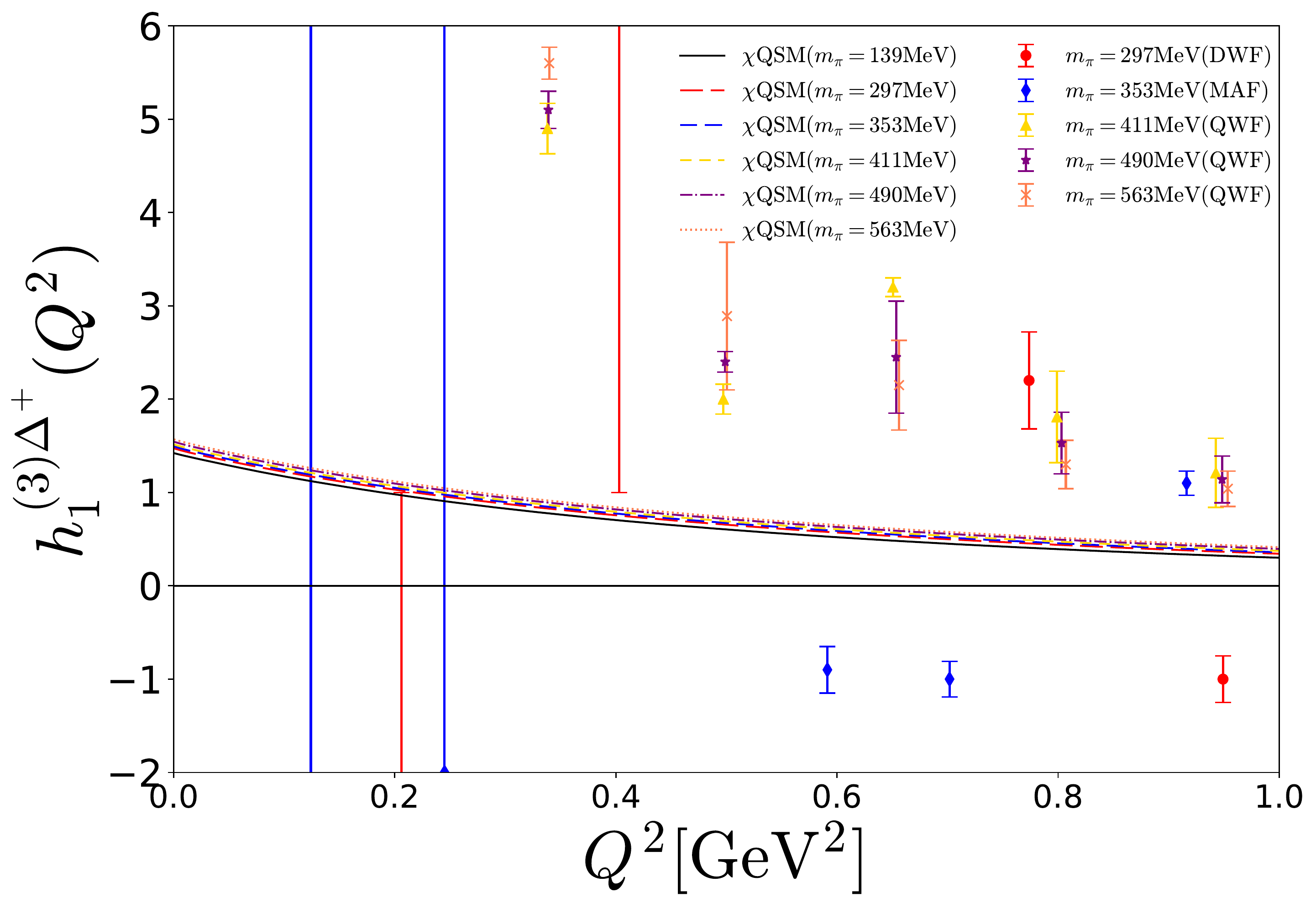}\hspace{0.3cm}
  \includegraphics[scale=0.29]{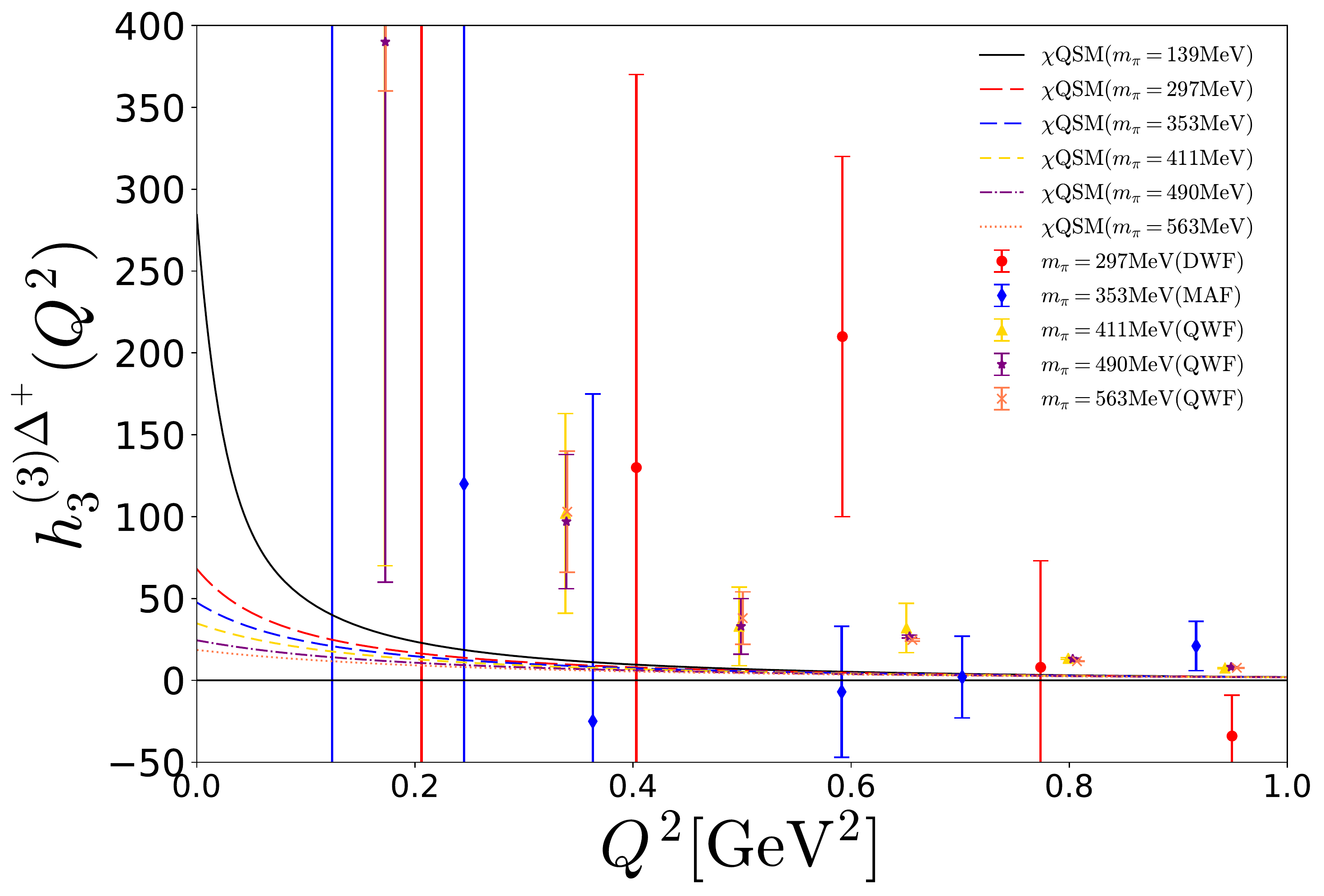}
\caption{Numerical results of the triplet axial-vector form factors 
  of the $\Delta^{+}$ baryon in comparison with the data taken from
  lattice QCD~\cite{Alexandrou:2013opa}.}
\label{fig:4}
\end{figure}
Since there are only the results from lattice QCD on the triplet 
axial-vector form factors of the $\Delta^+$~\cite{Alexandrou:2013opa}, 
it is of great importance to compare the present results with them. 
However, before we make a comparison of the present results with the 
lattice data, we have to consider the unphysical pion mass used in 
the lattice simulation~\cite{Alexandrou:2013opa}. This means that we 
need to derive the profile function of the chiral soliton by solving 
the equation of motion again. This can be done as follows: In
Eq.~\eqref{eq:DiracHam}, we replace the physical value of the average
current quark mass of the up and down quarks with the unphysical one
that corresponds to the pion mass adopted by the lattice
calculation. Then the axial-vector form factors of the baryon decuplet
can be recalculated by using the profile function with the unphysical
value of the pion mass. In fact, the pion mass dependence of baryonic
observables has been already investigated~\cite{Goeke:2005fs, 
Goeke:2007fq, Kim:2018nqf, Kim:2019wbg} in the context of the
comparison with lattice results.

In Fig~.\ref{fig:4}, we exhibit the results of the $\Delta^+$ triplet 
axial-vector form factors in comparison with those from the lattice
calculation, taking into account the values of the unphysical pion
mass, which are used in Ref.~\cite{Alexandrou:2013opa}. 
As mentioned previously, $h_1^{(3)}(Q^{2})$ and $h_3^{(3)}(Q^{2})$ 
are in effect the same respectively as $g_1^{(3)}(Q^{2})$ and
$g_3^{(3)}(Q^{2})$ except for the kinematical factors, which have been
expressed implicitly in Eq.~\eqref{eq:MatrixEl2}. As depicted in the
upper left panel of Fig.~\ref{fig:4}, the results of
$g_1^{(3)\Delta^+}(Q^2)$ increase as larger values of the unphysical
pion mass are used. The $Q^2$ dependence of the form factor tends to
be similar to that of the lattice calculation. However, the lattice
data seem to fall off rather slowly as $Q^2$ increases, compared with
the present results. The upper right panel of Fig.~\ref{fig:4} depicts
the results for $g_3^{(3)\Delta^+}(Q^2)$ with the pion mass varied. It
is interesting to see that the magnitude of the form factor decreases
as large pion masses are employed. The lattice data on
$g_3^{(3)\Delta^+}(Q^2)$ show relatively smaller fluctuations,
compared with those on other axial-vector form factors. The $Q^2$
dependence of the present results is in line with the lattice
data. Since there are large uncertainties in the lattice results and
a lack of data in smaller $Q^2$ regions, it is rather difficult to
make a clear comparison of the present results with those of the
lattice  calculation. In the lower panel of Fig.~\ref{fig:4}, we
compare the results of $h_1^{(3)\Delta^+}(Q^{2})$ and
$h_3^{(3)\Delta^+}(Q^{2})$ with those of lattice QCD. Since the  
lattice data show large fluctuations in general, we are not able to 
draw any meaningful conclusions.

\begin{figure}[ht]
  \includegraphics[scale=0.26]{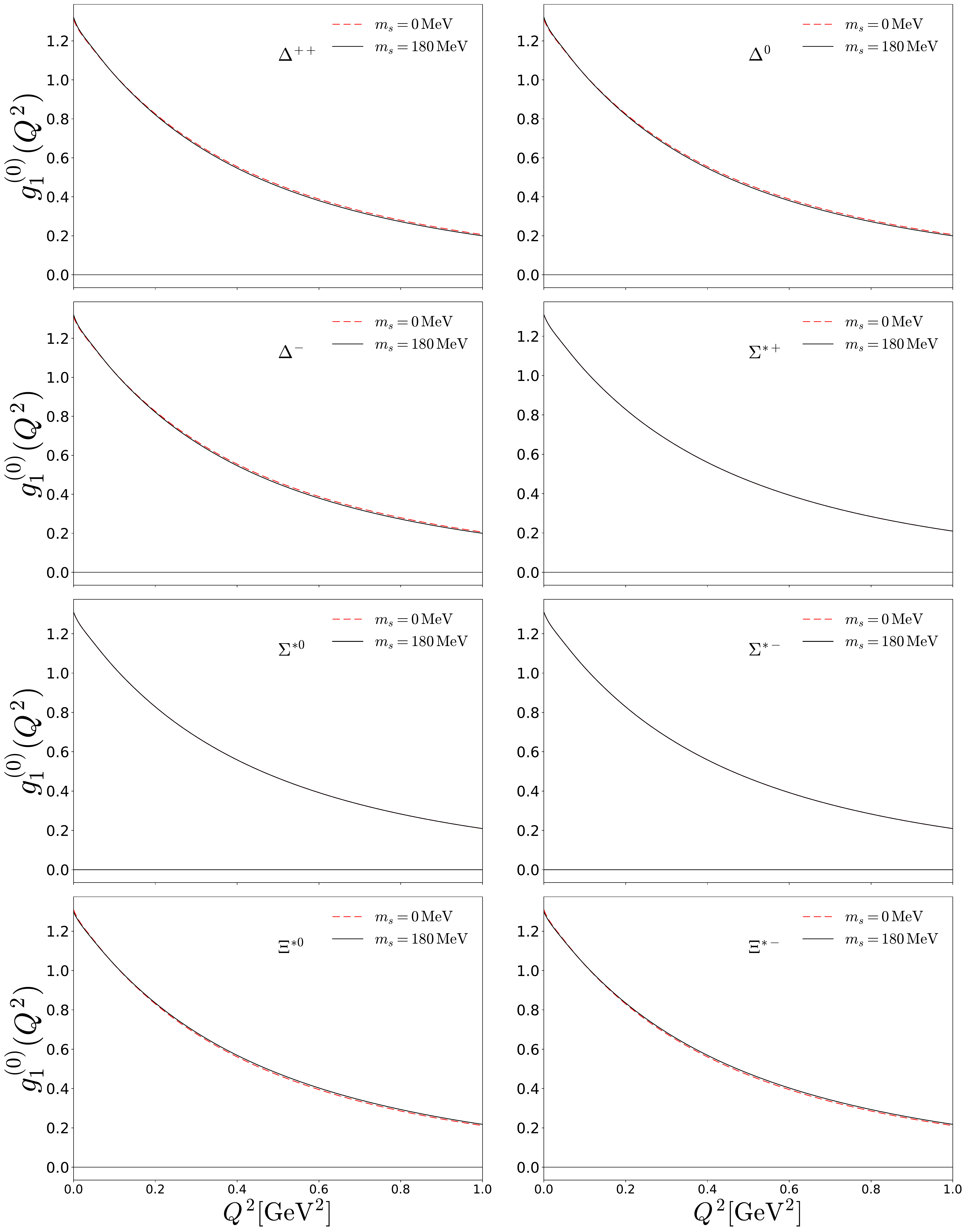}
\caption{Effects of the explicit flavor SU(3) symmetry breaking on
  $g^{(0)B}_{1}(Q^{2})$ of the baryon decuplet except for the 
  $\Delta^+$ and $\Omega^{-}$ baryons. Notations are the same as 
  in Fig.~\ref{fig:1}.}
\label{fig:7}
\end{figure}

\begin{figure}[ht]
  \includegraphics[scale=0.26]{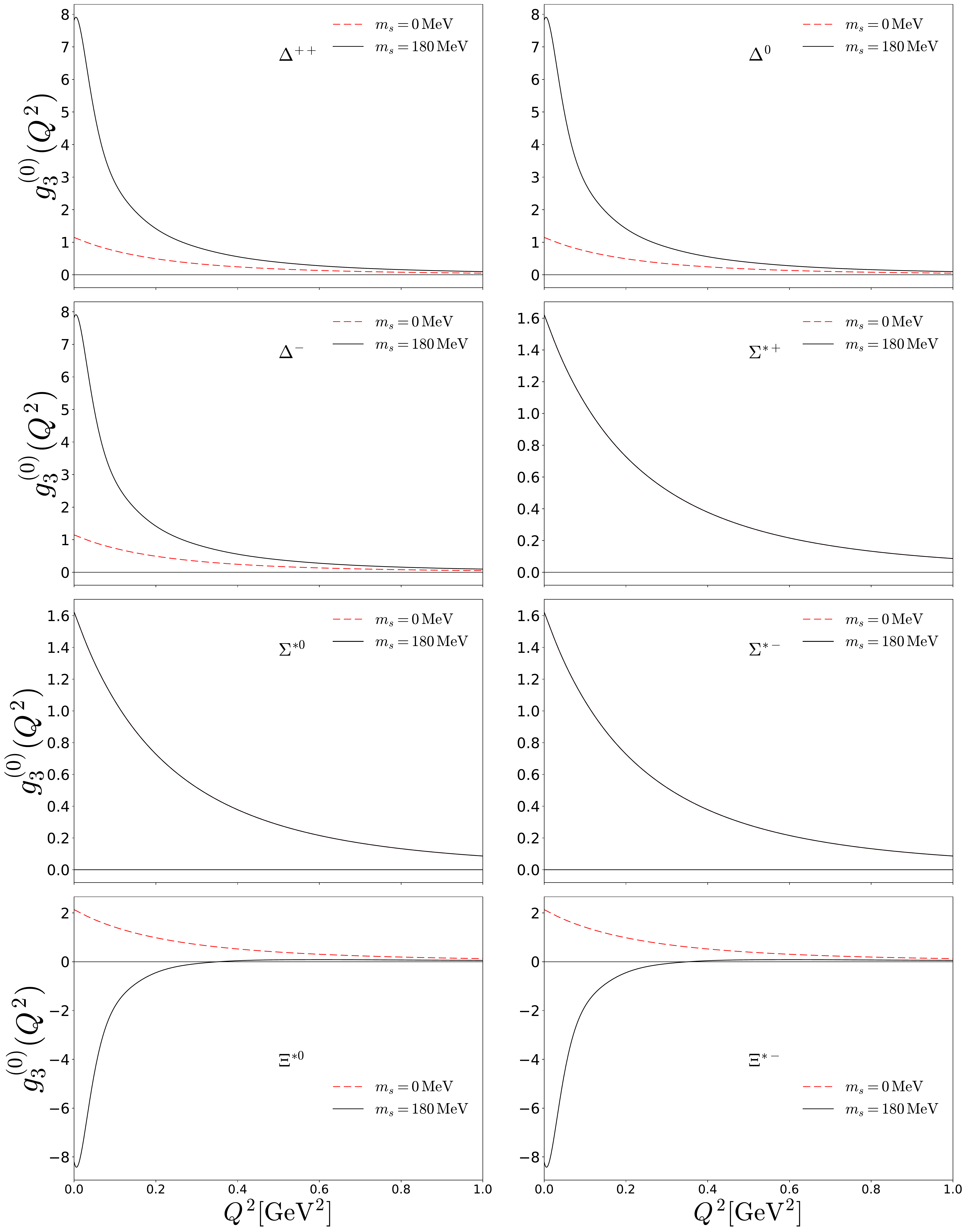}
\caption{Effects of the explicit flavor SU(3) symmetry breaking on
  $g^{(0)B}_{3}(Q^{2})$ of the baryon decuplet except for the 
  $\Delta^+$ and $\Omega^{-}$ baryons. Notations are the same as 
  in Fig.~\ref{fig:1}.}
\label{fig:8}
\end{figure}

\begin{figure}[ht]
  \includegraphics[scale=0.26]{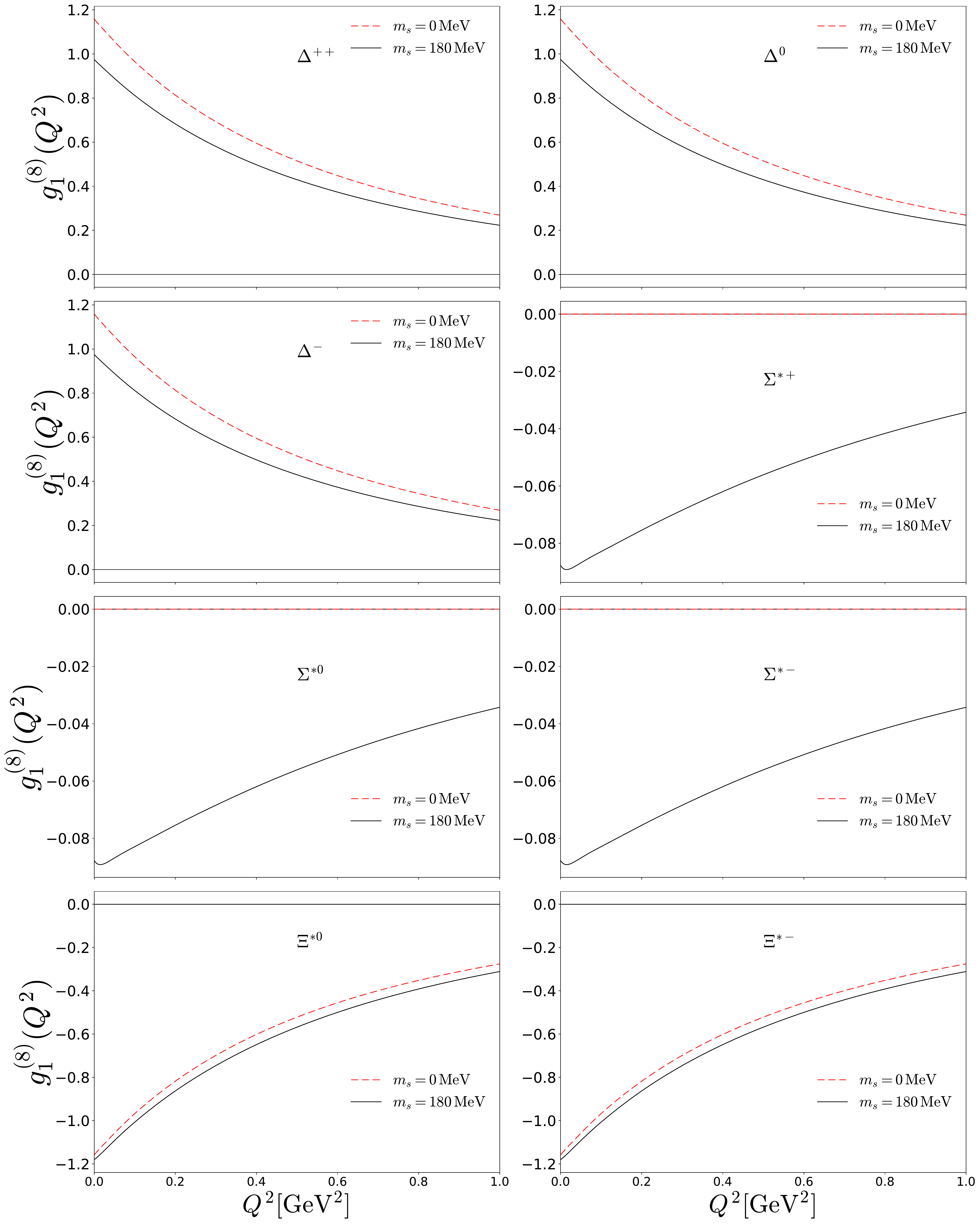}
\caption{Effects of the explicit flavor SU(3) symmetry breaking on
  $g^{(8)B}_{1}(Q^{2})$ of the baryon decuplet except for the 
  $\Delta^+$ and $\Omega^{-}$ baryons. Notations are the same as 
  in Fig.~\ref{fig:1}.}
\label{fig:9} 
\end{figure}

\begin{figure}[ht]
  \includegraphics[scale=0.26]{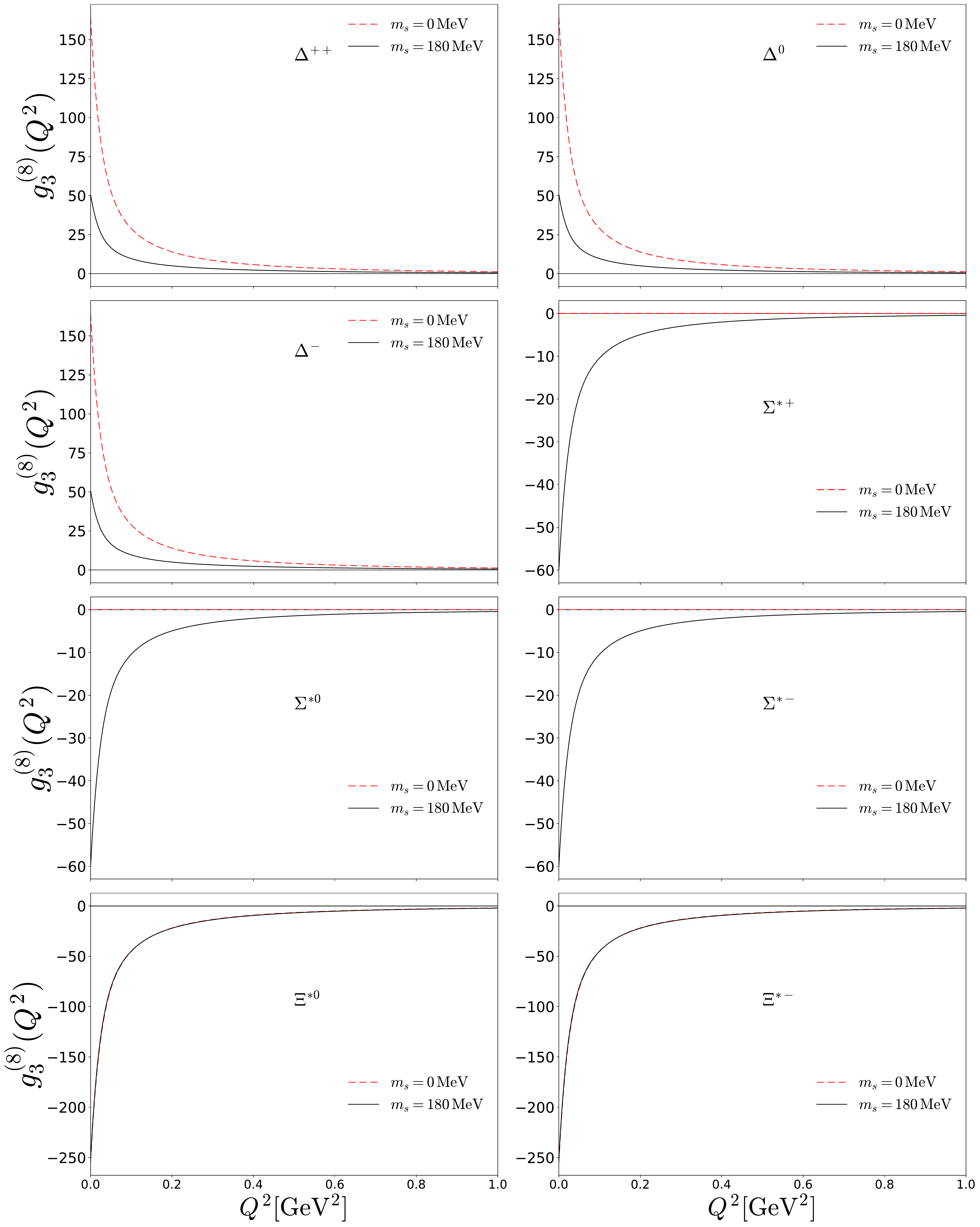}
\caption{Effects of the explicit flavor SU(3) symmetry breaking on
  $g^{(8)B}_{3}(Q^{2})$ of the baryon decuplet except for the 
  $\Delta^+$ and $\Omega^{-}$ baryons. Notations are the same as 
  in Fig.~\ref{fig:1}.}
\label{fig:10} 
\end{figure}
In Fig.~\ref{fig:5}, we show the numerical results for the triplet
axial-vector form factors $g_1^{(3)}(Q^2)$ of the baryon decuplet except
for $\Delta^+$ and $\Omega^-$. The effects of the flavor SU(3)
symmetry breaking are in general marginal. Figure~\ref{fig:6} presents
those for $g_3^{(3)}(Q^2)$ of all other members of the decuplet. The
tendency of the linear $m_{\mathrm{s}}$ corrections is basically the
same as that for $\Delta^+$ as discussed in Fig.~\ref{fig:1}. 
In Fig.~\ref{fig:7} we illustrate those for the singlet axial-vector form
factors $g_1^{(0)}(Q^2)$ for the other members of the decuplet. Again
the results look very similar to those of $\Delta^+$ and $\Omega^-$ as
explained in Fig.~\ref{fig:2}. As mentioned previously, if one
neglects the effects of flavor SU(3) symmetry breaking, then
$g_1^{(0)B} (Q^2)$ is flavor-independent as clearly shown in
Fig.~\ref{fig:7}. In Fig.~\ref{fig:8} we display the results for the
$g_3^{(0)}(Q^2)$ of the baryon decuplet except for the $\Delta^+$ and
$\Omega^-$. In particular, those of $\Sigma^*$ do not acquire any
contributions from the linear $m_{\mathrm{s}}$ corrections. This can
be easily understood by examining Eqs.~\eqref{eq:ga30opcorr} and
\eqref{eq:ga30wfcorr}. Since $\Sigma^*$ has hypercharge $Y=0$, the
linear $m_{\mathrm{s}}$ corrections from the effective chiral action
vanish. The wavefunction corrections do not exist at all for both
$g_1^{(0)}(Q^2)$ and $g_3^{(0)}(Q^2)$. Figures~\ref{fig:9} and
\ref{fig:10} depict respectively the results for $g_1^{(8)}(Q^2)$ and
$g_3^{(8)} (Q^2)$ for the other members of the baryon decuplet again
except for $\Delta^+$ and $\Omega^-$. Interestingly, there are 
no flavor SU(3) symmetric contributions to $g_1^{(8)}(Q^2)$ and
$g_3^{(8)}(Q^2)$ of $\Sigma^*$ because of the hypercharge of
$\Sigma^*$, as shown in Eqs.~\eqref{eq:ga18leading} and
\eqref{eq:ga38leading}. Thus, the linear $m_{\mathrm{s}}$ corrections
are considered to be the leading-order contributions. 

\begin{sidewaystable}[ht]
\caption{Numerical results for the triplet axial-vector constant 
  $g^{(3)B}_{1}(0)$ or the axial charge in comparison with those 
  from lattice QCD (LQCD)~\cite{Alexandrou:2013opa,Alexandrou:2016xok},
  the relativistic constituent quark model (RCQM)
  ~\cite{Choi:2013ysa,Choi:2010ty}, the Light-cone QCD sum rules (LCSR)
  ~\cite{Kucukarslan:2014bla}, chiral perturbation theory ($\chi$PT)
  ~\cite{Jiang:2008we} and the perturbarive chiral quark model (PCQM) 
  ~\cite{Liu:2018jiu}.} 
\label{tab:1}
 \begin{threeparttable}
 \renewcommand{\arraystretch}{1.4}
 {\setlength{\tabcolsep}{2pt}
 \begin{tabularx}{\linewidth}{ c | c c c c c c c c c c } 
  \hline 
  \hline 
$g^{(3)B}_{1}(0)$ & $\Delta^{++}$ & $\Delta^{+}$ & $\Delta^{0}$ 
& $\Delta^{-}$ & $\Sigma^{*+}$ & $\Sigma^{*0}$ & $\Sigma^{*-}$ 
& $\Xi^{*0}$ & $\Xi^{*-}$ & $\Omega^{-}$ \\
  \hline 
$m_{\mathrm{s}} = 0\;\mathrm{MeV}$ 
& $2.0064$ & $0.6688$ & $-0.6688$ & $-2.0064$ & $1.338$ & $0$ 
& $-1.338$ & $0.669$ & $-0.669$& $0$ \\ 
$m_{\mathrm{s}} = 180\;\mathrm{MeV}$ 
& $2.1333$ & $0.7111$ & $-0.7111$ & $-2.1333$ & $1.440$ & $0$ 
& $-1.440$ & $0.729$ & $-0.729$& $0$\\  
 \hline 
LQCD~\cite{Alexandrou:2016xok}\;($m_{\pi}=131.2(13)\;\mathrm{MeV}$)
& $-$ & $-$ & $-$ & $-$ & $1.1740(380)$ & $-$ & $-$ & $0.5891(198)$ 
& $-$ & $-$\\  
LQCD~\cite{Alexandrou:2016xok}\;($m_{\pi}=213\;\mathrm{MeV}$)
&$1.9777(1458)$ & $0.5181(981)$ & $-0.6499(973)$ & $-1.7090(1422)$ 
&$1.1929(521)$ & $-0.1367(685)$ & $-1.2633(516)$ & $0.5869(216)$ 
&$-0.6682(382)$ & $-$\\  
LQCD~\cite{Alexandrou:2016xok}\;($m_{\pi}=256\;\mathrm{MeV}$)
&$1.6956(1897)$ & $0.5670(1479)$ & $-0.5929(1167)$ & $-1.7322(1718)$ 
&$1.1462(720)$ & $0.0148(542)$ & $-1.0646(661)$ & $0.5785(278)$ 
& $-0.5424(303)$& $-$\\  
LQCD~\cite{Alexandrou:2013opa}\;($m_{\pi}=297\;\mathrm{MeV}$)
&$-$ & $0.604(38)$ & $-$ & $-$ & $-$ & $-$ & $-$ & $-$ & $-$ 
&$-$\\  
LQCD~\cite{Alexandrou:2016xok}\;($m_{\pi}=302\;\mathrm{MeV}$)
&$1.9574(1552)$ & $0.6374(976)$ & $-0.4798(1063)$ & $-1.4374(1331)$ 
&$1.2839(636)$ & $0.0654(444)$ & $-1.0423(619)$ & $0.6204(256)$ 
&$-0.5459(299)$ & $-$\\  
LQCD~\cite{Alexandrou:2013opa}\;($m_{\pi}=353\;\mathrm{MeV}$)
&$-$ & $0.640(26)$ & $-$ & $-$ & $-$ & $-$ & $-$ & $-$ & $-$ 
&$-$\\  
LQCD~\cite{Alexandrou:2016xok}\;($m_{\pi}=373\;\mathrm{MeV}$)
&$1.7602(1035)$ & $0.5215(639)$ & $-0.5676(635)$ & $-1.5872(1270)$ 
&$1.1478(558)$ & $-0.0130(323)$ & $-1.1139(485)$ & $0.5741(243)$ 
&$-0.5702(230)$ & $-$\\  
LQCD~\cite{Alexandrou:2013opa}\;($m_{\pi}=411\;\mathrm{MeV}$)
&$-$ & $0.571(18)$ & $-$ & $-$ & $-$ & $-$ & $-$ & $-$ & $-$ 
&$-$\\  
LQCD~\cite{Alexandrou:2016xok}\;($m_{\pi}=432\;\mathrm{MeV}$)
&$1.8520(875)$ & $0.6129(478)$ & $-0.5949(489)$ & $-1.8108(868)$ 
&$1.2228(473)$ & $0.0124(244)$ & $-1.1765(450)$ & $0.6059(213)$ 
&$-0.5885(223)$& $-$\\  
LQCD~\cite{Alexandrou:2013opa}\;($m_{\pi}=490\;\mathrm{MeV}$)
&$-$ & $0.578(13)$ & $-$ & $-$ & $-$ & $-$ & $-$ & $-$ & $-$ 
&$-$\\  
LQCD~\cite{Alexandrou:2013opa}\;($m_{\pi}=563\;\mathrm{MeV}$)
&$-$ & $0.5887(98)$ & $-$ & $-$ & $-$ & $-$ & $-$ & $-$ & $-$ 
&$-$\\  
\hline
$\chi$PT~\cite{Jiang:2008we} &$2.25^{*}$ & $-$ & $-$ & $-$ 
& $-$ & $-$ & $-$ & $-$ & $-$ & $-$\\
RCQM[GBE]~\cite{Choi:2013ysa,Choi:2010ty} &$2.24^{*}$ & $-$  
& $-$ & $-$ & $1.499^{\dagger}$ & $-$ & $-$ 
& $0.75^{\dagger}$ & $-$& $-$\\
LCSR~\cite{Kucukarslan:2014bla} &$2.70 \pm 0.6^{*}$ & $-$  
& $-$ & $-$ & $-$ & $-$ & $-$ & $-$ & $-$ & $-$\\
PCQM~\cite{Liu:2018jiu} & $1.863^{*}$ & $-$ & $-$ & $-$ 
& $1.242^{\dagger}$ & $-$ & $-$ & $0.621^{\dagger}$ 
& $-$ & $-$\\
 \hline 
 \hline
\end{tabularx}}
\begin{tablenotes}\footnotesize
\item[*] Since the expressions for the axial-vector constants in
  Refs.~\cite{Choi:2013ysa,Choi:2010ty,Jiang:2008we} and
  Refs.~\cite{Liu:2018jiu,Kucukarslan:2014bla} are different 
  from the present one by $-2$ and $-1$ respectively, 
  we have considered these factors for comparison.
\item[$\dagger$] Since the expressions for the axial-vector 
  constants of $\Sigma^{*}$ and $\Xi^{*}$ in
  Refs.~\cite{Choi:2010ty,Liu:2018jiu} are different from the present  
  definition by $-1/\sqrt{2}$ and $-1$ respectively, we have 
  considered these factors for comparison.
\end{tablenotes}
\end{threeparttable}
 \end{sidewaystable}
In Table~\ref{tab:1}, we list the results for the triplet axial-vector
constant, i.e. axial charge with the pion mass varied, intending to
compare them with those from lattice
QCD~\cite{Alexandrou:2013opa,Alexandrou:2016xok}. The third row 
presents the final results from this work with the value of the 
physical pion mass and the strange quark mass $m_{\mathrm{s}}=180$
MeV, whereas the second row gives those without linear
$m_{\mathrm{s}}$ corrections. The present value of the  
$\Delta^{++}$ axial charge is in good agreement with those from the RCQM 
and $\chi$PT. However, it is rather difficult to compare the present
results with those from lattice QCD. Interestingly, the present
results for the triplet axial-vector constants of the other members of
the baryon decuplet are in better agreement with the corresponding
lattice data. Note that the lattice data are consistently smaller than
the values obtained in the present work. Those of $\Sigma^{*+}$ and
$\Xi^{*0}$ from the RCQM are in very good agreement with the present 
results.

\begin{figure}[ht]
  \includegraphics[scale=0.3]{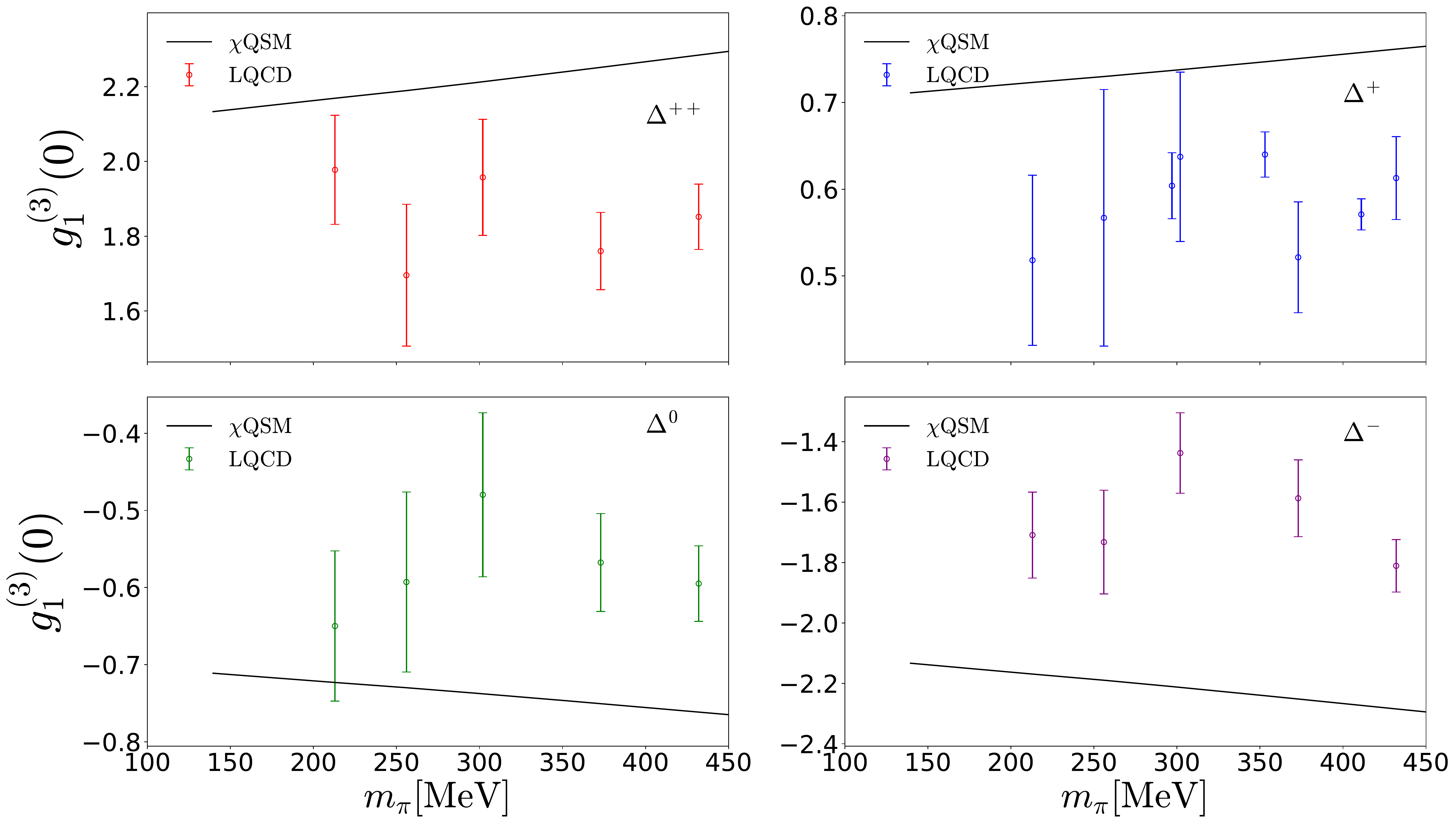}
\caption{Dependence of the triplet axial-vector constants
of the $\Delta$ isobars on the pion mass. The numerical results are
drawn in the solid curves, which are compared with those from lattice 
  QCD (LQCD)~\cite{Alexandrou:2013opa, Alexandrou:2016xok}.}  
\label{fig:11}
\end{figure}
\begin{figure}[ht]
  \includegraphics[scale=0.3]{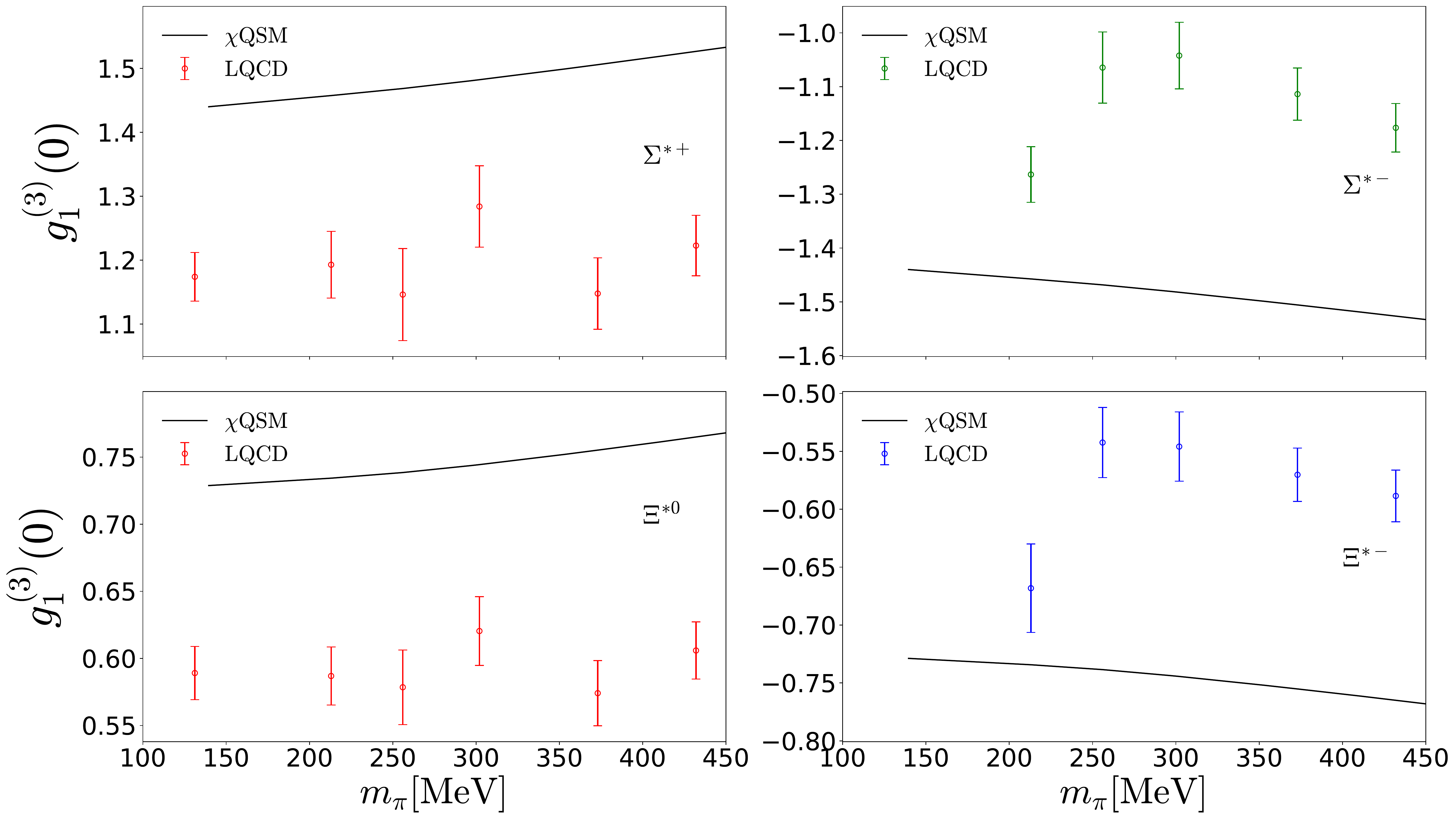}
\caption{Dependence of the triplet axial-vector constants
of the $\Sigma^*$ and $\Xi^*$ on the pion mass. The numerical results
are drawn in the solid curves, which are compared with those from
lattice   QCD (LQCD)~\cite{Alexandrou:2016xok}.}  
\label{fig:12}
\end{figure}
\begin{figure}[ht]
  \includegraphics[scale=0.3]{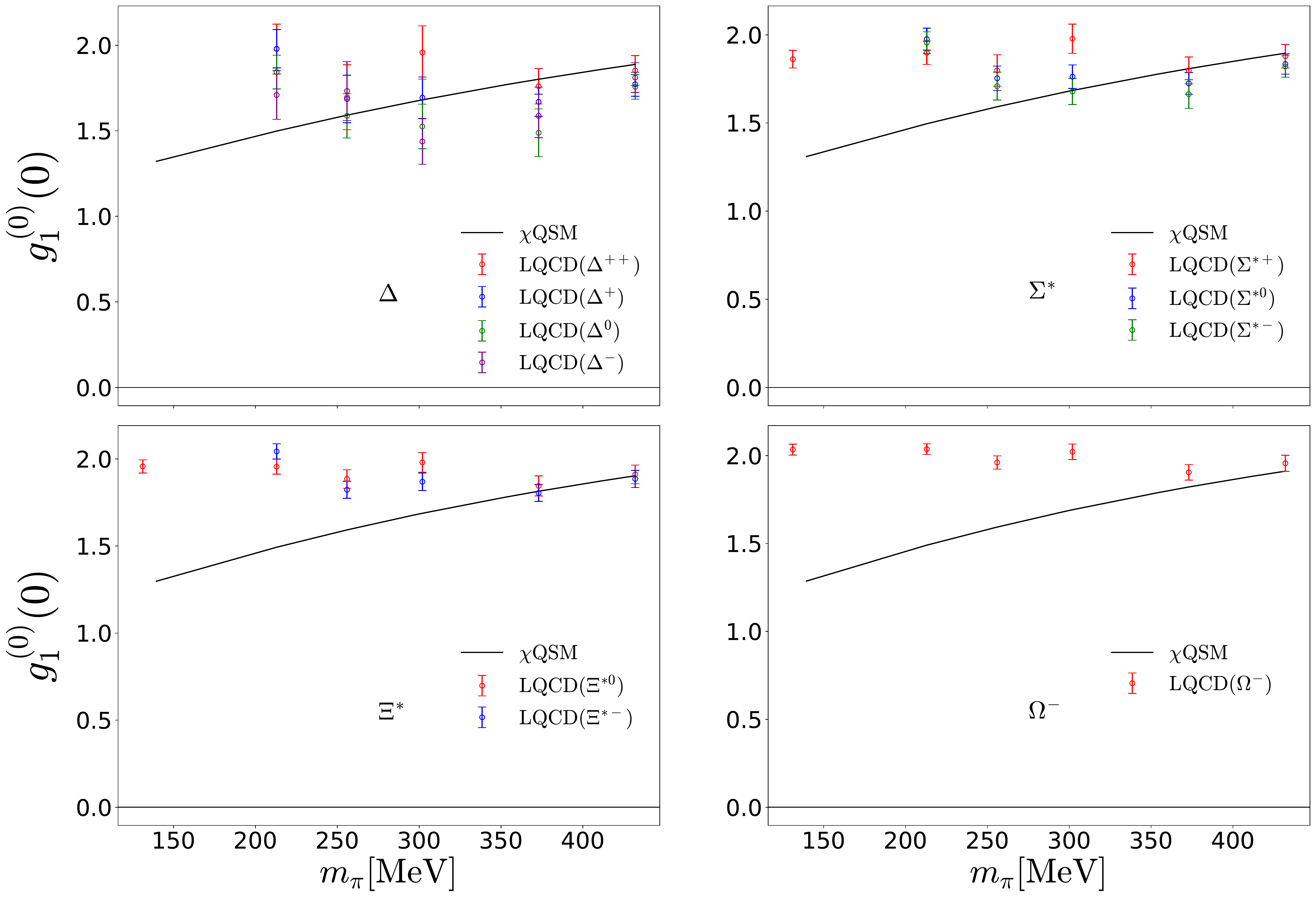}
\caption{Dependence of the singlet axial-vector constants
  $g^{(0)B}_{1}(0)$ on the pion mass. The numerical results are drawn
in the solid curves, which are compared with those from lattice
  QCD (LQCD)~\cite{Alexandrou:2016xok}.}  
\label{fig:13}
\end{figure}
\begin{figure}[ht]
  \includegraphics[scale=0.3]{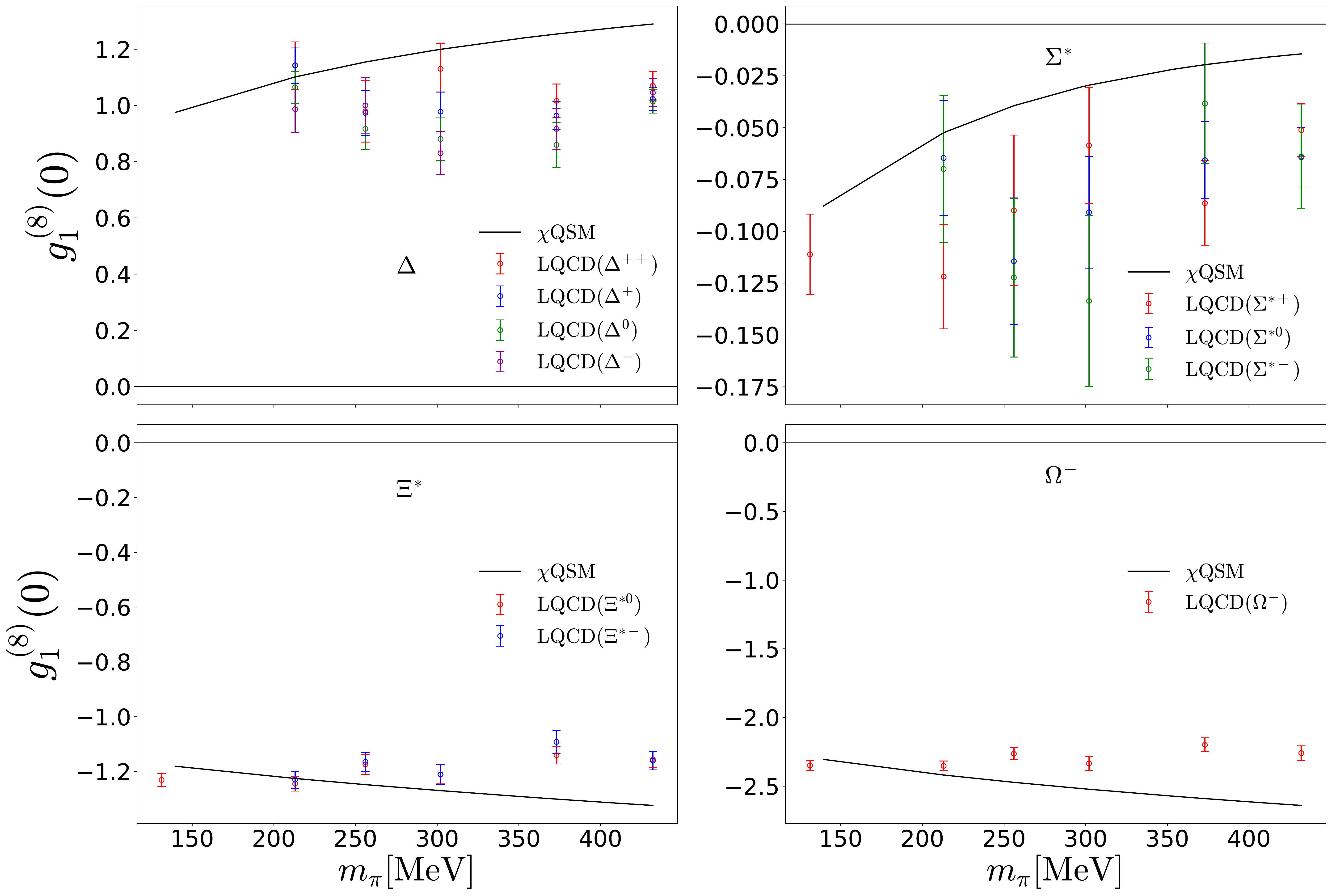}
\caption{Dependence of the octet axial-vector constant
  $g^{(8)B}_{1}(0)$ on the pion mass. The numerical results are drawn
in the solid curves, which are compared with those from lattice
  QCD (LQCD)~\cite{Alexandrou:2016xok}. Note that 
the expressions for the octet axial-vector constants in
  Ref.~\cite{Alexandrou:2016xok} are different from the present one 
 by $\sqrt{3}$, so we have considered it for comparison.}  
\label{fig:14}
\end{figure}
In Figs.~\ref{fig:11} and~\ref{fig:12}, we show respectively the
numerical results for the triplet axial-vector constants of the
$\Delta$ isobars, $\Sigma^*$ and $\Xi^*$ as functions of the pion
mass, compared them with the lattice data. The magnitudes of
$g_1^{(3)B}(0)$ generally increase as the value of $m_\pi$ increases.   
The present results turn out larger than those of lattice QCD.  
Figure~\ref{fig:13} depicts the numerical results for the singlet
axial-vector constants of the baryon decuplet as functions of the
pion mass in comparison with the lattice
data~\cite{Alexandrou:2016xok}. As we have mentioned already,
the values of $g_1^{(0)B}(0)$ of the baryon decuplet are almost the
same each other. As the pion mass grows larger, the magnitudes of
$g_1^{(0)B}(0)$ monotonically increase. When $m_\pi=432$ MeV is used,
those of $g_1^{(0)B}(0)$ become larger by about $30\,\%$. Interestingly,
the present results get closer to the lattice data as the value of
the pion mass increases. They are in very good agreement with the
lattice data at $m_\pi=432$ MeV. Note that in the present framework
the singlet axial-vector constants are isospin symmetric. In
Fig.~\ref{fig:14}, we compare the results for the octet axial-vector
constants of the baryon decuplet with the corresponding lattice data
with the pion mass varied. Again, the magnitudes of the octet
axial-vector constants also rise as the pion mass increases as in the
case of $g_1^{(0)B}(0)$. However, when we compare the present results
with the lattice data, the situation turns out opposite. That is, the 
present results tend generally to deviate, except for the $\Sigma^*$,
from the lattice ones as the pion mass increases. When it comes to the
case of $\Sigma^*$, $g_1^{(8)\Sigma^*}(0)$ exhibits dependence on
$m_\pi$ similar to the corresponding lattice one. The results for
$g_1^{(8)B}(0)$ are in good agreement with the lattice data at
$m_\pi=213$ MeV. 

\begin{table}[ht]
\renewcommand{\arraystretch}{1.7}
\caption{Numerical results for the flavor axial-vector constants except 
  for the axial-vector constants 
  $g_1^{(\com{a})}(0)$, axial masses and axial radii. All the results 
  are obtained with flavor SU(3) symmetry breaking taken into account.}
\label{tab:2}
{\setlength{\tabcolsep}{6pt}
 \begin{tabular}{ c | c c c c c c c c c c c }
  \hline 
  \hline 
 $m_{\mathrm{s}} = 180\;\mathrm{MeV}$ & $\Delta^{++}$ & $\Delta^{+}$ 
 & $\Delta^{0}$ & $\Delta^{-}$ & $\Sigma^{*+}$ & $\Sigma^{*0}$ 
 & $\Sigma^{*-}$ & $\Xi^{*0}$ & $\Xi^{*-}$ & $\Omega^{-}$ \\
  \hline 
$g^{(3)B}_{3}(0)$ & $346.1$ & $115.4$ & $-115.4$ & $-346.1$ 
& $303.9$ & $0$ & $-303.9$ & $193.7$ & $-193.7$ & $0$\\  
$ g^{(0)B}_{3}(0) $ & $7.822$ & $7.822$ & $7.822$ & $7.822$
& $1.622$ & $1.622$ & $1.622$ & $-8.204$ & $-8.204$ & $-21.936$\\  
$ g^{(8)B}_{3}(0) $ & $50.8$ & $50.8$ & $50.8$ & $50.8$
& $-60.0$ & $-60.0$ & $-60.0$ & $-251.9$ & $-251.9$ & $-542.8$\\  
 \hline
$\langle r^{2}_{A}\rangle_B\;[\mathrm{fm^{2}}]$ & $0.447$ & $0.447$ 
& $0.447$ & $0.447$ & $0.438$ & $-$ & $0.438$ & $0.431$ & $0.431$ & $-$\\  
$M_{A}\;[\mathrm{GeV}]$ & $1.023$ & $1.023$ & $1.023$ & $1.023$
& $1.033$ & $-$ & $1.033$ & $1.041$ & $1.041$ & $-$\\  
\hline
 \hline
\end{tabular}}
\end{table}
Table~\ref{tab:2} lists the numerical results for $g_3^{(3)B}(0)$,
$g_3^{(0)B}(0)$ and $g_3^{(8)B}(0)$, respectively, from the second row
till the fourth row. Since there are no lattice data and no results
from other works, they are the very first results for the second set
of the axial-vector constants. The fifth row lists the results for the
axial radii, which can be derived from the results for the triplet
axial-vector form factors of the baryon decuplet as follows
\begin{align}
\langle r^{2}_{A} \rangle_B & = \frac{-6}{g^{(3)B}_{1}(0)} \left. 
\frac{\partial{g^{(3)B}_{1}(Q^{2})}}{\partial{Q^{2}}}
\right|_{Q^{2}=0}.
\label{eq:axradii}
\end{align}
Note that when the strangeness of a decuplet baryon increases, the
value of the axial radius becomes smaller, as shown in
Table~\ref{tab:2}. This can be understood, since the corresponding
mass becomes larger due to the strange-quark component. It is of great
interest to compare the 
axial radius of the $\Delta^+$ with that of the proton, since
the axial radius of the proton is experimentally
known. In a recent review~\cite{Hill:2017wgb}, the average value of
the proton axial radius is given as $\langle r_A^2
\rangle_p=0.46(22)\,\mathrm{fm}^2$. Interestingly, the result obtained
in the present work for the $\Delta^+$ axial radius is
$0.447\,\mathrm{fm}^2$, which is very similar to that of the
proton. We want to mention that $\langle
r_A^2\rangle_p=0.536\,\mathrm{fm}^2$ was obtained within the  
same framework, i.e. the $\chi$QSM~\cite{Silva:2005fa}. This 
indicates that the present results for the triplet axial-vector form 
factor of $\Delta^+$ fall off more slowly than the proton one.

A baryon form factor is often parametrized in terms of a dipole-type
parametrization given by 
\begin{align}
g_1^{(3)B}(Q^{2}) &=\frac{g^{(3)B}_{1}(0)}{\left(1
  +\frac{Q^{2}}{M_{A}^2} \right)^{2}},
\label{eq:dipoleansatz}
\end{align}
where $M_A$ is known as the axial mass. This parametrization relates
the axial mass to the axial radius by the following relation
\begin{align}
  \label{eq:axialmass}
\langle r_A^2\rangle = \frac{12}{M_A^2}.  
\end{align}
The value of $M_A$ for the proton is also known
experimentally~\cite{Liesenfeld:1999mv} whereas those of 
the baryon decuplet are unknown. Equation~\eqref{eq:axialmass} 
already implies that the present result for the $\Delta^+$ axial mass
should be larger than the proton one that was obtained also in
Ref.~\cite{Silva:2005fa}, $M_A(p)=0.934\,\mathrm{GeV}$. Indeed, the
present result $M_A(\Delta^+)=1.023\,\mathrm{GeV}$ is larger than
that.

Finally, we want to consider another type of the parametrization for
the axial-vector form factors. In lattice calculations, a $p$-pole
parametrization is often adopted~\cite{Gockeler:2006zu, 
Brommel:2006ww,Brommel:2007xd}, which can be expressed as 
\begin{align}
g_i^{(3)B}(Q^{2})=\frac{g_i^{(3)B}(0)}{\left(1
  +\frac{Q^{2}}{p_i\Lambda_{p_i}^{2}}\right)^{p_i}}.
\label{eq:ppoleansatz}
\end{align}
As drawn in the right panel of Fig.~\ref{fig:1},
it is rather difficult to fit $g_3^{(3)\Delta^+}(Q^2)$ by using the
dipole-type parametrization. On the other hand, if one parametrizes
$g_3^{(3)\Delta^+}(Q^2)$ by the $p$-pole type~\eqref{eq:ppoleansatz},
then we are able to parametrize $g_3^{(3)\Delta^+}(Q^2)$ by fixing the
values of $p_3=1.472$ and
$\Lambda_{p_3}=0.174\,\mathrm{GeV}$. Similarly, $g_3^{(8)}(Q^2)$ can 
be fitted by using the $p$-pole parametrization. 

\section{Summary and conclusion \label{sec:5}}
We aimed at investigating the axial-vector form factors of the baryon
decuplet within the framework of the self-consistent chiral
quark-soliton model. We consider the rotational $1/N_c$ corrections
and the linear $m_{\mathrm{s}}$ corrections. Since all the parameters
in the model were fixed by reproducing the proton properties, we did
not have any parameter to fit. We first computed the triplet
axial-vector form factors of the $\Delta^+$, because lattice QCD and
all other models concentrated on them. We found that the effects of
flavor SU(3) symmetry breaking turn out very small on the triplet form
factors of $\Delta^+$. We then proceeded to compute the singlet
axial-vector form factors $g_{1,3}^{(0)}(Q^{2})$. In this case, there is no
leading-order contribution, so that the rotational $1/N_c$ and linear 
$m_{\mathrm{s}}$ corrections are only involved. Concerning the 
$g_1^{(0)B}(Q^{2})$ form factors, the linear $m_{\mathrm{s}}$ 
corrections are almost negligible.
On the other hand, $g_3^{(0)}(Q^{2})$ form factors acquire in
general large contributions from the $m_s$ corrections. 
We then derived the octet axial-vector form factors of the baryon 
decuplet. The effects of flavor SU(3) symmetry breaking on 
$g_1^{(8)}(Q^{2})$ are in general very small. However, these linear 
$m_{\mathrm{s}}$ corrections come into play as leading-order 
contributions in the case of the $\Sigma^*$ octet axial-vector 
form factors, since the symmetric parts vanish because of the values 
of their hypercharges. The octet form factors $g_3^{(8)}(Q^{2})$ of 
the $\Delta$ isobars get large $m_{\mathrm{s}}$ contributions whereas
those of $\Xi^*$ and $\Omega^-$ receive tiny $m_{\mathrm{s}}$
corrections.  We have carefully inspected the dependence of the
axial-vector constants as functions of the pion mass to compare the
present results with those from lattice QCD. We found that the results
of the axial-vector constants turn out larger than the physical ones,
when the unphysical values of the pion mass are employed. The
magnitudes of the triplet axial-vector constants are in general larger
than the lattice data. When it comes to the singlet axial-vector
constants, the present results are in very good agreement with the
lattice data with $m_\pi=432$ MeV used. On the other hand, the results
for the octet axial-vector constants are in better agreement with
the lattice ones at $m_\pi=213$ MeV.

We also presented the results for the axial radii and axial masses of
the baryon decuplet. We found that the axial radius of $\Delta^+$ is
very close to the experimental data on the proton axial
radius. Compared with the value of the proton axial radius derived
from the same model, we observed that the $\Delta^+$ axial radius is
smaller than that of the proton. It indicates that the triplet
axial-vector form factor of $\Delta^+$ falls off more slowly than the
proton one. When the strangeness content of a
decuplet baryon becomes larger, the corresponding axial radius gets
smaller. It indicates that the mass of a baryon may be connected to
the axial radius. We also obtained the axial masses, which can be
regarded as the inverse of the axial radii. Since the $p$-pole
parametrization of hadronic form factors is often employed 
in lattice calculations, we parametrized the present results of the 
axial-vector form factors, in particular, of the triplet ones, and 
determined the $p$ and cutoff mass $\Lambda_p$, hoping that results 
from lattice QCD will appear in the near future. 

Since we have computed all possible axial-vector form factors with
three different flavors, we are able to express the axial-vector form
factors in terms of the flavor-decomposed form factors. This is also
very interesting, because we can scrutinize the strange-quark spin
content of the $\Delta$ isobars and the up- and down-quark spin content
of the $\Omega^-$ hyperon. The corresponding work will appear
elsewhere.  In addition, we can also investigate the transition
axial-vector form factors of the baryon decuplet, which will provide
further understanding of the structure of the baryon decuplet. The
relevant investigation is under way. 

\begin{acknowledgments}
The authors are very grateful to J.-Y. Kim and Gh.-S. Yang for valuable
discussions and comments. The present work was supported by Basic
Science Research Program through the National Research Foundation of
Korea funded by the Ministry of Education, Science and Technology 
(2018R1A2B2001752 and 2018R1A5A1025563).
\end{acknowledgments} 

\appendix
\section{Components of the axial-vector form factors} \label{app:A}
In this Appendix, the $Q^2$-dependent functions in
Eqs.~\eqref{eq:gatrioct1} and~\eqref {eq:gatrioct3} will be expressed
explicitly. $\mathcal{A}_0^B(Q^2)$, $\cdots$, $\mathcal{J}_0^B(Q^2)$
are defined by
\begin{align}
\mathcal{A}^{B}_{0}(Q^{2}) &= \frac{N_{c} M_{B}}{E_{B}} 
  \int d^{3} r j_{0}(Q |\bm{r}|) \left[ \phi^{\dagger}_{\mathrm{val}}
  (\bm{r}) \bm{\sigma} \cdot \bm{\tau} \phi_{\mathrm{val}}(r)
  + \sum_{n} \phi^{\dagger}_{n}(\bm{r}) \bm{\sigma} \cdot \bm{\tau} 
  \phi_{n}(\bm{r}) \mathcal{R}_{1}(E_n) \right] ,\\
\mathcal{B}^{B}_{0}(Q^{2}) &= \frac{N_{c} M_{B}}{E_{B}} 
  \int d^{3} r j_{0}(Q |\bm{r}|) \left[ \sum_{n \ne
  \mathrm{val} } \frac{1}{E_{\mathrm{val}}-E_{n}} 
  \phi^{\dagger}_{\mathrm{val}}(\bm{r}) \bm{\sigma} 
  \phi_{n}(\bm{r}) \cdot \langle n | \bm{\tau} | \mathrm{val} \rangle 
  \right. \cr
& \left. \hspace{3.9cm} 
  -\frac{1}{2} \sum_{n,m} \phi^{\dagger}_{n}(\bm{r}) \bm{\sigma} 
  \phi_{m}(\bm{r}) \cdot \langle m | \bm{\tau} | n \rangle 
  \mathcal{R}_{5}(E_n,E_m) \right],\\ 
\mathcal{C}^{B}_{0}(Q^{2}) &= \frac{N_{c} M_{B}}{E_{B}} 
  \int d^{3} r j_{0}(Q |\bm{r}|) \left[ \sum_{n_{0} \ne
  \mathrm{val} } \frac{1}{E_{\mathrm{val}}-E_{n_{0}}} 
  \phi^{\dagger}_{\mathrm{val}}(\bm{r}) \bm{\sigma} \cdot \bm{\tau} 
  \phi_{n_{0}}(\bm{r}) \langle n_{0} | \mathrm{val} \rangle \right. \cr
& \left. \hspace{3.9cm} 
  -\sum_{n,m_{0}} \phi^{\dagger}_{n}(\bm{r}) \bm{\sigma} \cdot \bm{\tau} 
  \phi_{m_{0}}(\bm{r}) \langle m_{0} | n \rangle 
  \mathcal{R}_{5}(E_n,E_{m_{0}}) \right], \\
\mathcal{D}^{B}_{0}(Q^{2}) &= \frac{N_{c} M_{B}}{E_{B}} 
  \int d^{3} r j_{0}(Q |\bm{r}|) \left[ \sum_{n \ne
  \mathrm{val} } \frac{\mathrm{sgn}(E_{n})}{E_{\mathrm{val}}-E_{n}} 
  \phi^{\dagger}_{\mathrm{val}}(\bm{r}) (\bm{\sigma} \times \bm{\tau})
  \phi_{n}(\bm{r}) \cdot \langle n | \bm{\tau} | \mathrm{val} 
  \rangle \right. \cr
& \left. \hspace{3.9cm} 
  + \frac{1}{2} \sum_{n,m} \phi^{\dagger}_{n}(\bm{r}) \bm{\sigma} 
  \times \bm{\tau} \phi_{m}(\bm{r}) \cdot \langle m | \bm{\tau} 
  | n \rangle \mathcal{R}_{4}(E_n,E_m) \right],\\
\mathcal{H}^{B}_{0}(Q^{2}) &= \frac{N_{c} M_{B}}{E_{B}} 
  \int d^{3} r j_{0}(Q |\bm{r}|) \left[ \sum_{n \ne
  \mathrm{val} } \frac{1}{E_{\mathrm{val}}-E_{n}} 
  \phi^{\dagger}_{\mathrm{val}}(\bm{r}) \bm{\sigma} \cdot \bm{\tau} 
  \langle \bm{r} | n \rangle \langle n | \gamma^{0}| \mathrm{val} 
  \rangle \right. \cr
& \left. \hspace{3.9cm} 
  + \frac{1}{2} \sum_{n,m} \phi^{\dagger}_{n}(\bm{r}) \bm{\sigma} \cdot 
  \bm{\tau} \phi_{m}(\bm{r}) \langle m | \gamma^{0} | n \rangle 
  \mathcal{R}_{2}(E_n,E_m) \right], \\
\mathcal{I}^{B}_{0}(Q^{2}) &= \frac{N_{c} M_{B}}{E_{B}} 
  \int d^{3} r j_{0}(Q |\bm{r}|) \left[ \sum_{n \ne
  \mathrm{val} } \frac{1}{E_{\mathrm{val}}-E_{n}} 
  \phi^{\dagger}_{\mathrm{val}}(\bm{r}) \bm{\sigma} \phi_{n}(\bm{r}) 
  \cdot \langle n | \gamma^{0} \bm{\tau} | \mathrm{val} \rangle 
  \right. \cr
& \left. \hspace{3.9cm} +\frac{1}{2} \sum_{n,m} \phi^{\dagger}_{n}
  (\bm{r}) \bm{\sigma} \phi_{m}(\bm{r}) \cdot \langle m | \gamma^{0} 
  \bm{\tau} | n \rangle \mathcal{R}_{2}(E_n,E_m)\right],\\
\mathcal{J}^{B}_{0}(Q^{2}) &= \frac{N_{c} M_{B}}{E_{B}} 
  \int d^{3} r j_{0}(Q |\bm{r}|) \left[ \sum_{n_{0} \ne
  \mathrm{val} } \frac{N_{c}}{E_{\mathrm{val}}-E_{n_{0}}} 
  \phi^{\dagger}_{\mathrm{val}}(\bm{r}) \bm{\sigma} \cdot \bm{\tau} 
  \phi_{n_{0}}(\bm{r}) \langle n_{0}| \gamma^{0} | \mathrm{val} \rangle 
  \right. \cr
& \left. \hspace{3.9cm} +N_{c} \sum_{n,m_{0}} \phi^{\dagger}_{n}(\bm{r}) 
  \bm{\sigma} \cdot \bm{\tau} \phi_{m_{0}}(\bm{r}) 
  \langle m_{0}| \gamma^{0} | n \rangle \mathcal{R}_{2}(E_n,E_{m_{0}}) 
  \right].
\label{AxComp10}
\end{align}
where the regularization functions are defined as 
\begin{align}
\mathcal{R}_{1}(E_{n}) &= \frac{-E_{n}}{2 \sqrt{\pi}} \int^{\infty}_{0}
  \phi(u) \frac{du}{\sqrt{u}} e^{-u E_{n}^{2}}, \\
\mathcal{R}_{2}(E_{n},E_{m}) &= \frac{1}{2 \sqrt{\pi}} \int^{\infty}_{0}
  \phi(u) \frac{du}{\sqrt{u}} \frac{ E_{m} e^{-u E_{m}^{2}} 
  -E_{n}e^{-uE_{n}^{2}}}{E_{n} - E_{m}}, \\
\mathcal{R}_{4}(E_{n},E_{m}) &= \frac{1}{2 \pi} \int^{\infty}_{0} du
  \, \phi(u) \int^{1}_{0} d\alpha e^{-\alpha uE^{2}_{m} 
  -(1-\alpha)uE^{2}_{n}} \frac{(1-\alpha)E_{n}-\alpha E_{m}}
  {\sqrt{\alpha(1-\alpha)}}, \\
\mathcal{R}_{5}(E_{n},E_{m}) &=
  \frac{\mathrm{sgn}(E_{n})-\mathrm{sgn}(E_{m})}{2(E_{n}-E_{m})}.
\end{align}
Here, $|\mathrm{val}\rangle$ and $|n\rangle$ denote the state of 
the valence and sea quarks with the corresponding eigenenergies
$E_{\mathrm{val}}$ and $E_n$ of the one-body Dirac 
Hamiltonian $h(U)$, respectively.

$\mathcal{A}_2^B (Q^2)$, $\cdots$, $\mathcal{J}_2^B (Q^2)$ are
defined by 
\begin{align}
\mathcal{A}^{B}_{2}(Q^{2}) &= \frac{N_{c} M_{B}}{E_{B}} 
  \int d^{3} r j_{2}(Q |\bm{r}|) \left[ \phi^{\dagger}_{\mathrm{val}}
  (\bm{r}) \left\{ \sqrt{2\pi}Y_{2} \otimes \sigma_{1}\right\}_{1} \cdot 
  \bm{\tau} \phi_{\mathrm{val}}(\bm{r}) \right. \cr
& \left. \hspace{3.9cm} 
  +\sum_{n} \phi^{\dagger}_{n}(\bm{r}) \left\{ \sqrt{2\pi}Y_{2} 
  \otimes \sigma_{1}\right\}_{1} \cdot \bm{\tau}
  \phi_{n}(\bm{r}) \mathcal{R}_{1}(E_n) \right] ,\\
\mathcal{B}^{B}_{2}(Q^{2}) &= \frac{N_{c} M_{B}}{E_{B}} 
  \int d^{3} r j_{2}(Q |\bm{r}|) \left[ \sum_{n \ne
  \mathrm{val} } \frac{1}{E_{\mathrm{val}}-E_{n}} 
  \phi^{\dagger}_{\mathrm{val}}(\bm{r}) \left\{ \sqrt{2\pi}Y_{2} \otimes 
  \sigma_{1}\right\}_{1} \phi_{n}(\bm{r}) \cdot 
  \langle n | \bm{\tau} | \mathrm{val} \rangle \right. \cr
& \left. \hspace{3.9cm} 
  -\frac{1}{2} \sum_{n,m} \phi^{\dagger}_{n}(\bm{r}) 
  \left\{ \sqrt{2\pi}Y_{2} \otimes \sigma_{1}\right\}_{1} 
  \phi_{m}(\bm{r}) \cdot \langle m | \bm{\tau} | n \rangle 
  \mathcal{R}_{5}(E_n,E_m) \right],\\ 
\mathcal{C}^{B}_{2}(Q^{2}) &= \frac{N_{c} M_{B}}{E_{B}} 
  \int d^{3} r j_{2}(Q |\bm{r}|) \left[ \sum_{n_{0} \ne
  \mathrm{val} } \frac{1}{E_{\mathrm{val}}-E_{n_{0}}} 
  \phi^{\dagger}_{\mathrm{val}}(\bm{r}) \left\{ \sqrt{2\pi}Y_{2} \otimes 
  \sigma_{1}\right\}_{1} \cdot \bm{\tau} \phi_{n_{0}}(\bm{r}) 
  \langle n_{0} | \mathrm{val} \rangle \right. \cr
& \left. \hspace{3.9cm} 
  -\sum_{n,m_{0}} \phi^{\dagger}_{n}(\bm{r}) \left\{ \sqrt{2\pi}Y_{2} 
  \otimes \sigma_{1}\right\}_{1} \cdot \bm{\tau} 
  \phi_{m_{0}}(\bm{r}) \langle m_{0} | n \rangle 
  \mathcal{R}_{5}(E_n,E_{m_{0}}) \right], \\
\mathcal{D}^{B}_{2}(Q^{2}) &= \frac{N_{c} M_{B}}{E_{B}} 
  \int d^{3} r j_{2}(Q |\bm{r}|) \left[ \sum_{n \ne
  \mathrm{val} } \frac{\mathrm{sgn}(E_{n})}{E_{\mathrm{val}}-E_{n}} 
  \phi^{\dagger}_{\mathrm{val}}(\bm{r}) \left\{ \sqrt{2\pi}Y_{2} \otimes 
  \sigma_{1}\right\}_{1} \times \bm{\tau} 
  \phi_{n}(\bm{r}) \cdot \langle n | \bm{\tau} | \mathrm{val} 
  \rangle \right. \cr
& \left. \hspace{3.9cm} 
  + \frac{1}{2} \sum_{n,m} \phi^{\dagger}_{n}(\bm{r}) 
  \left\{ \sqrt{2\pi}Y_{2} \otimes \sigma_{1}\right\}_{1} 
  \times \bm{\tau} \phi_{m}(\bm{r}) \cdot \langle m | \bm{\tau} 
  | n \rangle \mathcal{R}_{4}(E_n,E_m) \right],\\
\mathcal{H}^{B}_{2}(Q^{2}) &= \frac{N_{c} M_{B}}{E_{B}} 
  \int d^{3} r j_{2}(Q |\bm{r}|) \left[ \sum_{n \ne
  \mathrm{val} } \frac{1}{E_{\mathrm{val}}-E_{n}} 
  \phi^{\dagger}_{\mathrm{val}}(\bm{r}) \left\{ \sqrt{2\pi}Y_{2} \otimes 
  \sigma_{1}\right\}_{1} \cdot \bm{\tau} \langle \bm{r} 
  | n \rangle \langle n | \gamma^{0}| \mathrm{val} \rangle \right. \cr
& \left. \hspace{3.9cm} 
  + \frac{1}{2} \sum_{n,m} \phi^{\dagger}_{n}(\bm{r}) 
  \left\{ \sqrt{2\pi}Y_{2} \otimes \sigma_{1}\right\}_{1} \cdot 
  \bm{\tau} \phi_{m}(\bm{r}) \langle m | \gamma^{0} | n \rangle 
  \mathcal{R}_{2}(E_n,E_m) \right], \\
\mathcal{I}^{B}_{2}(Q^{2}) &= \frac{N_{c} M_{B}}{E_{B}} 
  \int d^{3} r j_{2}(Q |\bm{r}|) \left[ \sum_{n \ne
  \mathrm{val} } \frac{1}{E_{\mathrm{val}}-E_{n}} 
  \phi^{\dagger}_{\mathrm{val}}(\bm{r}) \left\{ \sqrt{2\pi}Y_{2} \otimes 
  \sigma_{1}\right\}_{1} \phi_{n}(\bm{r}) \cdot 
  \langle n | \gamma^{0} \bm{\tau} | \mathrm{val} \rangle \right. \cr
& \left. \hspace{3.9cm} +\frac{1}{2} \sum_{n,m} 
  \phi^{\dagger}_{n}(\bm{r}) \left\{ \sqrt{2\pi}Y_{2} \otimes 
  \sigma_{1}\right\}_{1} \phi_{m}(\bm{r}) \cdot \langle m | \gamma^{0} 
  \bm{\tau} | n \rangle \mathcal{R}_{2}(E_n,E_m)\right], \\
\mathcal{J}^{B}_{2}(Q^{2}) &= \frac{N_{c} M_{B}}{E_{B}} 
  \int d^{3} r j_{2}(Q |\bm{r}|) \left[ \sum_{n_{0} \ne
  \mathrm{val} } \frac{N_{c}}{E_{\mathrm{val}}-E_{n_{0}}} 
  \phi^{\dagger}_{\mathrm{val}}(\bm{r}) 
  \left\{ \sqrt{2\pi}Y_{2} \otimes \sigma_{1}\right\}_{1} \cdot \bm{\tau} 
  \phi_{n_{0}}(\bm{r}) \langle n_{0}| \gamma^{0} | \mathrm{val} \rangle 
  \right. \cr
& \left. \hspace{3.9cm}
  +N_{c} \sum_{n,m_{0}} \phi^{\dagger}_{n}(\bm{r}) 
  \left\{ \sqrt{2\pi}Y_{2} \otimes \sigma_{1}\right\}_{1} \cdot \bm{\tau} 
  \phi_{m_{0}}(\bm{r}) \langle m_{0}| \gamma^{0} | n \rangle 
  \mathcal{R}_{2}(E_n,E_{m_{0}}) \right].
\label{AxComp12}
\end{align}

$\mathcal{A}_0^{\prime B}(Q^2)$, $\cdots$, $\mathcal{J}_0'^B(Q^2)$ are
defined by 
\begin{align} 
\mathcal{A}^{\prime B}_{0}(Q^{2}) &= -\frac{4 N_{c} M^{2}_{B}}{Q^{2}}
  \frac{E_{B}-M_{B}}{E_{B}} \int d^{3} r j_{0}(Q |\bm{r}|) 
  \left[ \phi_{\mathrm{val}}(\bm{r}) \bm{\sigma} \cdot \bm{\tau} 
  \phi_{\mathrm{val}}(r) + \sum_{n} \phi^{\dagger}_{n}(\bm{r}) 
  \bm{\sigma} \cdot \bm{\tau} \phi_{n}(\bm{r}) \mathcal{R}_{1}(E_n) 
  \right] ,
\end{align}
\begin{align}
\mathcal{B}^{\prime B}_{0}(Q^{2}) &= -\frac{4 N_{c} M^{2}_{B}}{Q^{2}}
  \frac{E_{B}-M_{B}}{E_{B}} \int d^{3} r j_{0}(Q |\bm{r}|) \left[ \sum_{n \ne
  \mathrm{val} } \frac{1}{E_{\mathrm{val}}-E_{n}} 
  \phi^{\dagger}_{\mathrm{val}}(\bm{r}) \bm{\sigma} \phi_{n}(\bm{r}) 
  \cdot \langle n | \bm{\tau} | \mathrm{val} \rangle \right. \cr
& \left. \hspace{6.0cm} 
  -\frac{1}{2} \sum_{n,m} \phi^{\dagger}_{n}(\bm{r}) \bm{\sigma} 
  \phi_{m}(\bm{r}) \cdot \langle m | \bm{\tau} | n \rangle 
  \mathcal{R}_{5}(E_n,E_m) \right],\\ 
\mathcal{C}^{\prime B}_{0}(Q^{2}) &= -\frac{4 N_{c} M^{2}_{B}}{Q^{2}}
  \frac{E_{B}-M_{B}}{E_{B}} \int d^{3} r j_{0}(Q |\bm{r}|) \left[ \sum_{n_{0} \ne
  \mathrm{val} } \frac{1}{E_{\mathrm{val}}-E_{n_{0}}} 
  \phi^{\dagger}_{\mathrm{val}}(\bm{r}) \bm{\sigma} \cdot \bm{\tau} 
  \phi_{n_{0}}(\bm{r}) \langle n_{0} | \mathrm{val} \rangle \right. \cr
& \left. \hspace{6.0cm} 
  -\sum_{n,m_{0}} \phi^{\dagger}_{n}(\bm{r}) \bm{\sigma} \cdot \bm{\tau} 
  \phi_{m_{0}}(\bm{r}) \langle m_{0} | n \rangle 
  \mathcal{R}_{5}(E_n,E_{m_{0}}) \right], \\
\mathcal{D}^{\prime B}_{0}(Q^{2}) &= -\frac{4 N_{c} M^{2}_{B}}{Q^{2}}
  \frac{E_{B}-M_{B}}{E_{B}} \int d^{3} r j_{0}(Q |\bm{r}|) \left[ \sum_{n \ne
  \mathrm{val} } \frac{\mathrm{sgn}(E_{n})}{E_{\mathrm{val}}-E_{n}} 
  \phi^{\dagger}_{\mathrm{val}}(\bm{r}) (\bm{\sigma} \times \bm{\tau})
  \phi_{n}(\bm{r}) \cdot \langle n | \bm{\tau} | \mathrm{val} 
  \rangle \right. \cr
& \left. \hspace{6.0cm} 
  + \frac{1}{2} \sum_{n,m} \phi^{\dagger}_{n}(\bm{r}) \bm{\sigma} 
  \times \bm{\tau} \phi_{m}(\bm{r}) \cdot \langle m | \bm{\tau} 
  | n \rangle \mathcal{R}_{4}(E_n,E_m) \right],\\
\mathcal{H}^{\prime B}_{0}(Q^{2}) &= -\frac{4 N_{c} M^{2}_{B}}{Q^{2}}
  \frac{E_{B}-M_{B}}{E_{B}} \int d^{3} r j_{0}(Q |\bm{r}|) \left[ \sum_{n \ne
  \mathrm{val} } \frac{1}{E_{\mathrm{val}}-E_{n}} 
  \phi^{\dagger}_{\mathrm{val}}(\bm{r}) \bm{\sigma} \cdot \bm{\tau} 
  \langle \bm{r} | n \rangle \langle n | \gamma^{0}| \mathrm{val} 
  \rangle \right. \cr
& \left. \hspace{6.0cm} 
  + \frac{1}{2} \sum_{n,m} \phi^{\dagger}_{n}(\bm{r}) \bm{\sigma} \cdot 
  \bm{\tau} \phi_{m}(\bm{r}) \langle m | \gamma^{0} | n \rangle 
  \mathcal{R}_{2}(E_n,E_m) \right], \\
\mathcal{I}^{\prime B}_{0}(Q^{2}) &= -\frac{4 N_{c} M^{2}_{B}}{Q^{2}}
  \frac{E_{B}-M_{B}}{E_{B}} \int d^{3} r j_{0}(Q |\bm{r}|) \left[ \sum_{n \ne
  \mathrm{val} } \frac{1}{E_{\mathrm{val}}-E_{n}} 
  \phi^{\dagger}_{\mathrm{val}}(\bm{r}) 
  \bm{\sigma} \phi_{n}(\bm{r}) \cdot 
  \langle n | \gamma^{0} \bm{\tau} | \mathrm{val} \rangle \right. \cr
& \left. \hspace{6.0cm} +\frac{1}{2} \sum_{n,m} 
  \phi^{\dagger}_{n}(\bm{r}) \bm{\sigma} \phi_{m}(\bm{r}) \cdot 
  \langle m | \gamma^{0} \bm{\tau} | n \rangle 
  \mathcal{R}_{2}(E_n,E_m)\right],\\
\mathcal{J}^{\prime B}_{0}(Q^{2}) &= -\frac{4 N_{c} M^{2}_{B}}{Q^{2}}
  \frac{E_{B}-M_{B}}{E_{B}} \int d^{3} r j_{0}(Q |\bm{r}|) \left[ \sum_{n_{0} \ne
  \mathrm{val} } \frac{N_{c}}{E_{\mathrm{val}}-E_{n_{0}}} 
  \phi^{\dagger}_{\mathrm{val}}(\bm{r}) \bm{\sigma} \cdot \bm{\tau} 
  \phi_{n_{0}}(\bm{r}) \langle n_{0}| \gamma^{0} | \mathrm{val} 
  \rangle \right. \cr
& \left. \hspace{6.0cm}
  +N_{c} \sum_{n,m_{0}} \phi^{\dagger}_{n}(\bm{r}) \bm{\sigma} \cdot 
  \bm{\tau} \phi_{m_{0}}(\bm{r}) \langle m_{0}| \gamma^{0} | n \rangle 
  \mathcal{R}_{2}(E_n,E_{m_{0}}) \right]
\label{AxComp30}
\end{align}
and $\mathcal{A}_2'^B (Q^2)$, $\cdots$, $\mathcal{J}_2'^B (Q^2)$ are defined by
\begin{align}
\mathcal{A}^{\prime B}_{2}(Q^{2}) &= -\frac{4 N_{c} M^{2}_{B}}{Q^{2}} 
  \frac{2E_{B}+M_{B}}{E_{B}} \int d^{3} r j_{2}(Q |\bm{r}|) \left[ 
  \phi^{\dagger}_{\mathrm{val}}(\bm{r}) \left\{ \sqrt{2\pi}Y_{2} \otimes 
  \sigma_{1}\right\}_{1} \cdot \bm{\tau} \phi_{\mathrm{val}}(\bm{r}) 
  \right. \cr
& \left. \hspace{6.1cm} 
  +\sum_{n} \phi^{\dagger}_{n}(\bm{r}) \left\{ \sqrt{2\pi}Y_{2} 
  \otimes \sigma_{1}\right\}_{1} \cdot \bm{\tau}
  \phi_{n}(\bm{r}) \mathcal{R}_{1}(E_n) \right],\\
\mathcal{B}^{\prime B}_{2}(Q^{2}) &= -\frac{4 N_{c} M^{2}_{B}}{Q^{2}} 
  \frac{2E_{B}+M_{B}}{E_{B}} \int d^{3} r j_{2}(Q |\bm{r}|) \cr
& \hspace{3.8cm} \times \left[ \sum_{n \ne
  \mathrm{val} } \frac{1}{E_{\mathrm{val}}-E_{n}} 
  \phi^{\dagger}_{\mathrm{val}}(\bm{r}) \left\{ \sqrt{2\pi}Y_{2} \otimes 
  \sigma_{1}\right\}_{1} \phi_{n}(\bm{r}) \cdot 
  \langle n | \bm{\tau} | \mathrm{val} \rangle \right. \cr
& \left. \hspace{4.3cm} 
  -\frac{1}{2} \sum_{n,m} \phi^{\dagger}_{n}(\bm{r}) 
  \left\{ \sqrt{2\pi}Y_{2} \otimes \sigma_{1}\right\}_{1} 
  \phi_{m}(\bm{r}) \cdot \langle m | \bm{\tau} | n \rangle 
  \mathcal{R}_{5}(E_n,E_m) \right],
\end{align}
\begin{align}
\mathcal{C}^{\prime B}_{2}(Q^{2}) &= -\frac{4 N_{c} M^{2}_{B}}{Q^{2}} 
  \frac{2E_{B}+M_{B}}{E_{B}} \int d^{3} r j_{2}(Q |\bm{r}|) \cr
& \hspace{3.8cm} \times \left[ \sum_{n_{0} \ne \mathrm{val} } 
  \frac{1}{E_{\mathrm{val}}-E_{n_{0}}} 
  \phi^{\dagger}_{\mathrm{val}}(\bm{r}) \left\{ \sqrt{2\pi}Y_{2} \otimes 
  \sigma_{1}\right\}_{1} \cdot \bm{\tau} \phi_{n_{0}}(\bm{r}) 
  \langle n_{0} | \mathrm{val} \rangle \right. \cr
& \left. \hspace{4.3cm} -\sum_{n,m_{0}} \phi^{\dagger}_{n}(\bm{r}) 
  \left\{ \sqrt{2\pi}Y_{2} \otimes \sigma_{1}\right\}_{1} \cdot \bm{\tau} 
  \phi_{m_{0}}(\bm{r}) \langle m_{0} | n \rangle 
  \mathcal{R}_{5}(E_n,E_{m_{0}}) \right], \\
\mathcal{D}^{\prime B}_{2}(Q^{2}) &= -\frac{4 N_{c} M^{2}_{B}}{Q^{2}} 
  \frac{2E_{B}+M_{B}}{E_{B}} \int d^{3} r j_{2}(Q |\bm{r}|) \cr
& \hspace{3.8cm} \times \left[ \sum_{n \ne \mathrm{val} } 
  \frac{\mathrm{sgn}(E_{n})}{E_{\mathrm{val}}-E_{n}} 
  \phi^{\dagger}_{\mathrm{val}}(\bm{r}) \left\{ \sqrt{2\pi}Y_{2} \otimes 
  \sigma_{1}\right\}_{1} \times \bm{\tau} \phi_{n}(\bm{r}) \cdot 
  \langle n | \bm{\tau} | \mathrm{val} \rangle \right. \cr
& \hspace{4.3cm} \left. 
  + \frac{1}{2} \sum_{n,m} \phi^{\dagger}_{n}(\bm{r}) 
  \left\{ \sqrt{2\pi}Y_{2} \otimes \sigma_{1}\right\}_{1} 
  \times \bm{\tau} \phi_{m}(\bm{r}) \cdot \langle m | \bm{\tau} 
  | n \rangle \mathcal{R}_{4}(E_n,E_m) \right],\\
\mathcal{H}^{\prime B}_{2}(Q^{2}) &= -\frac{4 N_{c} M^{2}_{B}}{Q^{2}} 
  \frac{2E_{B}+M_{B}}{E_{B}} \int d^{3} r j_{2}(Q |\bm{r}|) \cr
& \hspace{3.8cm} \times \left[ \sum_{n \ne
  \mathrm{val} } \frac{1}{E_{\mathrm{val}}-E_{n}} 
  \phi^{\dagger}_{\mathrm{val}}(\bm{r}) \left\{ \sqrt{2\pi}Y_{2} \otimes 
  \sigma_{1}\right\}_{1} \cdot \bm{\tau} \langle \bm{r} | n \rangle 
  \langle n | \gamma^{0}| \mathrm{val} \rangle \right. \cr
& \left. \hspace{4.3cm} + \frac{1}{2} \sum_{n,m} 
  \phi^{\dagger}_{n}(\bm{r}) \left\{ \sqrt{2\pi}Y_{2} \otimes 
  \sigma_{1}\right\}_{1} \cdot \bm{\tau} \phi_{m}(\bm{r}) 
  \langle m | \gamma^{0} | n \rangle \mathcal{R}_{2}(E_n,E_m) 
  \right], \\
\mathcal{I}^{\prime B}_{2}(Q^{2}) &= -\frac{4 N_{c} M^{2}_{B}}{Q^{2}} 
  \frac{2E_{B}+M_{B}}{E_{B}} \int d^{3} r j_{2}(Q |\bm{r}|) \cr
& \hspace{3.8cm} \times \left[ \sum_{n \ne \mathrm{val} } 
  \frac{1}{E_{\mathrm{val}}-E_{n}} \phi^{\dagger}_{\mathrm{val}}(\bm{r}) 
  \left\{ \sqrt{2\pi}Y_{2} \otimes \sigma_{1}\right\}_{1} 
  \phi_{n}(\bm{r}) \cdot \langle n | \gamma^{0} \bm{\tau} | \mathrm{val} 
  \rangle \right. \cr
& \left. \hspace{4.3cm} +\frac{1}{2} \sum_{n,m} 
  \phi^{\dagger}_{n}(\bm{r}) \left\{ \sqrt{2\pi}Y_{2} \otimes 
  \sigma_{1}\right\}_{1} \phi_{m}(\bm{r}) \cdot \langle m | \gamma^{0} 
  \bm{\tau} | n \rangle \mathcal{R}_{2}(E_n,E_m)\right],\\
\mathcal{J}^{\prime B}_{2}(Q^{2}) &= -\frac{4 N_{c} M^{2}_{B}}{Q^{2}} 
  \frac{2E_{B}+M_{B}}{E_{B}} \int d^{3} r j_{2}(Q |\bm{r}|) \cr
& \hspace{3.8cm} \times \left[ \sum_{n_{0} \ne \mathrm{val} } 
  \frac{N_{c}}{E_{\mathrm{val}}-E_{n_{0}}} 
  \phi^{\dagger}_{\mathrm{val}}(\bm{r}) \left\{ \sqrt{2\pi}Y_{2} \otimes 
  \sigma_{1}\right\}_{1} \cdot \bm{\tau} 
  \phi_{n_{0}}(\bm{r}) \langle n_{0}| \gamma^{0} | \mathrm{val} \rangle 
  \right. \cr
& \left. \hspace{4.3cm} +N_{c} \sum_{n,m_{0}} \phi^{\dagger}_{n}(\bm{r}) 
  \left\{ \sqrt{2\pi}Y_{2} \otimes \sigma_{1}\right\}_{1} \cdot \bm{\tau} 
  \phi_{m_{0}}(\bm{r}) \langle m_{0}| \gamma^{0} | n \rangle 
  \mathcal{R}_{2}(E_n,E_{m_{0}}) \right].
\label{AxComp32}
\end{align}

\section{Matrix elements of the SU(3) Wigner ${D}$ function}\label{app:B} 
In the following we list the results of the matrix elements of the
relevant collective operators for the axial-vector form factors of 
the baryon decuplet in Table~\ref{tab:3} and \ref{tab:4}.
\begin{table}[ht]
\renewcommand{\arraystretch}{2.1}
  \caption{The matrix elements of the single and double 
  Wigner $D$ function operators.}
  \label{tab:3}
\begin{center}
\begin{tabular}{ c | c c c c } 
 \hline 
  \hline 
$J_{3}=3/2$ & $\Delta$ & $\Sigma^{*}$ & $\Xi^{*}$ & $\Omega$ \\  
 \hline
$\langle B_{{\cal{R}}} |D^{(8)}_{33} | B_{{\cal{R}}}\rangle$  
& $-\frac{1}{4} T_{3}$ & $-\frac{1}{4} T_{3}$
& $-\frac{1}{4} T_{3}$ & $-\frac{1}{4} T_{3}$ \\  
$\langle B_{{\cal{R}}} |D^{(8)}_{83} | B_{{\cal{R}}}\rangle$  
& $-\frac{\sqrt{3}}{8} Y$ & $-\frac{\sqrt{3}}{8} Y$
& $-\frac{\sqrt{3}}{8} Y$ & $-\frac{\sqrt{3}}{8} Y$ \\  
$\langle B_{{\cal{R}}} |D^{(8)}_{38} \hat{J}_{3} | B_{{\cal{R}}}\rangle$  
& $\frac{\sqrt{3}}{8} T_{3}$ & $\frac{\sqrt{3}}{8} T_{3}$
& $\frac{\sqrt{3}}{8} T_{3}$ & $\frac{\sqrt{3}}{8} T_{3}$ \\  
$\langle B_{{\cal{R}}} |D^{(8)}_{88} \hat{J}_{3} | B_{{\cal{R}}}\rangle$  
& $\frac{3}{16} Y$ & $\frac{3}{16} Y$
& $\frac{3}{16} Y$ & $\frac{3}{16} Y$ \\  
$\langle B_{{\cal{R}}} |d_{bc3} D^{(8)}_{3b} \hat{J}_{c} 
| B_{{\cal{R}}}\rangle$  
& $\frac{1}{8} T_{3}$ & $\frac{1}{8} T_{3}$
& $\frac{1}{8} T_{3}$ & $\frac{1}{8} T_{3}$ \\  
$\langle B_{{\cal{R}}} |d_{bc3} D^{(8)}_{8b} \hat{J}_{c} 
  | B_{{\cal{R}}}\rangle$  
& $\frac{\sqrt{3}}{16} Y$ & $\frac{\sqrt{3}}{16} Y$
& $\frac{\sqrt{3}}{16} Y$ & $\frac{\sqrt{3}}{16} Y$ \\  
 \hline 
 \hline
\end{tabular}
\hspace{1cm}
\begin{tabular}{ c | c c c c } 
 \hline 
  \hline 
$J_{3}=3/2$ & $\Delta$ & $\Sigma^{*}$ & $\Xi^{*}$ & $\Omega$ \\  
 \hline
$\langle B_{{\cal{R}}} |D^{(8)}_{83}D^{(8)}_{38} | B_{{\cal{R}}}\rangle$
& $-\frac{5}{84} T_{3}$ & $-\frac{1}{28} T_{3}$
& $-\frac{1}{84} T_{3}$ & $0$ \\  
$\langle B_{{\cal{R}}} |D^{(8)}_{83}D^{(8)}_{88} | B_{{\cal{R}}}\rangle$
& $\frac{\sqrt{3}}{56} $ & $\frac{\sqrt{3}}{84}$
& $-\frac{\sqrt{3}}{56}$ & $-\frac{\sqrt{3}}{14}$ \\  
$\langle B_{{\cal{R}}} |D^{(8)}_{88}D^{(8)}_{33} | B_{{\cal{R}}}\rangle$
& $-\frac{5}{84} T_{3}$ & $-\frac{1}{28} T_{3}$
& $-\frac{1}{84} T_{3}$ & $0$ \\  
$\langle B_{{\cal{R}}} |D^{(8)}_{88}D^{(8)}_{83} | B_{{\cal{R}}}\rangle$
& $\frac{\sqrt{3}}{56} $ & $\frac{\sqrt{3}}{84}$
& $-\frac{\sqrt{3}}{56}$ & $-\frac{\sqrt{3}}{14}$ \\  
$\langle B_{{\cal{R}}} |d_{bc3} D^{(8)}_{8c}D^{(8)}_{3b}| 
  B_{{\cal{R}}} \rangle$  
& $-\frac{11\sqrt{3}}{252} T_{3}$ & $-\frac{5\sqrt{3}}{84} T_{3}$ 
& $-\frac{19\sqrt{3}}{252} T_{3}$ & $0$  \\  
$\langle B_{{\cal{R}}} |d_{bc3} D^{(8)}_{8c}D^{(8)}_{8b}| 
  B_{{\cal{R}}} \rangle$  
& $\frac{5}{56} $ & $-\frac{1}{42}$ 
& $-\frac{5}{56}$ & $-\frac{3}{28}$  \\  
 \hline 
 \hline
\end{tabular}
\end{center}
\end{table}

\begin{table}[ht]
\renewcommand{\arraystretch}{2.2}
  \caption{The transition matrix elements of the single Wigner 
  $D$ function operators coming from the 27-plet and 35-plet 
  component of the baryon wavefunctions.}
  \label{tab:4}
\begin{center}
\begin{tabular}{ c | c c c c } 
 \hline 
  \hline 
$J_{3}=3/2$ & $\Delta$ & $\Sigma^{*}$ & $\Xi^{*}$ & $\Omega$ \\  
\hline
$\langle B_ {\bm{27}} |D^{(8)}_{33} | B_{\cal{R}} \rangle$
& $-\frac{1}{12} \sqrt{\frac{5}{6}} T_{3}$ & $-\frac{1}{8}T_{3}$
&  $-\frac{7}{12} \sqrt{\frac{1}{6}} T_{3}$ &  $0$  \\  
$\langle B_ {\bm{27}}  |D^{(8)}_{83} | B_{\cal{R}} \rangle$
& $\frac{1}{8} \sqrt{\frac{5}{2}} $ &  $\frac{1}{4} \sqrt{\frac{1}{3}} $
&$\frac{1}{8} \sqrt{\frac{1}{2}} $ & $0$  \\  
$\langle B_ {\bm{27}}  |D^{(8)}_{38}J_{3} | B_{\cal{R}}\rangle$  
& $-\frac{1}{8} \sqrt{\frac{5}{2}}T_{3} $ & $-\frac{3\sqrt{3}}{16} T_{3}$
& $-\frac{7}{8} \sqrt{\frac{1}{2}}T_{3} $ & $0$   \\  
$\langle B_ {\bm{27}}  |D^{(8)}_{88}J_{3} | B_{\cal{R}}\rangle$ 
& $\frac{3}{16} \sqrt{\frac{15}{2}} $ & $\frac{3}{8} $ 
& $\frac{3}{16} \sqrt{\frac{3}{2}} $ &$0$ \\  
$\langle B_ {\bm{27}}  |d_{ab3}D^{(8)}_{3a}J_{b} | B_{\cal{R}}\rangle$ 
& $-\frac{1}{24} \sqrt{\frac{5}{6}} T_{3} $ &$-\frac{1}{16} T_{3} $ 
& $-\frac{7}{24} \sqrt{\frac{1}{6}} T_{3} $ &$0$   \\   
$\langle B_ {\bm{27}}  |d_{ab3}D^{(8)}_{8a}J_{b} | B_{\cal{R}}\rangle$ 
& $\frac{1}{16} \sqrt{\frac{5}{2}} $ & $\frac{1}{8} \sqrt{\frac{1}{3}} $ 
&$\frac{1}{16} \sqrt{\frac{1}{2}} $ &$0$   \\   
 \hline 
 \hline
\end{tabular}
\quad
\begin{tabular}{ c | c  c c c  } 
 \hline 
  \hline 
$J_{3}=3/2$ & $\Delta$ & $\Sigma^{*}$ & $\Xi^{*}$ & $\Omega$ \\  
 \hline
$\langle B_ {\bm{35}} |D^{(8)}_{33} | B_{\cal{R}} \rangle$
& $-\frac{1}{20} \sqrt{\frac{1}{14}} T_{3}$ &
$-\frac{1}{8}\sqrt{\frac{1}{35}}T_{3}$&  $-\frac{1}{4}
 \sqrt{\frac{1}{70}} T_{3}$ &  $0$  \\    
$\langle B_ {\bm{35}}  |D^{(8)}_{83} | B_{\cal{R}} \rangle$
& $-\frac{1}{8}  \sqrt{\frac{3}{14}}$&$-\frac{1}{4}\sqrt{\frac{3}{35}}$
&$-\frac{3}{8}\sqrt{\frac{3}{70}}$&$-\frac{1}{4}\sqrt{\frac{3}{35}}$\\ 
$\langle B_ {\bm{35}}  |D^{(8)}_{38}J_{3} | B_{\cal{R}}\rangle$  
& $\frac{1}{8} \sqrt{\frac{3}{14}}T_{3} $ & $\frac{1}{16}
\sqrt{\frac{15}{7}} 
T_{3}$ & $\frac{1}{8}\sqrt{\frac{15}{14}} T_{3}$ & $0$   \\  
$\langle B_ {\bm{35}}  |D^{(8)}_{88}J_{3} | B_{\cal{R}}\rangle$ 
& $\frac{15}{16} \sqrt{\frac{1}{14}} $ & $\frac{3}{8}\sqrt{\frac{5}{7}} $ 
& $\frac{9}{16} \sqrt{\frac{5}{14}} $ & $\frac{3}{8} \sqrt{\frac{5}{7}}$ 
\\  
$\langle B_ {\bm{35}}  |d_{ab3}D^{(8)}_{3a}J_{b} | B_{\cal{R}}\rangle$ 
& $-\frac{1}{8}\sqrt{\frac{1}{14}}T_{3}$ & $-\frac{1}{16}
\sqrt{\frac{5}{7}}T_{3}$
& $-\frac{1}{8}\sqrt{\frac{5}{14}}T_{3}$ & $0$   \\   
$\langle B_ {\bm{35}}  |d_{ab3}D^{(8)}_{8a}J_{b} | B_{\cal{R}}\rangle$ 
& $-\frac{5}{16} \sqrt{\frac{3}{14}}$ & $-\frac{1}{8}\sqrt{\frac{15}{7}}$ 
&$-\frac{3}{16} \sqrt{\frac{15}{14}}$ & $-\frac{1}{8}\sqrt{\frac{15}{7}}$
\\   
 \hline 
 \hline
\end{tabular}
\end{center}
\end{table}


\end{document}